\documentclass[letterpaper,journal]{IEEEtran}
\usepackage{amsmath,amsfonts}
\usepackage{algorithm}
\usepackage{algorithmic}
\usepackage{array}
\usepackage[caption=false,font=normalsize,labelfont=sf,textfont=sf]{subfig}
\usepackage{textcomp}
\usepackage{stfloats}
\usepackage{url}
\usepackage{verbatim}
\usepackage{graphicx}
\usepackage{cite}
\usepackage{booktabs}
\usepackage{enumitem}
\usepackage{multirow}
\usepackage{pifont}
\usepackage{soul}
\usepackage{xcolor}
\usepackage{colortbl}
\usepackage{xspace}
\usepackage[hidelinks]{hyperref}
\usepackage{booktabs}
\usepackage{makecell}
\usepackage{multirow}
\usepackage{graphicx}
\usepackage{xcolor}

\definecolor{DBBest}{HTML}{B42318}
\definecolor{DBSecond}{HTML}{175CD3}

\makeatletter
\def\onedot{\futurelet\@let@token\@onedot}
\def\@onedot{\ifx\@let@token.\else.\null\fi\xspace}
\makeatother
\def\eg{\emph{e.g}\onedot}
\def\ie{\emph{i.e}\onedot}

\definecolor{lightblue}{RGB}{235,245,255}
\definecolor{lightorange}{RGB}{255,245,230}
\definecolor{lightgreen}{RGB}{235,255,235}
\definecolor{highlightgreen}{RGB}{210,255,210}

\begin{document}

\title{Mega-MMDF and DeepfakeBench-MM: Benchmarking Multimodal Deepfake Detection Across Diverse Forgery Pipelines}

\author{Kangran Zhao\textsuperscript{*},
        Xiaoyu Zhang\textsuperscript{*},
        Yupeng Chen\textsuperscript{*},
        Yize Chen,
        Weinan Guan,
        Bo Zheng,
        Yufan Liao,
        % Baicheng Chen,
        Chengzhe Sun,
        Soumyya Kanti Datta,
        Qingshan Liu,
        Siwei Lyu,
        and Baoyuan Wu\textsuperscript{\(\dagger\)}%
        
\thanks{\textsuperscript{*}Kangran Zhao (kangranzhao@link.cuhk.edu.cn), Xiaoyu Zhang (xiaoyuzhang2@link.cuhk.edu.cn), and Yupeng Chen (yupeng.chen@eng.ox.ac.uk) contributed equally to this work. \textsuperscript{\(\dagger\)}Corresponding author: Baoyuan Wu (e-mail: wubaoyuan@cuhk.edu.cn).}%
\thanks{Kangran Zhao, Xiaoyu Zhang, Yize Chen, Weinan Guan, Bo Zheng, Yufan Liao, and Baoyuan Wu are with the School of Artificial Intelligence, The Chinese University of Hong Kong, Shenzhen (CUHK(SZ)), Shenzhen, China. Kangran Zhao, Yize Chen, and Bo Zheng are also with Shenzhen Loop Area Institute (SLAI), Shenzhen, China.}%
\thanks{
Yupeng Chen is with the Department of Engineering Science, University of Oxford, Oxford, UK. His contribution to this work was made while he was with the School of Data Science, CUHK(SZ).}%
\thanks{Chengzhe Sun, Soumyya Kanti Datta, and Siwei Lyu are with the Department of Computer Science and Engineering, University at Buffalo, Buffalo, NY, USA.}%
\thanks{Qingshan Liu is with the School of Artificial Intelligence, Nanjing University of Posts and Telecommunications, Nanjing, China.}%
}

% Preprint header for the arXiv version; IEEEtran retains the page number on the right.
\markboth{P\MakeLowercase{reprint}}{P\MakeLowercase{reprint}}

\maketitle

\begin{abstract}
Audio-visual deepfakes can spread misinformation and enable impersonation, undermining trust in digital media. However, existing datasets provide limited coverage of forgery operations and their combinations, constraining detector development and systematic evaluation. We introduce Mega-MMDF, containing 100,000 real and 1.2 million forged clips across 22 pipelines spanning modality-specific manipulations, stacked forgeries, and joint audio-video generation. To assess its training value and enable fair comparisons, we develop DeepfakeBench-MM, which standardizes training and evaluation across 8 datasets and 10 detectors. Training-based detectors achieve their highest mean AUC when trained on Mega-MMDF, with gains of up to 0.0784 over the best alternative training dataset for each detector. Experiments on data scale and pipeline coverage further support the value of large, diverse training data. Cross-pipeline evaluations reveal bidirectional and asymmetric generalization. Training on composite forgeries does not necessarily enable detection of their constituent operations in isolation. Our analysis of altered inputs shows that modality masking can increase false alarms and missed detections, and temporal misalignment can raise mean predicted forgery probabilities. Together, these resources provide a foundation for understanding and improving multimodal deepfake detection, with publicly available code and dataset access subject to application and usage terms.
\end{abstract}
\begin{IEEEkeywords}
Multimodal deepfake detection, audio-visual deepfakes, deepfake benchmark, deepfake dataset.
\end{IEEEkeywords}

\section{Introduction}
\label{sec:introduction}

\IEEEPARstart{D}{eepfake} technologies, powered by advances in generative AI, enable the synthesis and manipulation of realistic human voices~\cite{XTTS, YourTTS, FreeVC}, facial images and videos~\cite{BlendFace, SDXL, PhotoMaker}, and audio-visual content~\cite{Diff2Lip, SadTalker, LTX-2}. Their misuse poses risks of misinformation, impersonation, and financial fraud, underscoring the need for effective multimodal deepfake detection (MM-DFD)\footnote{In this work, multimodal content refers specifically to audio-visual content.}. While unimodal detectors examine forgery artifacts in audio~\cite{AASIST, RawNet2, wav2vec2-AASIST, RWM} or visual content~\cite{FaceXray, LipForensics, FTCN, SBI, UCF, AltFreezing, LSDA, ForensicsAdapter} separately, multimodal approaches~\cite{MDS, AVFF, MRDF, FGMDF, SpeechForensics, LTI, MAVPL, AVH} can additionally exploit relationships between the two modalities.

Existing audio-visual deepfake datasets provide resources for detector training and evaluation~\cite{FakeAVCeleb, LAV-DF, PolyGlotFake, IDForge, DigiFakeAV&DigiShield}. In particular, the AVDeepfake1M series~\cite{AVDeepfake1M, AVDeepfake1M++} has expanded these resources to the million scale. However, existing datasets cover a limited range of forgery methods and pipelines, and many are also limited in scale. Specifically, a forgery method denotes a specific generation or manipulation algorithm, whereas a forgery pipeline specifies the selection and ordering of forgery operations, each implemented by a corresponding method. As audio-visual content can undergo individual operations or compositions of multiple operations, limited forgery pipeline coverage restricts the conditions represented in training and evaluation. These limitations motivate a large-scale dataset with broader coverage of forgery methods and pipelines.

To address this need, we introduce \textbf{Mega-MMDF}, a million-scale audio-visual deepfake dataset containing 100,000 real and 1.2 million forged clips. Its construction uses 10 audio forgery methods, 12 visual forgery methods, 6 audio-driven face reenactment methods, and 2 joint audio-video generation methods. We organize these components into pipelines with the aim of maintaining correspondence between speech content and visible articulation, applying audio-driven face reenactment where needed. The resulting 22 forgery pipelines cover manipulations of individual modalities, compositions of multiple operations, and joint audio-video generation. Beyond providing training data, pipeline identities and modality-specific authenticity annotations support systematic evaluation of detection coverage and cross-pipeline generalization.

To evaluate Mega-MMDF’s training value and characterize detector behavior, we need reproducible comparisons under consistent experimental settings. However, limited access to training implementations and differences in evaluation protocols in existing works make transparent and reproducible comparisons difficult. We therefore develop \textbf{DeepfakeBench-MM}, a unified and extensible benchmark supporting eight datasets and ten multimodal deepfake detectors. Its modular framework integrates detector implementations with standardized data processing, explicit training configurations, and shared evaluation protocols while accommodating detector-specific input requirements. Our evaluations based on DeepfakeBench-MM show that training on Mega-MMDF achieves the highest average clip-level AUC score over eight test datasets for all seven training-based detectors. Data-scale experiments show that larger subsets of Mega-MMDF generally improve detection performance. Pipeline-wise evaluations further show that training on individual pipelines does not provide consistently strong detection across forgery conditions, highlighting the value of the diverse pipeline coverage provided by Mega-MMDF.

Within DeepfakeBench-MM, Mega-MMDF also supports systematic analysis of how detectors generalize across forgery pipelines and respond to changes in input conditions. Cross-pipeline evaluations reveal bidirectional and asymmetric transfer, showing that training on a composite pipeline does not necessarily enable detection of its constituent operations in isolation. Zero-filling a missing modality can increase false alarms on authentic clips and missed detections of forgeries in the retained modality. When both modalities remain available, disrupting their temporal correspondence in authentic clips produces detector-dependent prediction changes. These findings motivate broader forgery coverage, reliable handling of missing modalities, and further investigation of how audio-visual correspondence should complement modality-specific authenticity evidence.

Our main contributions are threefold:
\begin{itemize}
    \item \textbf{Mega-MMDF}\footnote{\href{https://huggingface.co/datasets/SCLBD/Mega-MMDF}{https://huggingface.co/datasets/SCLBD/Mega-MMDF}}. We introduce a million-scale audio-visual deepfake dataset with 22 forgery pipelines combining diverse audio, visual, audio-driven reenactment, and joint generation methods. Its broad pipeline coverage, pipeline identities, and modality-specific authenticity annotations provide a foundation for detector training and systematic analysis across forgery pipelines and different input conditions.
    \item \textbf{DeepfakeBench-MM}\footnote{\href{https://github.com/SCLBD/DeepfakeBench-MM}{https://github.com/SCLBD/DeepfakeBench-MM}}. We develop a unified and extensible benchmark supporting 8 datasets and 10 detectors. Through standardized data processing, explicit training configurations, and shared evaluation protocols, it enables reproducible assessment of Mega-MMDF’s training value and systematic investigation of detector behavior.
    \item \textbf{Systematic Analysis of Multimodal Deepfake Detection}. Using Mega-MMDF within DeepfakeBench-MM, we demonstrate training benefits of Mega-MMDF and reveal bidirectional and asymmetric cross-pipeline generalization, showing that training on composite pipelines does not necessarily enable detection of their constituent operations in isolation. We further identify missing-modality failures and detector-dependent responses to temporal misalignment, motivating more reliable integration of modality-specific and cross-modal evidence.
\end{itemize}

\section{Related Work}
\label{sec:related_work}

\subsection{Multimodal Deepfake Detectors}
Unimodal deepfake detectors identify synthesis artifacts in either audio~\cite{RawNet2,AASIST,wav2vec2-AASIST,RWM} or visual content~\cite{FaceXray,MultiAttention,LipForensics,FTCN,SBI,UCF,AltFreezing,TALL,LSDA,ForensicsAdapter}. In contrast, multimodal detectors use both modalities to capture intra-modal artifacts and cross-modal relations. Existing approaches can be broadly grouped into four categories.

\paragraph{Aggregation of modality-specific representations and predictions} These methods combine information extracted from audio and video for deepfake detection. AVTS~\cite{AVTS} concatenates modality-specific temporal embeddings for Transformer-based classification, while MRDF~\cite{MRDF} introduces cross- and within-modality regularization. MAVPL~\cite{MAVPL} uses learnable prompt tokens to adapt frozen CLIP~\cite{CLIP} and Whisper~\cite{Whisper} encoders before aggregating their temporally aligned representations. KLASSify~\cite{KLASSify} calibrates modality-specific scores and combines them through max-out fusion.

\paragraph{Explicit cross-modal relation modeling} 
These methods examine audio-visual correspondence at different temporal and semantic levels. MDS~\cite{MDS} measures modality dissonance, AVAD~\cite{AVAD} models synchronization anomalies, and LTI~\cite{LTI} examines attention-weighted local temporal distances. Other approaches focus on voice-face homogeneity~\cite{VFD}, audio-visual speech agreement~\cite{SpeechForensics}, and frame-level alignment~\cite{AVH}. 

\paragraph{Cross-modal representation fusion} 
Another direction uses information from one modality to update, predict, or reconstruct representations of the other. JAV-DFD~\cite{JAV-DFD} constructs a synchronization stream through multi-level connections, while AVoiD-DF~\cite{AVoiD-DF} and AVA-CL~\cite{AVA-CL} employ bidirectional cross-attention. FGMDF~\cite{FGMDF} models neighboring audio-visual segments using a heterogeneous graph. HOLA~\cite{HOLA} and DigiShield~\cite{DigiFakeAV&DigiShield} explore hierarchical contextual aggregation and spatiotemporal attention, respectively. FRADE~\cite{FRADE} distills modality-agnostic audio information into the visual stream, while MCL~\cite{MCL} combines attention-based feature composition, contrastive learning, and frame-to-video distillation. In addition, AVFF~\cite{AVFF}, inspired by CAV-MAE~\cite{CAV-MAE}, learns audio-visual representations through complementary masking and cross-modal reconstruction. PVASS-MDD~\cite{PVASS-MDD} similarly uses cross-modal prediction and reconstruction, while AVAM~\cite{AVAM} focuses on asynchrony-robust alignment and reconstruction.

\paragraph{Explicit forensic-prior modeling}
These methods incorporate semantic, behavioral, or generator-related evidence based on the properties of audio-visual synthesis. Specifically, CCFD~\cite{CCFD} compares transcripts from audio and visual speech recognition to identify content inconsistencies. PIA~\cite{PIA} analyzes phoneme-conditioned articulation and temporal identity drift, while FauForensics~\cite{FauForensics} exploits facial-action-unit dynamics. To exploit generator-related evidence, X-AVDT~\cite{MMDF&X-AVDT} uses an inverted audio-conditioned diffusion model to extract reconstruction discrepancies and internal cross-attention patterns. In addition, EMO-Boost~\cite{EMOBoost} models intra- and inter-modal emotional dynamics and integrates them with an existing detector.

\begin{table*}[t]
    \centering
    \caption{Comparison of deepfake datasets.}
    \label{tab:dataset-comparison}
    \begingroup
    \scriptsize
    \setlength{\tabcolsep}{2.4pt}
    \renewcommand{\arraystretch}{1.10}
    \resizebox{\textwidth}{!}{%
    \begin{tabular}{@{}llcrrc*{8}{c}cc@{}}
    \toprule
    \multirow[c]{2}{*}{Dataset}
    & \multirow[c]{2}{*}{\shortstack{Publication\\\& Year}}
    & \multirow[c]{2}{*}{Modality}
    & \multirow[c]{2}{*}{\shortstack{Real\\Samples}}
    & \multirow[c]{2}{*}{\shortstack{Fake\\Samples}}
    & \multirow[c]{2}{*}{\shortstack{\# Forgery\\Methods}}
    & \multicolumn{8}{c}{Forgery Types}
    & \multirow[c]{2}{*}{\shortstack{Stacked\\Forgery}}
    & \multirow[c]{2}{*}{\shortstack{Latest\\Method Year}} \\
    \cmidrule(lr){7-14}
    & & & & & & TTS & VC & PF & FS & EFS & FE & FR & T2AV & & \\
    \midrule
    
    % Audio manipulation
    \rowcolor{lightblue} ASVspoof2019LA~\cite{ASVspoof2019} & CSL20 & A & 12,483 & 108,978 & 17 & 10 & 7 & - & - & - & - & - & - & \ding{51} & 19 \\
    \rowcolor{lightblue} PartialSpoof~\cite{PartialSpoof} & TASLP23 & A & 12,483 & 108,978 & 17 & - & - & 17 & - & - & - & - & - & \ding{51} & 19 \\
    \rowcolor{lightblue} ASVspoof2021LA~\cite{ASVspoof2021} & TASLP23 & A & 16,492\rlap{\textsuperscript{*}} & 148,148 & 13 & 7 & 6 & - & - & - & - & - & - & \ding{51} & 19 \\
    \rowcolor{lightblue} CD-ADD~\cite{CD-ADD} & EMNLP24 & A & 25,111 & 120,459 & 5 & 5 & - & - & - & - & - & - & - & \ding{55} & 24 \\ 
    \rowcolor{lightblue} ASVspoof5 (Track 1)~\cite{ASVspoof5} & CSL26 & A & 188,819 & 815,262 & 24 & 19 & 5 & - & - & - & - & - & - & \ding{51} & 24 \\
    \rowcolor{lightblue} VoiceWukong~\cite{VoiceWukong} & USENIX Sec.25 & A & 413,400 & 413,400 & 34 & 29 & 5 & - & - & - & - & - & - & \ding{55} & 24 \\
    
    \midrule
    % Visual manipulation
    \rowcolor{lightorange}
    FaceForensics++~\cite{FaceForensics++}
    & ICCV19 & V & 1,000 & 4,000 & 4
    & - & - & - & 2 & - & - & 2 & -
    & \ding{55} & 19 \\
    
    \rowcolor{lightorange}
    Celeb-DF~\cite{Celeb-DF}
    & CVPR20 & V & 590 & 5,639 & 1
    & - & - & - & 1 & - & - & - & -
    & \ding{55} & 20 \\
    
    \rowcolor{lightorange}
    DeeperForensics-1.0~\cite{DeeperForensics}
    & CVPR20 & V
    & 48,475\rlap{\textsuperscript{*}}
    & 11,000\rlap{\textsuperscript{*}} & 1
    & - & - & - & 1 & - & - & - & -
    & \ding{55} & 20 \\
    
    \rowcolor{lightorange}
    ForgeryNet~\cite{ForgeryNet}
    & CVPR21 & V & 99,630 & 121,617 & 8
    & - & - & - & 5 & - & 1 & 3 & -
    & \ding{51} & 21 \\
    
    \rowcolor{lightorange}
    KoDF~\cite{KoDF}
    & ICCV21 & V & 62,166 & 175,776 & 6
    & - & - & - & 3 & - & - & 3 & -
    & \ding{55} & 20 \\
    
    \rowcolor{lightorange}
    FFIW-10K~\cite{FFIW}
    & CVPR21 & V
    & 9,988\rlap{\textsuperscript{*}}
    & 9,988\rlap{\textsuperscript{*}} & 3
    & - & - & - & 3 & - & - & - & -
    & \ding{55} & 20 \\
    
    \rowcolor{lightorange}
    DF-Platter~\cite{DF-Platter}
    & CVPR23 & V & 764 & 132,496 & 3
    & - & - & - & 3 & - & - & 1 & -
    & \ding{55} & 21 \\
    
    \rowcolor{lightorange}
    DF40~\cite{DF40}
    & NeurIPS24 D\&B & V & 1,890 & 30,257\rlap{\textsuperscript{*}} & 40
    & - & - & - & 10 & 12 & 5 & 13 & -
    & \ding{55} & 24 \\
    
    \rowcolor{lightorange}
    TalkingHeadBench~\cite{TalkingHeadBench}
    & WACV26 & V & 2,312 & 2,994 & 7
    & - & - & - & - & 1 & - & 7 & -
    & \ding{55} & 25 \\
    
    \rowcolor{lightorange}
    Celeb-DF++~\cite{Celeb-DF++}
    & arXiv25 & V & 590 & 53,196 & 22
    & - & - & - & 8 & - & - & 14 & -
    & \ding{55} & 25 \\
    
    \rowcolor{lightorange}
    MMDF~\cite{MMDF&X-AVDT}
    & CVPR26 & V & 14,400 & 14,400 & 6
    & - & - & - & 1 & - & - & 5 & -
    & \ding{55} & 25 \\
    
    \midrule
    % Audio-visual manipulation
    \rowcolor{lightgreen}
    DFDC~\cite{DFDC}
    & arXiv20 & AV & 23,654 & 104,500 & 7
    & - & 1 & - & 6 & - & - & 2 & -
    & \ding{55} & 20 \\
    
    \rowcolor{lightgreen}
    FakeAVCeleb~\cite{FakeAVCeleb}
    & NeurIPS21 D\&B & AV & 500
    & 21,044\rlap{\textsuperscript{*}} & 4
    & 1 & - & - & 2 & - & - & 1 & -
    & \ding{51} & 20 \\
    
    \rowcolor{lightgreen}
    LAV-DF~\cite{LAV-DF}
    & CVIU23 & AV & 36,431 & 99,873 & 2
    & - & - & 1 & - & - & - & 1 & -
    & \ding{55} & 20 \\
    
    \rowcolor{lightgreen}
    AVDeepfake1M~\cite{AVDeepfake1M}
    & MM24 & AV & 286,721 & 860,039 & 3
    & - & - & 2 & - & - & - & 1 & -
    & \ding{55} & 23 \\
    
    \rowcolor{lightgreen}
    IDForge~\cite{IDForge}
    & MM24 & AV
    & 80,000\rlap{\textsuperscript{*}}
    & 170,000\rlap{\textsuperscript{*}} & 6
    & 1 & 1 & - & 3 & - & - & 1 & -
    & \ding{51} & 23 \\
    
    \rowcolor{lightgreen}
    PolyGlotFake~\cite{PolyGlotFake}
    & ICPR24 & AV & 766 & 14,472 & 8
    & 5 & 1 & - & - & - & - & 2 & -
    & \ding{51} & 24 \\
    
    \rowcolor{lightgreen}
    DeepSpeak~\cite{DeepSpeak}
    & CVPR26F & AV & 16,043 & 14,005 & 13
    & 3 & - & - & 3 & - & - & 7 & -
    & \ding{55} & 25 \\
    
    \rowcolor{lightgreen}
    AVDeepfake1M++~\cite{AVDeepfake1M++}
    & MM25 & AV & 627,936 & 1,423,218 & 7
    & - & - & 4 & - & - & - & 3 & -
    & \ding{55} & 25 \\
    
    \rowcolor{lightgreen}
    DigiFakeAV~\cite{DigiFakeAV&DigiShield}
    & ICASSP26 & AV & 10,000 & 50,000 & 6
    & 1 & - & - & - & - & - & 5 & -
    & \ding{55} & 25 \\
    
    \rowcolor{lightgreen}
    ArEnAV~\cite{ArEnAV}
    & ICLR26 & AV & 96,768 & 290,304 & 6
    & 3 & 1 & - & - & - & - & 2 & -
    & \ding{51} & 25 \\
    
    \midrule
    \rowcolor{highlightgreen}
    \textbf{Mega-MMDF}
    & \textbf{Ours}
    & \textbf{AV}
    & \textbf{100,000}
    & \textbf{1,200,000}
    & \textbf{30}
    & \textbf{5}
    & \textbf{3}
    & \textbf{2}
    & \textbf{4}
    & \textbf{5}
    & \textbf{3}
    & \textbf{6}
    & \textbf{2}
    & \textbf{\ding{51}}
    & \textbf{26} \\
    \bottomrule
    \end{tabular}%
    }
    
    \par\smallskip
    \begin{minipage}{\textwidth}
    \scriptsize
    \raggedright
    \textit{Notes.}
    A: Audio; V: Visual; AV: Audio-Visual. TTS: Text-to-Speech; VC: Voice Conversion; PF: Partial Fake; FS: Face Swapping; EFS: Entire Face Synthesis; FE: Face Editing; FR: Face Reenactment; T2AV: Text-to-Audio-Video.
    \textsuperscript{*} indicates an updated sample count relative to the originally reported figure. \ding{51}/\ding{55} indicate presence/absence of stacked forgery.
\end{minipage}

\endgroup
\end{table*}

\subsection{Deepfake Datasets}
Unimodal audio and visual datasets have supported deepfake detection research and the development of practical countermeasures for synthetic speech, voice conversion, and facial manipulations~\cite{ASVspoof2019,ASVspoof2021,PartialSpoof, FaceForensics++,Celeb-DF,DeeperForensics,DF40}. As generation technologies advance, audio-visual forgeries pose a growing threat by manipulating both spoken content and visual appearance. FakeAVCeleb~\cite{FakeAVCeleb}, an early dataset dedicated to multimodal deepfake detection, combines speech synthesis, face swapping, and lip synchronization to support research on manipulations affecting either or both modalities. Subsequent datasets have developed along two complementary directions: temporal forgery localization and broader coverage for deepfake classification.

\paragraph{Temporal localization}
LAV-DF~\cite{LAV-DF} introduces localized, content-driven manipulations that change the meaning of spoken content. Following this line of research, AV-Deepfake1M~\cite{AVDeepfake1M} substantially expands the dataset scale and uses a large language model to generate diverse semantic modifications through word replacements, insertions, and deletions. AV-Deepfake1M++~\cite{AVDeepfake1M++} further broadens this coverage by expanding the manipulation strategies and introducing audio-visual perturbations. In addition, ArEnAV~\cite{ArEnAV} applies similar operations to Arabic-English code-switching and dialectal variation.

\paragraph{Deepfake classification} Datasets for deepfake classification cover an expanding range of languages and generation techniques. PolyGlotFake~\cite{PolyGlotFake} and ILLUSION~\cite{ILLUSION} broaden linguistic coverage through multilingual synthesis and multimodal manipulations. MAVOS-DD~\cite{MAVOS-DD} evaluates generalization to unseen languages and generators, while ArEnAV~\cite{ArEnAV} addresses Arabic-English code-switching and dialectal variation, supporting both classification and temporal localization. Alongside these linguistic extensions, DigiFakeAV~\cite{DigiFakeAV&DigiShield} extends coverage beyond conventional face swapping to diffusion-based digital humans by combining synthesized speech with audio-driven video generation. Related resources also broaden the range of generated facial videos available for audio-visual detection: TalkingHeadBench~\cite{TalkingHeadBench} focuses on modern talking-head generators and changes in speaker identity and generation method, whereas MMDF~\cite{MMDF&X-AVDT} incorporates diverse facial generation methods based on adversarial learning, diffusion, and flow matching.

Despite these advances, existing datasets remain limited in forgery coverage. Our Mega-MMDF broadens this coverage through diverse methods and pipeline compositions. In addition, some construction pipelines in existing datasets alter speech without corresponding changes in visible articulation. We design our pipeline compositions to avoid such mismatches and preserve audio–lip correspondence.

\begin{figure*}
    \centering
    \includegraphics[width=\linewidth]{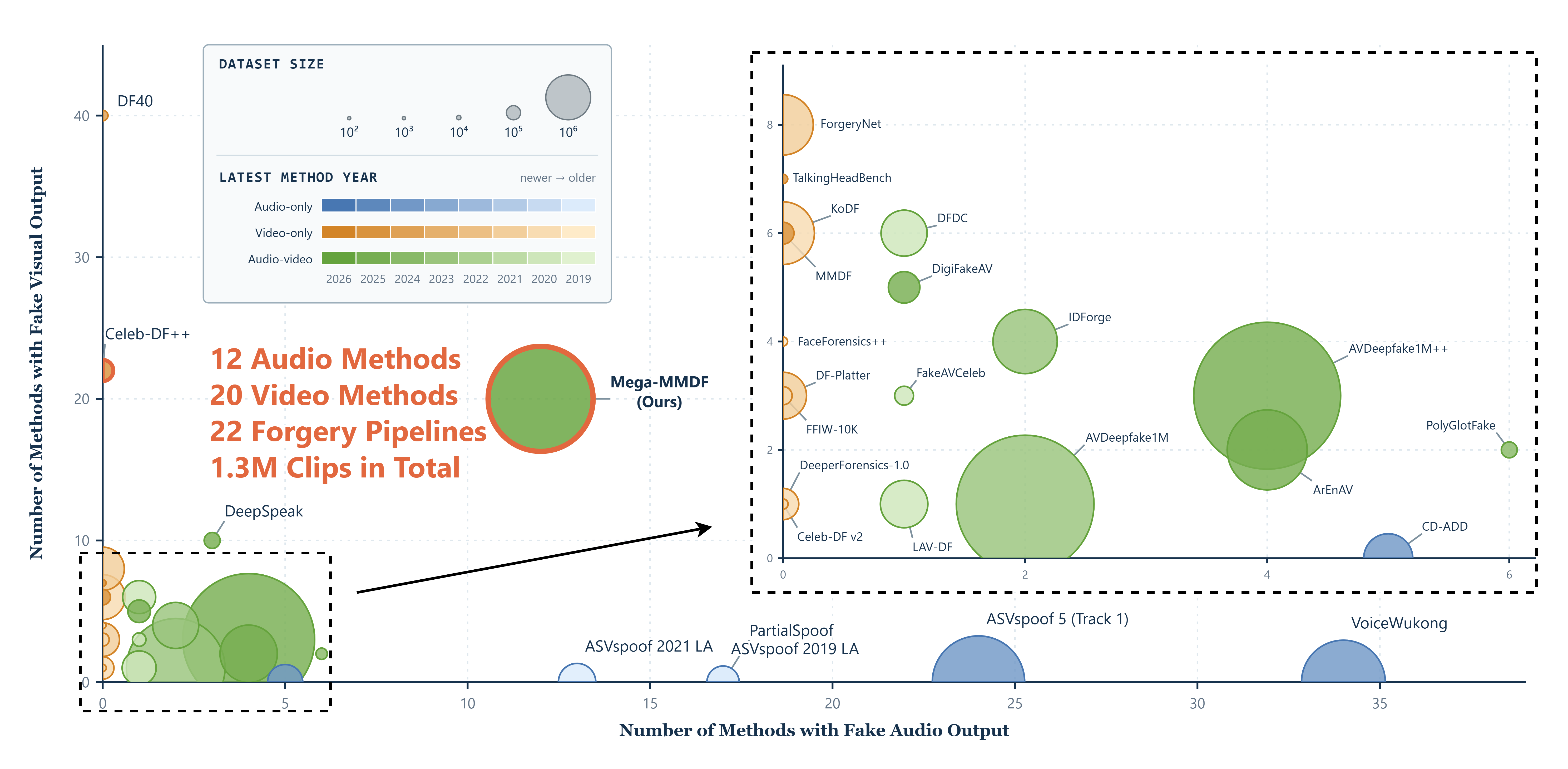}
    \caption{The horizontal and vertical axes indicate the numbers of methods with fake audio and fake visual output, respectively. The size of each circle represents the total dataset size. Circle color encodes the dataset modality, and the color gradient within each modality encodes the year of its latest forgery method. Mega-MMDF covers 12 methods with fake audio output and 20 with fake visual output. The T2AV methods are included in both counts, yielding 30 methods in total. Mega-MMDF contains 1.3 million audio-visual clips across 22 forgery pipelines.}
    \label{fig:dataset-comparison}
\end{figure*}

\subsection{Deepfake Detection Benchmarks}
Standardized benchmarks provide a common basis for reproducible deepfake detection research. In the visual domain, DeepfakeBench~\cite{DeepfakeBench} and ForensicHub~\cite{ForensicHub} provide unified implementations and evaluation procedures for deepfake detection and broader image forensic tasks. In the audio domain, the ASVspoof series~\cite{ASVspoof2019,ASVspoof2021,ASVspoof5} supports systematic comparisons through shared protocols, metrics, and baseline systems.

Audio-visual dataset releases also provide baseline comparisons and generalization evaluations~\cite{FakeAVCeleb,PolyGlotFake,MAVOS-DD, TalkingHeadBench,DigiFakeAV&DigiShield,MMDF&X-AVDT}. Beyond these dataset-specific evaluations, ILLUSION~\cite{ILLUSION} examines detection across languages and modalities, under real-world distortions and previously unseen attacks. AVFakeBench~\cite{AVFakeBench} and Omni-Fake~\cite{Omni-Fake} extend evaluation to multimodal detection and explanatory reasoning, with the former emphasizing the capabilities of audio-visual large multimodal models and the latter covering multiple media modalities and out-of-distribution generalization. Boldisor et al.~\cite{SSR-DFD} systematically analyze pretrained representations in terms of detection effectiveness, interpretability, and cross-modal complementarity.

Despite these advances, reproducible comparisons across heterogeneous audio-visual detectors still require consistent data handling, clearly specified training configurations, and shared evaluation procedures, while incomplete access to training implementations can limit retraining under common conditions. To address these needs, we develop DeepfakeBench-MM, an open and modular framework for transparent and extensible comparisons across methods and datasets.

\section{Mega-MMDF: A Diverse Million-Scale Multimodal Deepfake Dataset}
\label{datasets}

We introduce \textbf{Mega-MMDF}, a million-scale audio-visual deepfake dataset with diverse forgery methods and pipelines. It contains 100,000 real and 1.2 million forged clips, constructed using 30 generation methods spanning eight forgery types and organized into 22 forgery pipelines. Table~\ref{tab:dataset-comparison} compares its scale and forgery coverage with representative existing datasets. 

\subsection{Real Data Collection} 
\label{sec: subsec real data}

We collect real audio-visual clips from VoxCeleb2~\cite{VoxCeleb2}, MEAD~\cite{MEAD}, and CelebV-HQ~\cite{CelebV-HQ}, covering different recording environments, resolutions, and speaker demographics. To balance the representation of the selected demographic groups, we organize the clips into eight groups based on two gender categories and four skin-tone categories. These categories are used for dataset balancing and group-wise evaluation and do not exhaustively represent demographic diversity. The demographic labels were initially assigned using a multimodal large language model (MLLM) and subsequently inspected and corrected manually for apparent labeling errors. We then randomly select 12,500 clips from each group, obtaining 100,000 real clips with an average duration of 5.757 seconds. The demographic composition and detailed distributions are provided in Section~\ref{appendix:dataset_statistics}.

Within each demographic group, we organize the selected real clips into disjoint pairs, with each clip belonging to exactly one pair. Each pair provides two directional source–target assignments for constructing the forgery variants. We retain these pairs across pipelines and use them as the unit for dataset partitioning.

\subsection{Forgery Pipeline Construction}
\label{subsec:forgery_pipeline_construction}

\begin{table}[t]
    \centering
    \caption{Pipeline definitions in Mega-MMDF.}
    \label{tab:mega_mmdf_pipelines}
    \small
    \setlength{\tabcolsep}{4pt}
    \renewcommand{\arraystretch}{1.05}
    \begin{tabular*}{\columnwidth}{@{\extracolsep{\fill}}ccccc@{}}
        \toprule
        ID & Audio forgery & Visual forgery & \shortstack{Audio-driven FR} & Type \\
        \midrule
        P0  & Real & Real   & -- & RARV \\
        P1  & Real & FS     & -- & RAFV \\
        P2  & Real & --     & \ding{51} & RAFV \\
        P3  & VC   & Real   & -- & FARV \\
        P4  & VC   & FS     & -- & FAFV \\
        P5  & VC   & --     & \ding{51} & FAFV \\
        P6  & Real & EFS+FS & -- & RAFV \\
        P7  & Real & EFS    & \ding{51} & RAFV \\
        P8  & Real & FE     & \ding{51} & RAFV \\
        P9  & Real & FS     & \ding{51} & RAFV \\
        P10 & VC   & EFS+FS & -- & FAFV \\
        P11 & VC   & EFS    & \ding{51} & FAFV \\
        P12 & VC   & FE     & \ding{51} & FAFV \\
        P13 & VC   & FS     & \ding{51} & FAFV \\
        P14 & TTS  & --     & \ding{51} & FAFV \\
        P15 & TTS  & EFS    & \ding{51} & FAFV \\
        P16 & TTS  & FE     & \ding{51} & FAFV \\
        P17 & TTS  & FS     & \ding{51} & FAFV \\
        P18 & PF   & --     & \ding{51} & FAFV \\
        P19 & PF   & EFS    & \ding{51} & FAFV \\
        P20 & PF   & FE     & \ding{51} & FAFV \\
        P21 & PF   & FS     & \ding{51} & FAFV \\
        P22 & T2AV & T2AV   & -- & FAFV \\
        \bottomrule
    \end{tabular*}

    \vspace{3pt}
    \begin{minipage}{\columnwidth}
        \footnotesize
        \textit{Note:} The T2AV entries in the audio and visual columns refer to a single operation that jointly generates audio and video.
    \end{minipage}
\end{table}

Audio-visual deepfakes can be constructed by sequentially applying multiple generation or manipulation steps, resulting in stacked forgeries. For example, a synthesized face can be animated using generated speech, while a face-swapped video can undergo further reenactment to match a new audio track. We define a forgery method as a specific generation algorithm that performs an operation on given inputs to produce an output. A forgery pipeline specifies the selection and order of these operations, each implemented by a corresponding method.

We construct 22 forgery pipelines with the aim of producing audio-visual deepfakes without obvious mismatches between speech and lip movements. Depending on the pipeline, audio and visual content may be retained or manipulated, with audio-driven face reenactment applied when needed to establish the intended audio–lip correspondence. We also include joint audio-video generation, which produces both modalities directly from text prompts. Table~\ref{tab:mega_mmdf_pipelines} summarizes the pipeline compositions. Specifically, these pipelines draw on the following four components.

\textit{Audio forgery.} We employ 10 methods covering Text-to-Speech synthesis (TTS), Voice Conversion (VC), and Partial Fake (PF). In particular, TTS generates speech from text using a reference voice, whereas VC changes speaker characteristics while aiming to preserve the linguistic content. PF modifies selected portions of an utterance, such as individual words, while retaining the surrounding authentic speech. These methods comprise 5 TTS, 3 VC, and 2 PF methods.

\textit{Visual forgery.} We employ 12 methods covering Face Swapping (FS), Entire Face Synthesis (EFS), and Face Editing (FE). FS replaces facial identity while aiming to preserve the target video's expression, head pose, and surrounding content. EFS generates a synthetic facial image, while FE modifies selected facial attributes, such as glasses or hair color. We use 4 FS, 5 EFS, and 3 FE methods.

\textit{Audio-driven face reenactment.} We incorporate 6 audio-driven face reenactment (FR) methods to animate facial images or adapt facial motion in videos to the accompanying audio. This stage enables image-based synthesis and editing methods to produce audio-visual clips. It also supports pipelines in which generated speech requires corresponding changes in visual content.

\textit{Joint audio-video generation.} To cover recent joint generation systems, we incorporate 2 Text-to-Audio-Video (T2AV) methods that generate audio and video together from text prompts.

Together, these components support both individual forgery operations and stacked forgeries that combine multiple operations in sequence. For example, a synthesized face can be animated through FR, or face swapping can be followed by reenactment driven by generated speech. The appendix provides a complete list of generation methods (Table~\ref{tab:dataset_method}), construction diagrams for each pipeline (Figs.~\ref{fig:pipeline0-3}--\ref{fig:pipeline20-22}), and examples of generated clips (Figs.~\ref{fig:pipeline_samples_0_9}--\ref{fig:pipeline_samples_20_22}).

\subsection{Annotations and Dataset Partitioning}
\label{subsec:annotations_and_dataset_partitioning}
We assign each clip audio and visual authenticity labels, defining four combinations: real audio and real video (RARV), real audio and fake video (RAFV), fake audio and real video (FARV), and fake audio and fake video (FAFV). For clips generated using PF, the clip-level label indicates the presence of forgery in the corresponding modality, while temporal annotations specify the manipulated interval.

The accompanying metadata records the generation methods used for forged clips and the source-file references for real clips. Clip identifiers and the dataset organization also identify the forgery pipeline and demographic group. In addition, generation prompts are retained for T2AV clips. 

We partition the dataset at the pair level. Within each demographic group, pairs are assigned to the training, validation, and test sets in a 6:2:2 ratio. Each selected real clip is used without further subdivision and belongs to exactly one pair. Both real clips in a pair and all generated clips associated with that pair are kept in the same split across pipelines, preventing cross-split reuse of the selected source clips and their forgery variants.
The final dataset contains 1.3 million clips spanning 1,864 hours, with an average duration of 5.162 seconds. Detailed statistics by pipeline, demographic group, and partition are provided in Section~\ref{appendix:dataset_statistics}.

\subsection{Ethics and Responsible Release}
\label{subsec:ethics_and_responsible_release}
To respect the access and distribution terms of the source datasets, Mega-MMDF will provide source-file indices for real clips rather than redistribute the corresponding media. Users must obtain these clips through the official access channels of the source datasets and comply with their applicable terms. To access the released media, users will need to register with a valid email address and accept a research-use agreement that prohibits malicious impersonation and deceptive dissemination.

We will maintain a dedicated channel for privacy and rights-related removal requests. Upon verification, we will remove affected clips and associated derivatives from distribution channels under our control. Registered users will also be notified and requested to delete the corresponding local copies. We will version dataset releases and document changes to support the traceability of experimental results. Source-specific licensing information and further details of the release policy are provided in Sections~\ref{appendix:megammdf_ethics} and~\ref{appendix:license}.

\section{DeepfakeBench-MM: A Unified and Integrated Benchmark for MM-DFD}
\label{sec:benchmark}

\begin{table*}[t]
    \centering
    \caption{Clip-level AUC across eight test datasets. For each training-based detector, bold and underlining in the Average column indicate the highest and second-highest mean AUC across the five training datasets, respectively.}
    \label{tab:test-publication-full}
    \setlength{\tabcolsep}{2.5pt}
    \renewcommand{\arraystretch}{1.12}
    \scriptsize
    \resizebox{\textwidth}{!}{%
    \begin{tabular}{@{}ll*{8}{c}@{\hspace{7pt}}c@{}}
        \toprule
        \multirow{2}{*}{Detector} & \multirow{2}{*}{Training dataset} & \multicolumn{9}{c@{}}{Test datasets} \\
        \cmidrule(lr){3-11}
        & & FakeAVCeleb & LAV-DF & PolyGlotFake & IDForge & Mega-MMDF & TalkingHeadBench & DigiFakeAV & MMDF & Average \\
        \midrule
        \multirow{5}{*}{Baseline} & FakeAVCeleb~\cite{FakeAVCeleb} & 0.8260 & 0.6209 & 0.8094 & 0.6068 & 0.5449 & 0.4223 & 0.5366 & 0.4971 & 0.6080 \\
         & LAV-DF~\cite{LAV-DF} & 0.6968 & 0.9898 & 0.7958 & 0.5296 & 0.5829 & 0.5104 & 0.5518 & 0.5487 & \underline{0.6507} \\
         & PolyGlotFake~\cite{PolyGlotFake} & 0.5963 & 0.5891 & 0.9537 & 0.5285 & 0.6555 & 0.4844 & 0.5350 & 0.4456 & 0.5985 \\
         & IDForge~\cite{IDForge} & 0.6558 & 0.5253 & 0.7959 & 0.8358 & 0.6254 & 0.4645 & 0.5296 & 0.4552 & 0.6109 \\
         & Mega-MMDF (Ours) & 0.7715 & 0.5093 & 0.9100 & 0.4854 & 0.9958 & 0.8802 & 0.5360 & 0.4925 & \textbf{0.6976} \\
        \midrule
        \multirow{5}{*}{AVFF~\cite{AVFF}} & FakeAVCeleb~\cite{FakeAVCeleb} & 0.8015 & 0.6603 & 0.8478 & 0.6503 & 0.5559 & 0.4335 & 0.5490 & 0.5166 & 0.6269 \\
         & LAV-DF~\cite{LAV-DF} & 0.8076 & 0.9951 & 0.8797 & 0.5648 & 0.5760 & 0.5008 & 0.5454 & 0.5604 & \underline{0.6787} \\
         & PolyGlotFake~\cite{PolyGlotFake} & 0.6525 & 0.7120 & 0.9985 & 0.7047 & 0.7000 & 0.4077 & 0.5684 & 0.4780 & 0.6527 \\
         & IDForge~\cite{IDForge} & 0.7660 & 0.5279 & 0.8113 & 0.9242 & 0.5521 & 0.4128 & 0.5219 & 0.4863 & 0.6253 \\
         & Mega-MMDF (Ours) & 0.7017 & 0.6833 & 0.8843 & 0.5561 & 0.9819 & 0.9082 & 0.5058 & 0.4533 & \textbf{0.7093} \\
        \midrule
        \multirow{5}{*}{AVH~\cite{AVH}} & FakeAVCeleb~\cite{FakeAVCeleb} & 0.9844 & 0.9345 & 0.9665 & 0.8051 & 0.8113 & 0.7285 & 0.5971 & 0.8426 & 0.8338 \\
         & LAV-DF~\cite{LAV-DF} & 0.8661 & 0.9989 & 0.8585 & 0.6417 & 0.6396 & 0.6759 & 0.5645 & 0.7656 & 0.7514 \\
         & PolyGlotFake~\cite{PolyGlotFake} & 0.9413 & 0.7542 & 0.9978 & 0.8588 & 0.8315 & 0.7650 & 0.7333 & 0.7599 & 0.8302 \\
         & IDForge~\cite{IDForge} & 0.9684 & 0.7985 & 0.9811 & 0.9758 & 0.8194 & 0.8551 & 0.6635 & 0.8466 & \underline{0.8636} \\
         & Mega-MMDF (Ours) & 0.9597 & 0.8293 & 0.9835 & 0.8343 & 0.9802 & 0.8988 & 0.7080 & 0.7826 & \textbf{0.8721} \\
        \midrule
        \multirow{5}{*}{FauForensics~\cite{FauForensics}} & FakeAVCeleb~\cite{FakeAVCeleb} & 0.8673 & 0.6195 & 0.8896 & 0.7147 & 0.4977 & 0.5010 & 0.5694 & 0.5206 & 0.6475 \\
         & LAV-DF~\cite{LAV-DF} & 0.7482 & 0.9992 & 0.8516 & 0.4822 & 0.5719 & 0.6744 & 0.5774 & 0.6799 & \underline{0.6981} \\
         & PolyGlotFake~\cite{PolyGlotFake} & 0.8794 & 0.6804 & 0.9906 & 0.7705 & 0.6695 & 0.5055 & 0.5935 & 0.4791 & 0.6961 \\
         & IDForge~\cite{IDForge} & 0.8897 & 0.6240 & 0.9422 & 0.8502 & 0.6390 & 0.3002 & 0.5618 & 0.5082 & 0.6644 \\
         & Mega-MMDF (Ours) & 0.9220 & 0.7748 & 0.8595 & 0.6518 & 0.9886 & 0.6927 & 0.5255 & 0.4621 & \textbf{0.7346} \\
        \midrule
        \multirow{5}{*}{FGMDF~\cite{FGMDF}} & FakeAVCeleb~\cite{FakeAVCeleb} & 0.9158 & 0.6481 & 0.9153 & 0.6398 & 0.5722 & 0.4348 & 0.5703 & 0.5261 & 0.6528 \\
         & LAV-DF~\cite{LAV-DF} & 0.6944 & 0.9954 & 0.7999 & 0.4879 & 0.5704 & 0.5069 & 0.4769 & 0.5272 & 0.6324 \\
         & PolyGlotFake~\cite{PolyGlotFake} & 0.7485 & 0.5931 & 0.9988 & 0.6906 & 0.6785 & 0.3937 & 0.6320 & 0.5632 & 0.6623 \\
         & IDForge~\cite{IDForge} & 0.8381 & 0.5105 & 0.8776 & 0.9936 & 0.6763 & 0.4441 & 0.5426 & 0.4582 & \underline{0.6676} \\
         & Mega-MMDF (Ours) & 0.8019 & 0.4815 & 0.9781 & 0.6560 & 0.9939 & 0.7024 & 0.6158 & 0.4757 & \textbf{0.7132} \\
        \midrule
        \multirow{5}{*}{FRADE~\cite{FRADE}} & FakeAVCeleb~\cite{FakeAVCeleb} & 0.9341 & 0.5744 & 0.9094 & 0.7592 & 0.5969 & 0.4086 & 0.6140 & 0.5291 & 0.6657 \\
         & LAV-DF~\cite{LAV-DF} & 0.7415 & 0.9968 & 0.8916 & 0.5451 & 0.5785 & 0.5192 & 0.5730 & 0.5918 & 0.6797 \\
         & PolyGlotFake~\cite{PolyGlotFake} & 0.8224 & 0.7230 & 0.9810 & 0.7303 & 0.6177 & 0.3751 & 0.6475 & 0.5775 & \underline{0.6843} \\
         & IDForge~\cite{IDForge} & 0.8644 & 0.5125 & 0.8610 & 0.8583 & 0.5595 & 0.3882 & 0.5714 & 0.4842 & 0.6374 \\
         & Mega-MMDF (Ours) & 0.9256 & 0.6874 & 0.9305 & 0.6279 & 0.9963 & 0.9075 & 0.5669 & 0.4597 & \textbf{0.7627} \\
        \midrule
        \multirow{5}{*}{MCL~\cite{MCL}} & FakeAVCeleb~\cite{FakeAVCeleb} & 0.7808 & 0.9388 & 0.9059 & 0.5224 & 0.5869 & 0.5304 & 0.5132 & 0.5693 & 0.6685 \\
         & LAV-DF~\cite{LAV-DF} & 0.7736 & 0.9946 & 0.8736 & 0.5539 & 0.5730 & 0.5288 & 0.5490 & 0.5645 & 0.6764 \\
         & PolyGlotFake~\cite{PolyGlotFake} & 0.7168 & 0.6693 & 0.9961 & 0.6362 & 0.7968 & 0.4966 & 0.6480 & 0.5116 & \underline{0.6839} \\
         & IDForge~\cite{IDForge} & 0.6110 & 0.5583 & 0.7163 & 0.9128 & 0.6839 & 0.4694 & 0.5638 & 0.5023 & 0.6272 \\
         & Mega-MMDF (Ours) & 0.7282 & 0.6743 & 0.9283 & 0.5885 & 0.9743 & 0.6025 & 0.6065 & 0.4957 & \textbf{0.6998} \\
        \midrule
        SpeechForensics~\cite{SpeechForensics} & \multirow{3}{*}{Training-free} & 0.8790 & 0.7414 & 0.5568 & 0.6969 & 0.7385 & 0.8031 & 0.5553 & 0.8696 & 0.7301 \\
        Qwen2.5-Omni~\cite{Qwen2.5-Omni} &  & 0.6574 & 0.6268 & 0.7445 & 0.6416 & 0.6071 & 0.7001 & 0.6527 & 0.5770 & 0.6509 \\
        VideoLLaMA2.1~\cite{VideoLLaMA2.1} &  & 0.6645 & 0.5828 & 0.6042 & 0.5842 & 0.5686 & 0.5669 & 0.5807 & 0.5693 & 0.5901 \\
        \bottomrule
    \end{tabular}%
    }
\end{table*}

To assess the training value of Mega-MMDF and compare detector behavior under a common experimental setup, we develop \textbf{DeepfakeBench-MM}, a unified benchmark supporting 8 datasets and 10 detectors. Its modular codebase provides a common experimental infrastructure while accommodating detector-specific requirements and configurations.

\subsection{Modular Codebase}
\label{sec:benchmark:modular_codebase}
\begin{figure*}
    \centering
    \includegraphics[width=\linewidth]{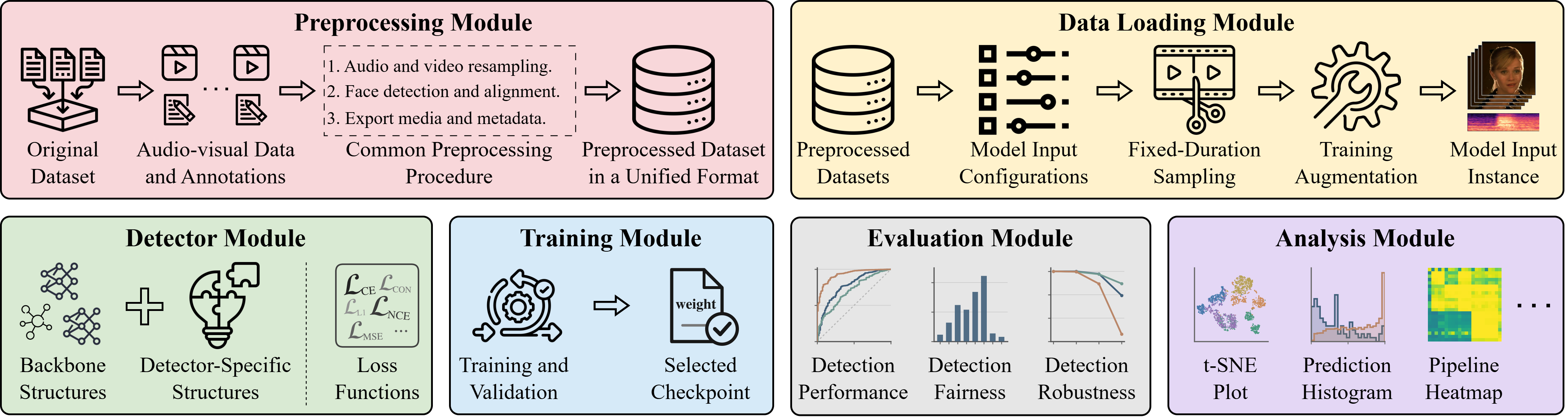}
    \caption{Modular Codebase of DeepfakeBench-MM}
    \label{fig:benchmark-codebase}
\end{figure*}

As illustrated in Fig.~\ref{fig:benchmark-codebase}, the codebase comprises six modules connected through shared interfaces. The preprocessing module standardizes data preprocessing and dataset organization, while the data-loading module constructs inputs according to detector-specific configurations. The detector module encapsulates model architectures and training objectives, and the training module manages optimization, validation, and checkpoint selection. The evaluation module implements common evaluation protocols, while the analysis module provides tools for inspecting predictions and learned representations. This organization supports extension through dataset-specific preprocessing adapters and loading configurations, detector implementations following the shared interface, and additional metrics and analysis tools within their respective modules. Implementation details and configuration options are provided in Appendix~\ref{appendix:benchmark}.

\subsection{Supported Datasets and Detectors}
\label{sec:benchmark:supported_datasets_and_detectors}

\paragraph{Datasets} DeepfakeBench-MM supports 8 multimodal deepfake datasets: FakeAVCeleb~\cite{FakeAVCeleb}, LAV-DF~\cite{LAV-DF}, PolyGlotFake~\cite{PolyGlotFake}, IDForge~\cite{IDForge}, TalkingHeadBench~\cite{TalkingHeadBench}, DigiFakeAV~\cite{DigiFakeAV&DigiShield}, MMDF~\cite{MMDF&X-AVDT}, and our Mega-MMDF. These datasets differ in storage formats, directory structures, media characteristics, and forgery annotations. 
To support consistent access across datasets, the preprocessing module standardizes their organization and annotations. We also exclude the initial interval corresponding to two video frames from both modalities to reduce potential reliance on the leading-silence artifacts identified in \cite{AVH}. For training and inference, the data-loading module samples fixed-length temporal windows from variable-length audio-visual clips. Each window forms an input instance, with its duration, video sampling rate, audio representation, and other settings determined by the requirements of each detector. Preprocessing and data-loading details are provided in Sections~\ref{appendix:benchmark_preprocessing} and~\ref{appendix:benchmark_data_loading}.

\paragraph{Detectors} The detector module integrates 10 multimodal detectors, divided into two categories. 
(1) \textit{Training-based detectors:} We implement a baseline comprising an audio backbone, a visual backbone, and an MLP classifier, together with 6 existing models: AVFF~\cite{AVFF}, AVH~\cite{AVH}, FRADE~\cite{FRADE},
FGMDF~\cite{FGMDF}, FauForensics~\cite{FauForensics}, and MCL~\cite{MCL}. For methods without complete public implementations, we request code or implementation guidance from the authors, or reproduce the models based on their published descriptions.
(2) \textit{Training-free detectors:} We include SpeechForensics~\cite{SpeechForensics}, Qwen2.5-Omni~\cite{Qwen2.5-Omni}, and VideoLLaMA2.1~\cite{VideoLLaMA2.1} to examine the applicability of pretrained audio-visual representations and multimodal large language models to deepfake detection without additional training on the benchmark datasets. The detector module connects these methods to the shared evaluation and analysis modules through standardized outputs. The details of the implementation are provided in Section~\ref{appendix:benchmark_models}.

\subsection{Unified Evaluation Protocols}
\label{sec:benchmark:unified_evaluation_protocols}

\paragraph{Training and test datasets} 
We use the training splits of FakeAVCeleb~\cite{FakeAVCeleb}, LAV-DF~\cite{LAV-DF}, PolyGlotFake~\cite{PolyGlotFake}, IDForge~\cite{IDForge}, and our Mega-MMDF as separate training sources, with their validation splits used for checkpoint selection. All eight datasets are used for testing.

\paragraph{Training settings}
For each training-based detector, we conduct separate training runs using each of the five training sources individually. We adopt a multi-dataset model-selection protocol, selecting the checkpoint with the highest mean instance-level AUC across the validation splits of the five datasets listed above. The selected checkpoint is used for subsequent evaluation and analysis. Unless otherwise specified, this criterion is applied consistently across detectors and training sources. Each training run uses a single NVIDIA H100 GPU. Detailed training configurations are provided in Section~\ref{appendix:benchmark_training} and Table~\ref{tab:training_optimizers}.

\paragraph{Evaluation and analysis} The evaluation module supports three complementary protocols.
(1) \textit{Detection performance:} We compute AUC, accuracy (ACC), equal error rate (EER), and average precision (AP) at both the instance and clip levels. Unless otherwise specified, we report clip-level AUC in this work. The complete AUC results are presented in Table~\ref{tab:test-publication-full}.
(2) \textit{Detection fairness:} We assess demographic disparities using the four fairness metrics adopted in~\cite{AIFace}, together with the standard deviation of group-wise AUC~\cite{Hu2024AttributeNeutral}.
(3) \textit{Robustness to post-processing:} We evaluate performance under practical post-processing perturbations, such as compression and random dropout. Detailed evaluation protocols are provided in Section~\ref{appendix:benchmark_evaluation}, and fairness and robustness results are presented in Section~\ref{sec:analysis:robustness_and_fairness}.
Beyond these evaluations, the analysis module uses standardized detector outputs to examine pipeline-specific performance, learned representations, and modality-related behavior. Details of the analysis tools are provided in Section~\ref{appendix:benchmark_analysis}.

\section{Analysis}
\label{sec:analysis}
\begin{table}[t]
    \centering
    \caption{Effect of the training dataset on overall detection performance. Each entry reports the average clip-level AUC over eight test datasets. Bold and underline indicate the training datasets yielding the highest and second-highest average AUCs for each detector, respectively. 
    Values in parentheses indicate the improvement over the second-best.}
    \label{tab:training-comparison}
    \setlength{\tabcolsep}{1.5pt}
    \renewcommand{\arraystretch}{1.15}
    \scriptsize
    \begin{tabular*}{\columnwidth}{@{\extracolsep{\fill}}lccccc@{}}
        \toprule
        \multirow{2}{*}[-2ex]{Detector}
        & \multicolumn{5}{c}{Training dataset} \\
        \cmidrule(lr){2-6}
        & \makecell{FakeAVCeleb\\\cite{FakeAVCeleb}}
        & \makecell{LAV-DF\\\cite{LAV-DF}}
        & \makecell{PolyGlotFake\\\cite{PolyGlotFake}}
        & \makecell{IDForge\\\cite{IDForge}}
        & \makecell{Mega-MMDF\\(Ours)} \\
        \midrule
        Baseline
        & 0.6080 & \underline{0.6507} & 0.5985 & 0.6109
        & \textbf{0.6976} (+0.0469) \\
        AVFF~\cite{AVFF}
        & 0.6269 & \underline{0.6787} & 0.6527 & 0.6253
        & \textbf{0.7093} (+0.0306) \\
        AVH~\cite{AVH}
        & 0.8338 & 0.7514 & 0.8302 & \underline{0.8636}
        & \textbf{0.8721} (+0.0085) \\
        FauForensics~\cite{FauForensics}
        & 0.6475 & \underline{0.6981} & 0.6961 & 0.6644
        & \textbf{0.7346} (+0.0365) \\
        FGMDF~\cite{FGMDF}
        & 0.6528 & 0.6324 & 0.6623 & \underline{0.6676}
        & \textbf{0.7132} (+0.0455) \\
        FRADE~\cite{FRADE}
        & 0.6657 & 0.6797 & \underline{0.6843} & 0.6374
        & \textbf{0.7627} (+0.0784) \\
        MCL~\cite{MCL}
        & 0.6685 & 0.6764 & \underline{0.6839} & 0.6272
        & \textbf{0.6998} (+0.0159) \\
        \bottomrule
    \end{tabular*}
\end{table}

We first examine the training value of Mega-MMDF through comparisons with existing datasets and analyses of data scale and forgery-pipeline coverage. Having established its training benefits, we use Mega-MMDF as a common training and testing ground to further examine detector capabilities and limitations, provide insights into the challenges facing current multimodal deepfake detectors and highlight directions for future research.

\subsection{Training Value of Mega-MMDF}
\label{sec:analysis:training_value_of_mega-mmdf}
We first evaluate the overall training benefits of Mega-MMDF compared with existing datasets. We then investigate the factors underlying these benefits by examining the effects of training data scale and the importance of forgery-pipeline coverage.

\ul{\textit{Q1: Does training on Mega-MMDF improve detection performance?}} 
To compare the effectiveness of different training datasets, we train each detector separately on FakeAVCeleb, LAV-DF, PolyGlotFake, IDForge, and Mega-MMDF. For each detector, we keep its training configuration consistent across training datasets, following the benchmark protocol described in Section~\ref{sec:benchmark}. We evaluate each resulting model on the same eight test datasets and report their mean clip-level AUCs. As shown in Table~\ref{tab:training-comparison}, training on Mega-MMDF achieves the highest mean AUC for all seven detectors. Compared with the strongest alternative training dataset for each detector, the improvements range from 0.0085 to 0.0784. These consistent improvements across different architectures demonstrate the overall effectiveness of Mega-MMDF as a training resource. We next examine the factors that may contribute to these benefits.

\begin{figure}[t]
    \centering
    \includegraphics[width=\columnwidth]{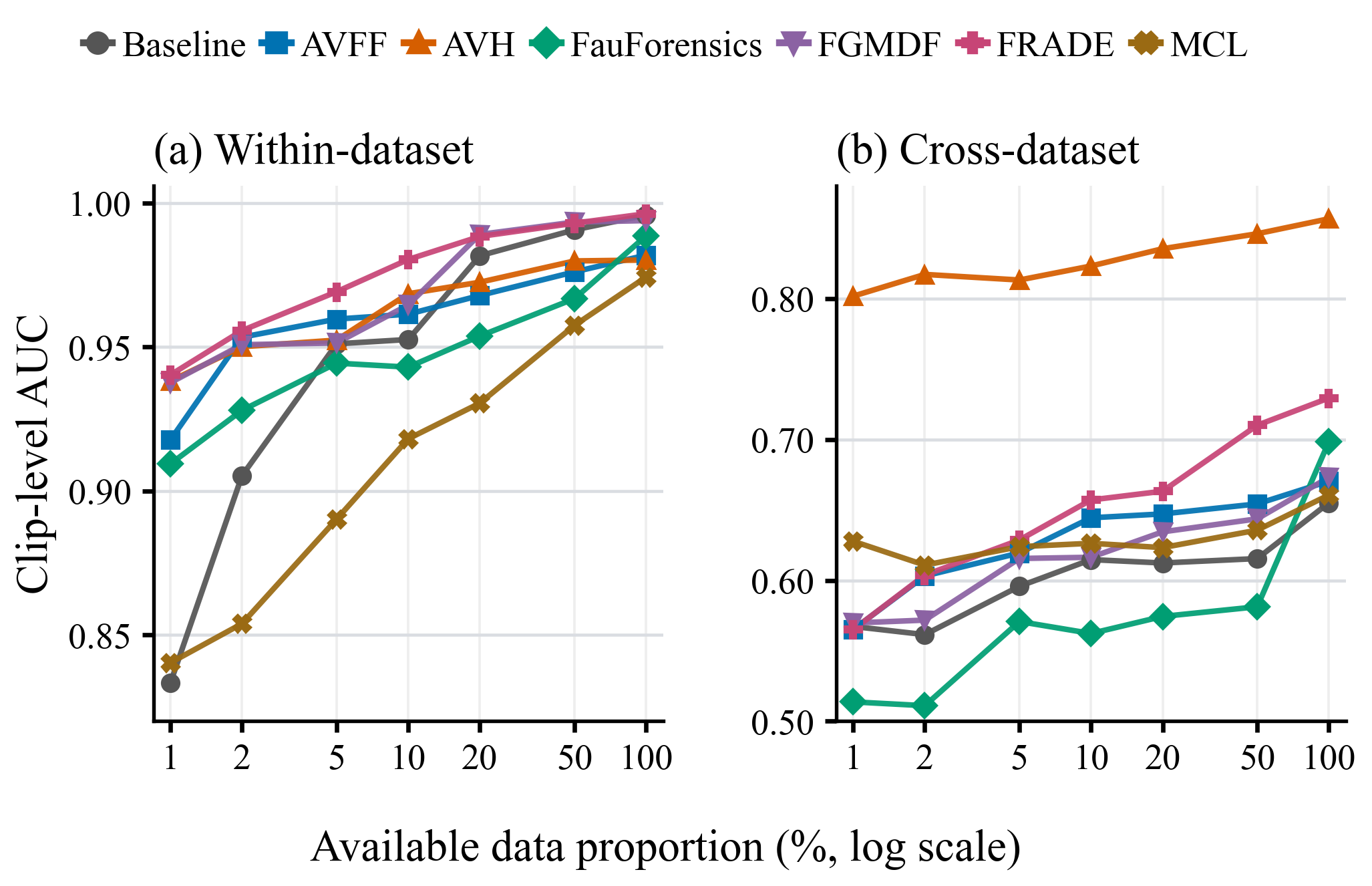}
    \caption{Effect of Mega-MMDF training data scale on (a) within-dataset detection and (b) cross-dataset performance.}
    \label{fig:data-scale}
\end{figure}

\begin{figure}[t]
    \centering
    \includegraphics[width=\columnwidth]{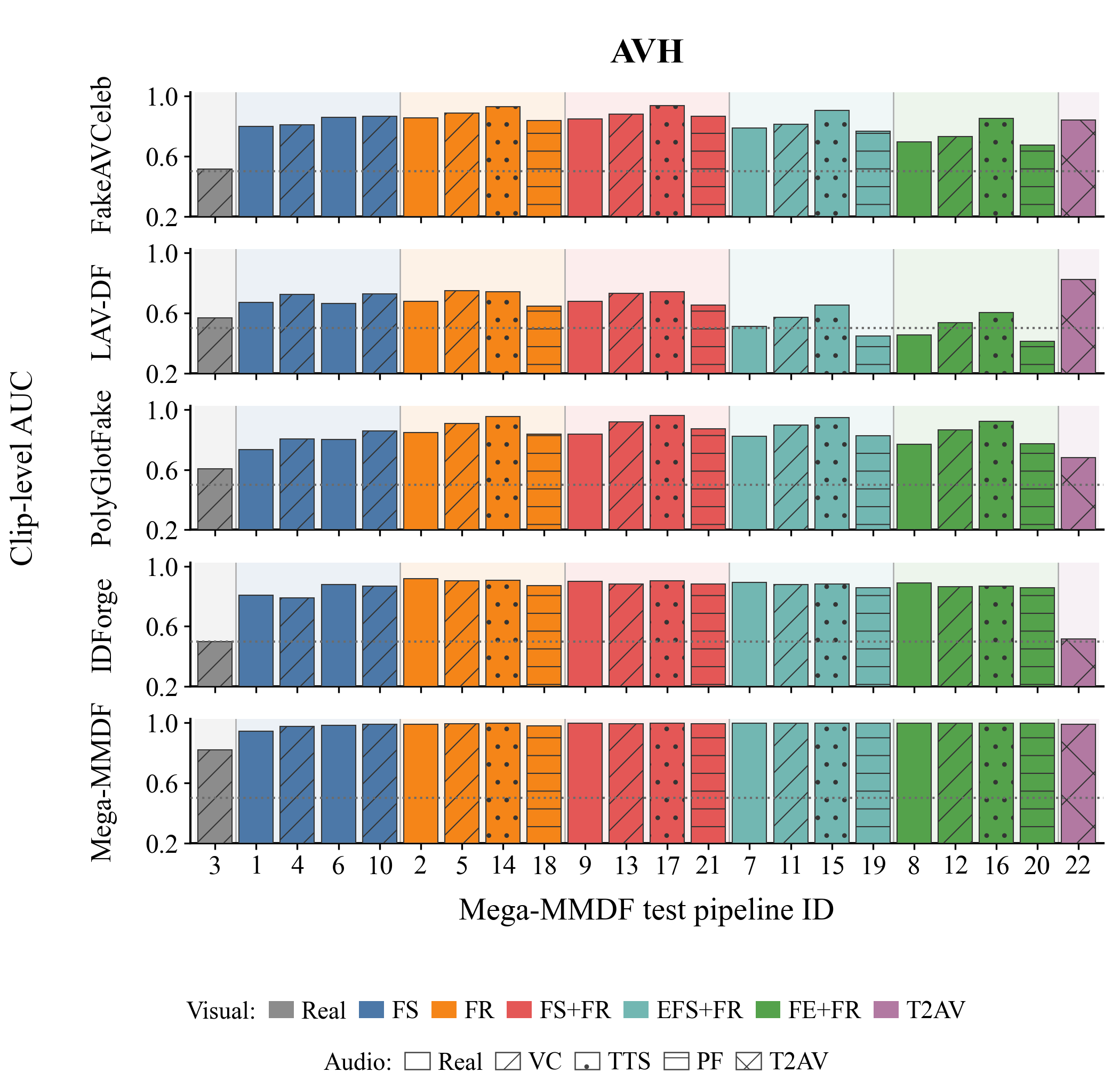}
    \caption{Pipeline-wise clip-level AUC of AVH trained on five datasets.}
    \label{fig:pipeline-coverage}
\end{figure}

\ul{\textit{Q2: How does training data scale affect detection performance?}} 
To examine the effect of dataset scale, we vary the amount of Mega-MMDF in both the training and validation splits. Within each demographic group, we sample 1\%, 2\%, 5\%, 10\%, 20\%, and 50\% of the pairs from the corresponding splits, retaining the real clips and associated forged clips of each selected pair. This pair-level sampling preserves coverage of forgery methods and pipelines across data scales. We train all seven detectors at each scale and compare them with the full-data setting. All resulting models are evaluated on the same full test sets, ensuring a consistent comparison across data scales. We report clip-level AUC on Mega-MMDF and mean clip-level AUC across the other seven datasets to measure within-dataset and cross-dataset performance, respectively. 
As shown in Fig.~\ref{fig:data-scale}, increasing the training data scale generally improves both within-dataset and cross-dataset performance. The full-data training achieves the highest performance in both settings for all training-based detectors. Increasing the training proportion from 1\% to 100\% improves within-dataset AUC by 0.0420–0.1625 and cross-dataset mean AUC by 0.0328–0.1846. These results indicate that the benefits of additional Mega-MMDF training data extend beyond its own test set, supporting the value of its large scale for more generalizable detection.

\begin{figure*}
    \includegraphics[width=\textwidth]{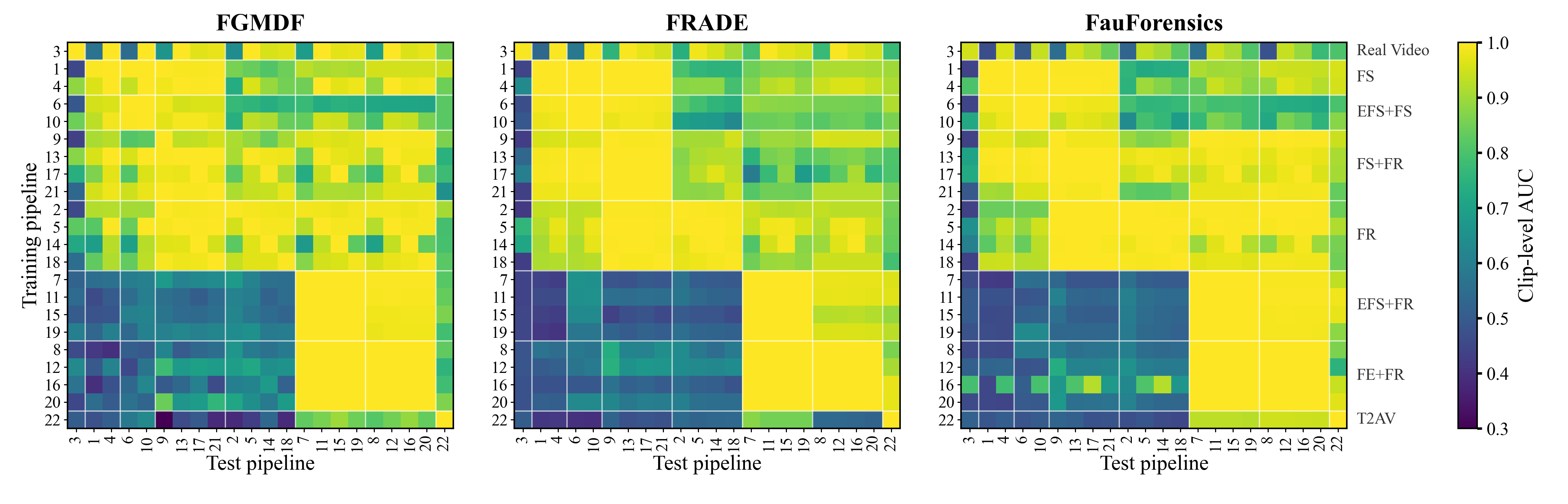}
    \caption{Cross-pipeline evaluation of FGMDF, FRADE, and FauForensics on Mega-MMDF. Rows denote training pipelines and columns denote test pipelines.}
    \label{fig:pipeline-transfer}
\end{figure*}

\ul{\textit{Q3: How important is forgery-pipeline coverage?}} 
Beyond data scale, we investigate whether limited forgery-pipeline coverage leaves gaps in detection performance. We first examine how detectors trained on existing datasets perform across the pipelines in Mega-MMDF. We then conduct a complementary experiment within Mega-MMDF, using a common source of real data to test whether training on a single pipeline provides broad cross-pipeline coverage.

In the first experiment, we evaluate the previously trained detectors separately on each Mega-MMDF pipeline. Each evaluation pairs the fake test clips from one pipeline with the same real test clips and reports clip-level AUC. Fig.~\ref{fig:pipeline-coverage} presents AVH as an example; results for the other detectors are provided in Figs.~\ref{fig:appendix-pipelines-01} and~\ref{fig:appendix-pipelines-02} in the appendix. Training on existing datasets produces uneven performance across pipelines, with clear weaknesses that are not apparent from the aggregate scores alone. For example, AVH trained on LAV-DF obtains AUCs of 0.510 and 0.456 on P7 and P8, and 0.446 and 0.411 on P19 and P20. Although IDForge includes additional face-swapping and voice-conversion pipelines, training on it only partially addresses these gaps, with weaknesses remaining on P1, P3, and P4. Performance on the emerging T2AV pipeline also varies considerably across training datasets. These results indicate that existing training resources provide incomplete coverage of the diverse forgery conditions represented in Mega-MMDF.

To examine whether training on an individual forgery pipeline can provide broad detection coverage while reducing differences between datasets, we next conduct a complementary experiment within Mega-MMDF using a common real-data source and consistent training settings. We train FGMDF, FRADE, and FauForensics separately on each forgery pipeline, combining its fake training clips with real training clips from P0. For each single-pipeline training run, we select one checkpoint with the highest clip-level AUC on an in-domain validation set comprising fake clips from the same pipeline and real clips from P0. The selected checkpoint is fixed for evaluation on all 22 fake pipelines, each paired with the same real test clips.
As shown in Fig.~\ref{fig:pipeline-transfer}, training on P13 yields the highest mean AUC across the 22 test pipelines for FGMDF and FauForensics, while training on P5 yields the highest mean for FRADE. However, these training pipelines still leave detection gaps: FGMDF trained on P13 obtains an AUC of 0.747 on P22, FRADE trained on P5 obtains 0.760 on P3, and FauForensics trained on P13 obtains 0.704 on P3. These results suggest that a single forgery pipeline is insufficient for broad cross-pipeline generalization and highlight the need for training datasets that encompass diverse forgery pipelines.

Together, the data-scale and pipeline-coverage experiments support the large-scale and diverse-pipeline design of Mega-MMDF. The cross-pipeline results also reveal structured generalization relationships among forgery pipelines, which we examine next.

\subsection{Generalization across Forgery Pipelines}
\label{sec:analysis:generalization_across_forgery_pipelines}

Building on the preceding cross-pipeline evaluation, we observe recurring generalization patterns across the three detectors. Beyond the overall coverage of a single pipeline, these results allow us to examine which forgery pipelines generalize to one another and whether such relationships hold in both directions. This motivates the following question.

\ul{\textit{Q4: What generalization relationships exist between forgery pipelines?}}
We analyze the complete cross-pipeline AUC results for FGMDF, FRADE, and FauForensics, complemented by selected instance-level t-SNE visualizations of FauForensics fusion features. For clarity, each visualization identifies the corresponding training pipeline. For each checkpoint, we compute the t-SNE embedding once and display the relevant categories without changing their projection coordinates.

\begin{itemize}[leftmargin=*,itemsep=3pt,topsep=3pt,parsep=0pt]
\item \textbf{Bidirectional generalization within two pipeline groups.}
The heatmaps show strong bidirectional generalization within two groups: (i) FS, EFS+FS, and FS+FR; and (ii) EFS+FR and FE+FR. For each detector, we average AUCs over the training-test pipeline pairs corresponding to each ordered pair of distinct visual forgery types. Across the three detectors, these averages range from 0.912 to 1.000 within the first group and from 0.959 to 1.000 within the second. Fig.~\ref{fig:transfer_mutual} shows the corresponding feature distributions for FauForensics models trained on individual pipelines from these groups. Each visual forgery type includes instances from its corresponding test pipelines. Within each group, different forgery types occupy neighboring or overlapping regions that are largely separated from real instances, although some limited local overlap remains. These qualitative observations are consistent with the strong within-group generalization shown by AUC.

\begin{figure}[t]
    \centering
    \includegraphics[width=\columnwidth]{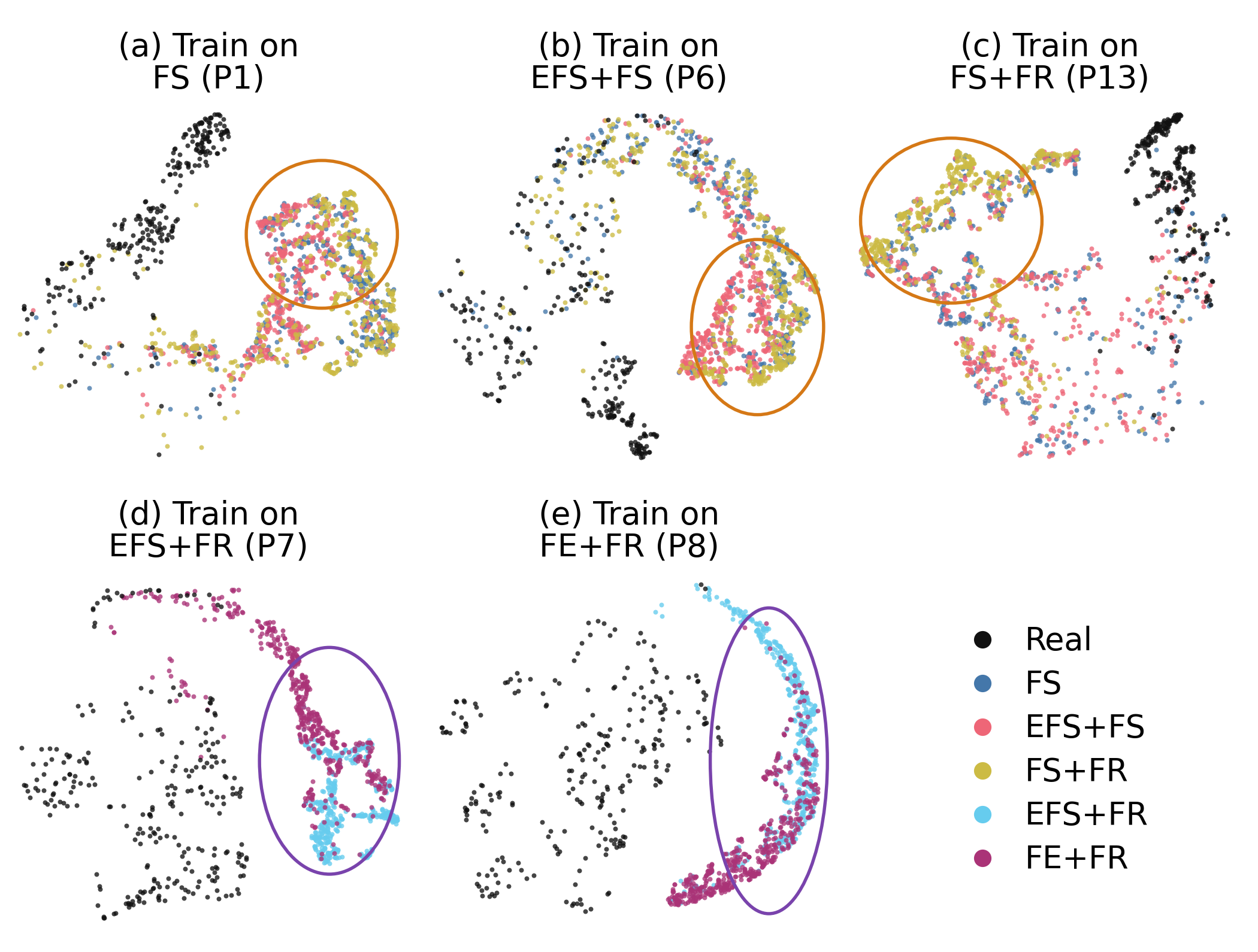}
    \caption{Distributions of fusion features extracted by the FauForensics detector~\cite{FauForensics}, associated with strong within-group generalization. (a)--(c): FS, EFS+FS, and FS+FR. (d)--(e): EFS+FR and FE+FR.}
    \label{fig:transfer_mutual}
\end{figure}

\begin{figure}[t]
    \centering
    \includegraphics[width=\columnwidth]{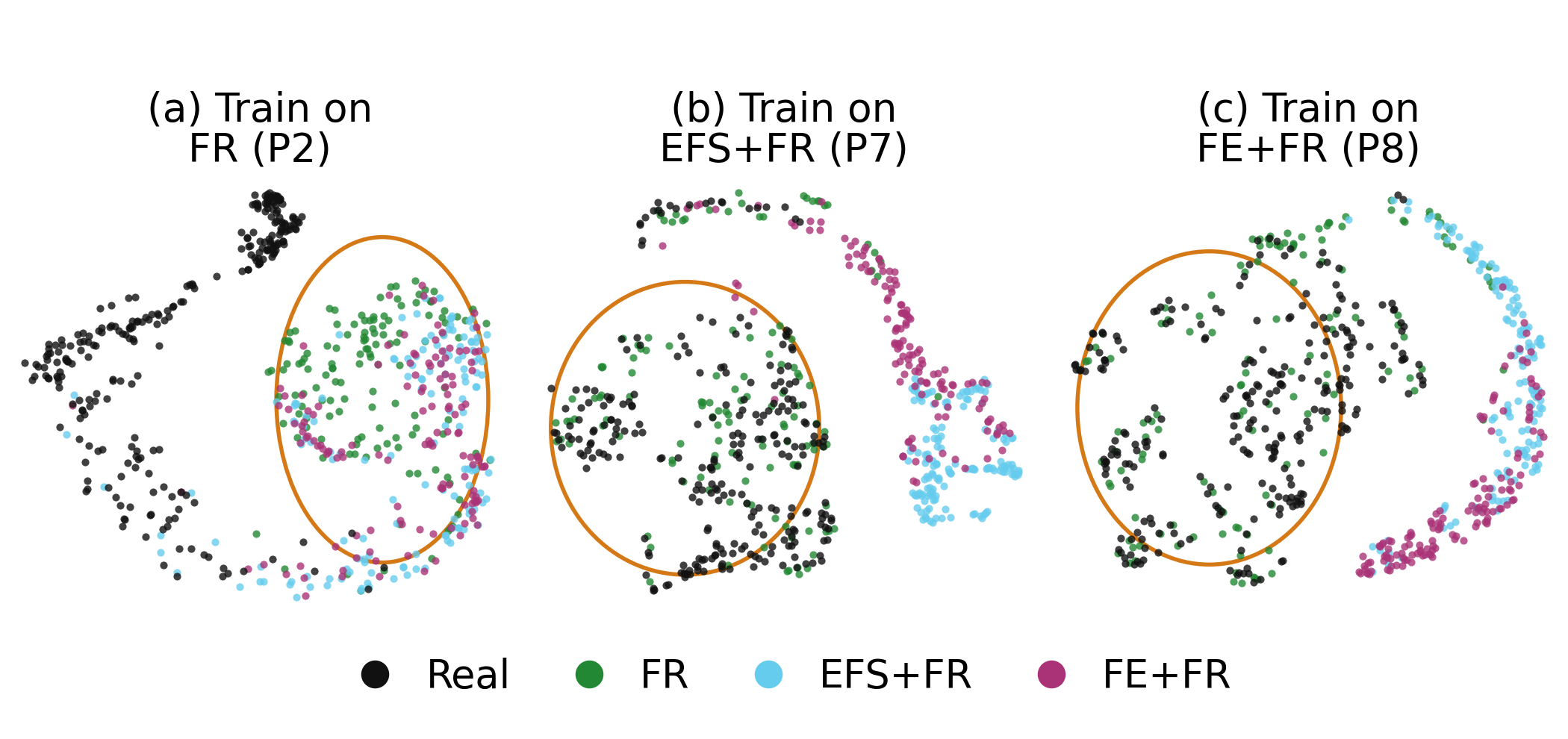}
    \caption{Distributions of fusion features extracted by the FauForensics detector~\cite{FauForensics}, associated with asymmetric generalization between FR (P2), EFS+FR (P7), and FE+FR (P8).}
    \label{fig:transfer_asymmetric}
\end{figure}

\item \textbf{Asymmetric generalization between certain pipelines.}
The heatmaps also reveal pronounced directional differences for some pipeline pairs. We use FR (P2), EFS+FR (P7), and FE+FR (P8) as examples to examine this asymmetry. Across the three detectors, training on P2 yields AUCs of 0.946--0.995 on P7 and 0.935--0.998 on P8. In contrast, training on P7 or P8 yields AUCs of only 0.559--0.677 and 0.601--0.666 on P2, respectively. Fig.~\ref{fig:transfer_asymmetric} shows the corresponding FauForensics features for fake instances from these three pipelines and real instances. When trained on P2, the model places many P7 and P8 instances in regions separate from real instances. When trained on P7 or P8, however, P2 instances overlap extensively with real instances. EFS+FS and FS exhibit strong bidirectional generalization, whereas models trained on EFS+FR or FE+FR generalize less effectively to FR than in the reverse direction. These patterns show that shared forgery operations do not guarantee reciprocal generalization, and training on a composite pipeline does not necessarily enable detection of its constituent operations in isolation. The observed asymmetry suggests that detectors may rely on pipeline-specific composite artifacts rather than evidence that generalizes across different uses of the same forgery operation.

\begin{figure}[t]
    \centering
    \includegraphics[width=\columnwidth]{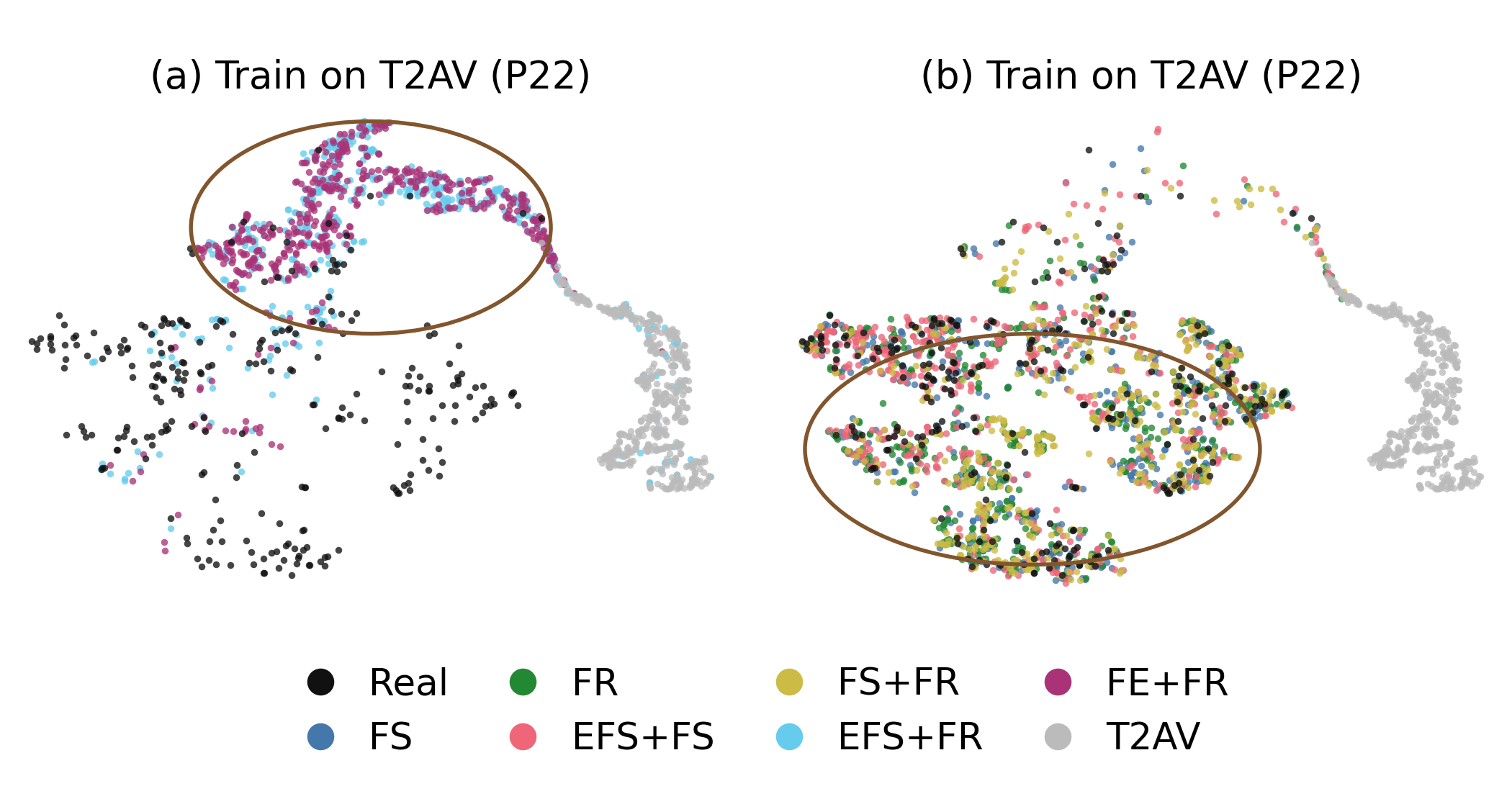}
    \caption{Selective generalization after T2AV (P22) training. Two views of the same t-SNE projection highlight (a) EFS+FR and FE+FR and (b) FS, FR, EFS+FS, and FS+FR.}
    \label{fig:transfer_t2av}
\end{figure}

\item \textbf{Limited coverage of established forgery types through T2AV training.}
Unlike pipelines composed of separate forgery operations, T2AV jointly generates audio and video. We therefore examine whether training on T2AV also enables detectors to recognize established forgery types and their compositions. After training on the T2AV pipeline (P22), FGMDF achieves mean AUCs of 0.858 and 0.848 across the EFS+FR and FE+FR pipelines, respectively, while FauForensics achieves 0.929 and 0.948. FRADE achieves 0.860 on EFS+FR but only 0.533 on FE+FR. For the remaining visual forgery groups, namely FS, FR, EFS+FS, and FS+FR, the group means range from 0.399 to 0.606 across the three detectors. We use FauForensics as an example to examine the corresponding feature distributions in Fig.~\ref{fig:transfer_t2av}. Many EFS+FR and FE+FR instances occupy regions separate from the main real distribution, whereas instances from the other four groups overlap more extensively with real instances. These qualitative observations are consistent with the group-wise AUC results for FauForensics. Together with the cross-pipeline evaluation, these results suggest that training on T2AV alone is insufficient for broad generalization across established forgery pipelines.

\end{itemize}

\subsection{Detection Behaviors}
\label{sec:analysis:detector_behavior}

Building on the training benefits established above, we use Mega-MMDF as a common training source to compare detectors and examine their responses to changes in audio-visual inputs. We first evaluate overall detection performance across all eight test datasets. We then investigate modality masking and temporal misalignment on Mega-MMDF, comparing predictions on the same test clips before and after each change to avoid introducing additional cross-dataset differences.

\begin{table*}[t]
    \centering
    \caption{Clip-level AUC (rank) across eight datasets. Bold and underline mark the best and second-best results in each column.}
    \label{tab:detector-ranking}
    \setlength{\tabcolsep}{3pt}
    \renewcommand{\arraystretch}{1.12}
    \scriptsize
    \resizebox{\textwidth}{!}{%
    \begin{tabular}{l*{8}{c} c |c}
        \toprule
        Detector
        & FakeAVCeleb
        & LAV-DF
        & PolyGlotFake
        & IDForge
        & Mega-MMDF
        & TalkingHeadBench
        & DigiFakeAV
        & MMDF
        & \shortstack{Average rank $\downarrow$}
        & \shortstack{Average AUC $\uparrow$} \\
        \midrule
        Baseline & 0.7715\,(6) & 0.5093\,(9) & 0.9100\,(5) & 0.4854\,(10) & \underline{0.9958\,(2)} & 0.8802\,(4) & 0.5360\,(8) & 0.4925\,(6) & 6.250 & 0.6976 (8) \\
        AVFF~\cite{AVFF} & 0.7017\,(8) & 0.6833\,(5) & 0.8843\,(6) & 0.5561\,(9) & 0.9819\,(5) & \textbf{0.9082\,(1)} & 0.5058\,(10) & 0.4533\,(10) & 6.750 & 0.7093 (6) \\
        AVH~\cite{AVH} & \textbf{0.9597\,(1)} & \textbf{0.8293\,(1)} & \textbf{0.9835\,(1)} & \textbf{0.8343\,(1)} & 0.9802\,(6) & 0.8988\,(3) & \textbf{0.7080\,(1)} & \underline{0.7826\,(2)} & \textbf{2.000} & \textbf{0.8721 (1)} \\
        FauForensics~\cite{FauForensics} & 0.9220\,(3) & \underline{0.7748\,(2)} & 0.8595\,(7) & 0.6518\,(4) & 0.9886\,(4) & 0.6927\,(8) & 0.5255\,(9) & 0.4621\,(8) & 5.625 & 0.7346 (3) \\
        FGMDF~\cite{FGMDF} & 0.8019\,(5) & 0.4815\,(10) & \underline{0.9781\,(2)} & 0.6560\,(3) & 0.9939\,(3) & 0.7024\,(6) & 0.6158\,(3) & 0.4757\,(7) & 4.875 & 0.7132 (5) \\
        FRADE~\cite{FRADE} & \underline{0.9256\,(2)} & 0.6874\,(4) & 0.9305\,(3) & 0.6279\,(6) & \textbf{0.9963\,(1)} & \underline{0.9075\,(2)} & 0.5669\,(6) & 0.4597\,(9) & \underline{4.125} & \underline{0.7627 (2)} \\
        MCL~\cite{MCL} & 0.7282\,(7) & 0.6743\,(6) & 0.9283\,(4) & 0.5885\,(7) & 0.9743\,(7) & 0.6025\,(9) & 0.6065\,(4) & 0.4957\,(5) & 6.125 & 0.6998 (7) \\
        SpeechForensics~\cite{SpeechForensics} & 0.8790\,(4) & 0.7414\,(3) & 0.5568\,(10) & \underline{0.6969\,(2)} & 0.7385\,(8) & 0.8031\,(5) & 0.5553\,(7) & \textbf{0.8696\,(1)} & 5.000 & 0.7301 (4) \\
        Qwen2.5-Omni~\cite{Qwen2.5-Omni} & 0.6574\,(10) & 0.6268\,(7) & 0.7445\,(8) & 0.6416\,(5) & 0.6071\,(9) & 0.7001\,(7) & \underline{0.6527\,(2)} & 0.5770\,(3) & 6.375 & 0.6509 (9) \\
        VideoLLaMA2.1~\cite{VideoLLaMA2.1} & 0.6645\,(9) & 0.5828\,(8) & 0.6042\,(9) & 0.5842\,(8) & 0.5686\,(10) & 0.5669\,(10) & 0.5807\,(5) & 0.5693\,(4) & 7.875 & 0.5901 (10) \\
        \bottomrule
    \end{tabular}%
    }
\end{table*}

\ul{\textit{Q5: How do detectors compare under the benchmark protocol?}} We compare the seven detectors trained on Mega-MMDF with three training-free methods: SpeechForensics~\cite{SpeechForensics}, Qwen2.5-Omni~\cite{Qwen2.5-Omni}, and VideoLLaMA2.1~\cite{VideoLLaMA2.1}. All methods are evaluated on the same eight test datasets using clip-level AUC. Table~\ref{tab:detector-ranking} reports per-dataset results, together with mean AUC and mean dataset-wise rank. Each dataset receives equal weight in both summaries.

AVH achieves the highest mean AUC and the best mean rank, followed by FRADE under both measures. However, the best-performing detector varies across datasets: FRADE leads on Mega-MMDF, AVFF on TalkingHeadBench, and SpeechForensics on MMDF. Among the training-free methods, SpeechForensics ranks fourth overall by mean AUC, outperforming several detectors trained on Mega-MMDF. Qwen2.5-Omni and VideoLLaMA2.1 rank ninth and tenth, respectively. These results establish a comparative baseline for both trained detectors and methods evaluated without benchmark-specific training.

The comparison above assumes complete audio-visual inputs. In practice, intentional stream removal or failures during transmission or decoding may leave only one modality available. Under these conditions, a detector should continue to identify forgeries in the remaining modality while correctly handling authentic content, without treating the missing modality itself as evidence of forgery. This motivates the following question.

\ul{\textit{Q6: Can multimodal detectors retain detection performance when one modality is unavailable?}}
To simulate a missing modality, we evaluate the seven detectors trained on Mega-MMDF using zero-filling to mask either audio or video. This preserves the detectors' input interfaces without modifying or retraining the models. Specifically, missing audio is replaced with a zero-valued waveform, and missing video with black frames. Each detector is evaluated on the same Mega-MMDF test clips before and after masking, avoiding additional cross-dataset differences.

We consider only conditions in which masking leaves the overall authenticity label unchanged. Audio masking is evaluated on RARV, RAFV, and FAFV clips, whereas video masking is evaluated on RARV, FARV, and FAFV clips. Thus, all evaluated fake clips retain a forged modality, while RARV clips remain real. Cases in which masking removes the only forged modality, such as video masking on RAFV clips, are excluded from this comparison.

\begin{figure}[t]
    \centering
    \includegraphics[width=\columnwidth]{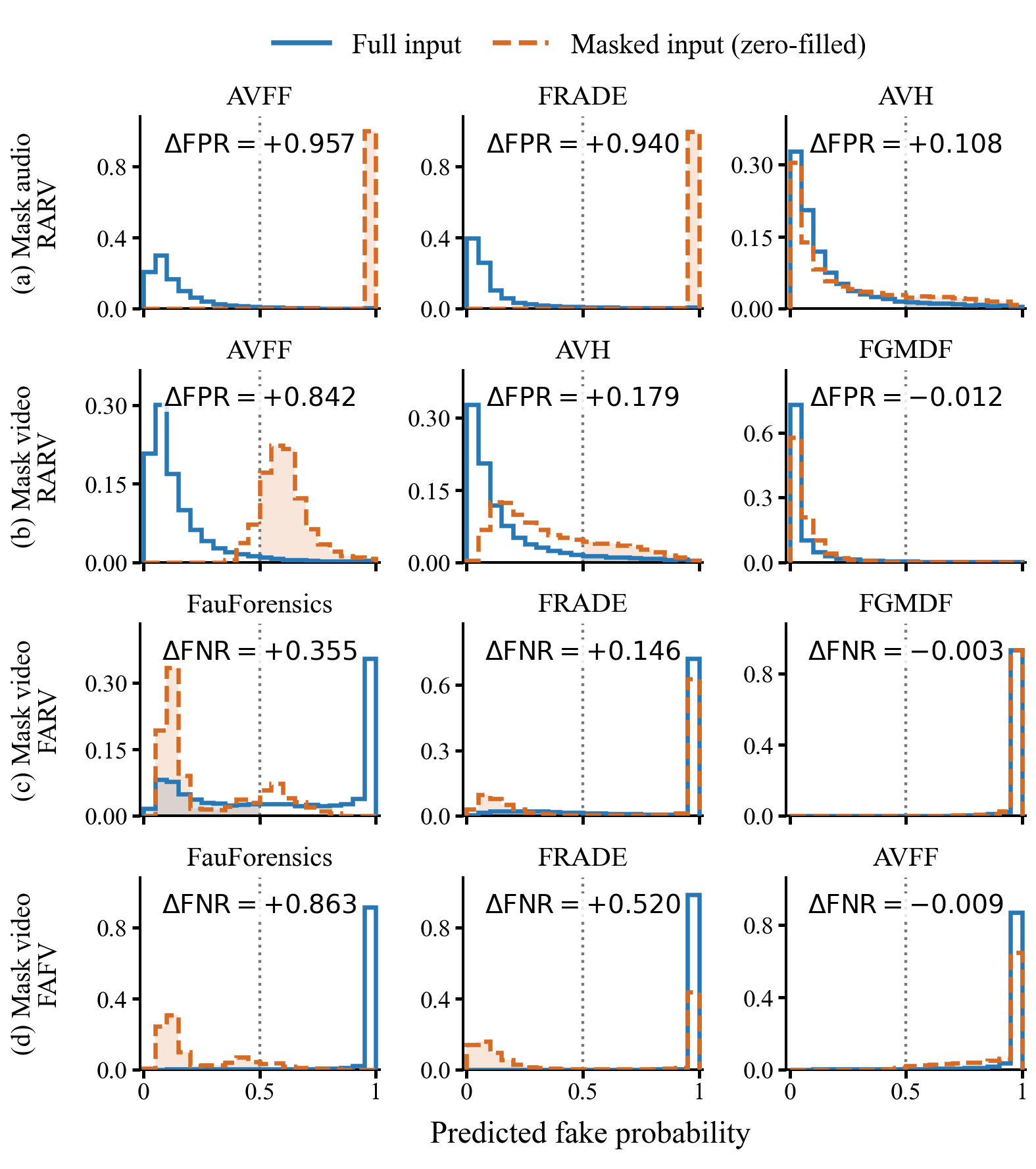}
    \caption{Selected clip-level fake-probability distributions before and after modality masking on Mega-MMDF. Vertical axes show within-class proportions with panel-specific scales, while horizontal axes show the predicted fake probability. Dotted lines mark the decision threshold of 0.5, and shaded areas indicate the error regions. Audio masking can shift authentic clips toward higher fake probabilities (a), while video masking can increase false alarms on authentic clips (b) and missed detections on clips with forged audio (c--d).}
    \label{fig:classwise_masking}
\end{figure}

For each condition, we report changes in the class-specific error rate and mean squared error (MSE). At a fixed decision threshold of 0.5, the false positive rate (FPR) measures the fraction of real clips classified as fake, while the false negative rate (FNR) measures the fraction of fake clips classified as real. MSE measures the average squared difference between the clip-level fake probability and the binary authenticity label:
\begin{equation}
    \mathrm{MSE}=\frac{1}{N}\sum_{i=1}^{N}(p_i-y_i)^2,
\end{equation}
where $p_i$ is the predicted fake probability and $y_i$ is 0 for real clips and 1 for fake clips. Changes are calculated as $\Delta=\mathrm{masked}-\mathrm{full}$, so positive changes indicate deterioration. More specifically, an increase in FPR or FNR indicates more false alarms or missed forgeries, respectively, while an increase in MSE indicates larger probability prediction errors. Table~\ref{tab:classwise_masking} highlights increases of at least 0.05 in red to facilitate identification of larger changes. Fig.~\ref{fig:classwise_masking} presents selected prediction histograms illustrating increased false alarms or missed detections, alongside examples with relatively small changes in error rates for comparison.

\begin{table*}[t]
    \centering
    \caption{Class-specific changes under modality masking on Mega-MMDF. Each entry reports $\Delta=\mathrm{masked}-\mathrm{full}$ on the same clips. FPR and FNR use a fixed threshold of 0.5. Red indicates increases of at least $+0.05$.}
    \label{tab:classwise_masking}
    \setlength{\tabcolsep}{2pt}
    \renewcommand{\arraystretch}{1.12}
    \scriptsize
    \begin{tabular*}{\textwidth}{@{\extracolsep{\fill}}lcccccc|cccccc@{}}
        \toprule
        & \multicolumn{6}{c|}{Mask audio: video retained}
        & \multicolumn{6}{c}{Mask video: audio retained} \\
        \cmidrule(lr){2-7}\cmidrule(lr){8-13}
        & \multicolumn{2}{c}{RARV}
        & \multicolumn{2}{c}{RAFV}
        & \multicolumn{2}{c|}{FAFV}
        & \multicolumn{2}{c}{RARV}
        & \multicolumn{2}{c}{FARV}
        & \multicolumn{2}{c}{FAFV} \\
        \cmidrule(lr){2-3}\cmidrule(lr){4-5}\cmidrule(lr){6-7}
        \cmidrule(lr){8-9}\cmidrule(lr){10-11}\cmidrule(lr){12-13}
        Detector
        & $\Delta$FPR~$\downarrow$ & $\Delta$MSE~$\downarrow$
        & $\Delta$FNR~$\downarrow$ & $\Delta$MSE~$\downarrow$
        & $\Delta$FNR~$\downarrow$ & $\Delta$MSE~$\downarrow$
        & $\Delta$FPR~$\downarrow$ & $\Delta$MSE~$\downarrow$
        & $\Delta$FNR~$\downarrow$ & $\Delta$MSE~$\downarrow$
        & $\Delta$FNR~$\downarrow$ & $\Delta$MSE~$\downarrow$ \\
        \midrule
        Baseline
        & $\textcolor{red}{+0.351}$ & $\textcolor{red}{+0.278}$
        & $-0.025$ & $-0.016$ & $-0.003$ & $-0.002$
        & $\textcolor{red}{+0.150}$ & $\textcolor{red}{+0.095}$
        & $\textcolor{red}{+0.118}$ & $\textcolor{red}{+0.093}$
        & $\textcolor{red}{+0.350}$ & $\textcolor{red}{+0.280}$ \\
        AVFF~\cite{AVFF}
        & $\textcolor{red}{+0.957}$ & $\textcolor{red}{+0.953}$
        & $-0.139$ & $-0.092$ & $-0.024$ & $-0.017$
        & $\textcolor{red}{+0.842}$ & $\textcolor{red}{+0.334}$
        & $-0.198$ & $-0.111$ & $-0.009$ & $+0.012$ \\
        AVH~\cite{AVH}
        & $\textcolor{red}{+0.108}$ & $\textcolor{red}{+0.063}$
        & $-0.006$ & $-0.003$ & $+0.014$ & $+0.012$
        & $\textcolor{red}{+0.179}$ & $\textcolor{red}{+0.111}$
        & $-0.052$ & $-0.080$
        & $\textcolor{red}{+0.325}$ & $\textcolor{red}{+0.192}$ \\
        FGMDF~\cite{FGMDF}
        & $\textcolor{red}{+0.231}$ & $\textcolor{red}{+0.140}$
        & $-0.137$ & $-0.093$ & $+0.018$ & $+0.016$
        & $-0.012$ & $-0.006$ & $-0.003$ & $-0.002$
        & $\textcolor{red}{+0.099}$ & $\textcolor{red}{+0.071}$ \\
        FRADE~\cite{FRADE}
        & $\textcolor{red}{+0.940}$ & $\textcolor{red}{+0.938}$
        & $-0.010$ & $-0.007$ & $-0.003$ & $-0.002$
        & $-0.053$ & $-0.043$
        & $\textcolor{red}{+0.146}$ & $\textcolor{red}{+0.135}$
        & $\textcolor{red}{+0.520}$ & $\textcolor{red}{+0.418}$ \\
        FauForensics~\cite{FauForensics}
        & $+0.012$ & $+0.002$ & $-0.012$ & $-0.008$
        & $\scalebox{0.8}{$\sim$}0.000$ & $-0.001$
        & $-0.033$ & $-0.024$
        & $\textcolor{red}{+0.355}$ & $\textcolor{red}{+0.297}$
        & $\textcolor{red}{+0.863}$ & $\textcolor{red}{+0.612}$ \\
        MCL~\cite{MCL}
        & $\textcolor{red}{+0.899}$ & $\textcolor{red}{+0.701}$
        & $-0.267$ & $-0.165$ & $-0.002$
        & $\scalebox{0.8}{$\sim$}0.000$
        & $-0.014$ & $+0.006$ & $+0.001$ & $+0.004$
        & $+0.013$ & $+0.015$ \\
        \bottomrule
    \end{tabular*}
\end{table*}

\begin{itemize}[leftmargin=*,itemsep=3pt,topsep=3pt,parsep=0pt]
\item \textbf{Audio masking primarily increases false alarms on authentic content.}
Audio masking increases FPR for all seven detectors. On RARV clips, six detectors show increases of at least 0.05 in both FPR and MSE. The FPR increases are particularly large for AVFF, FRADE, and MCL, reaching 0.957, 0.940, and 0.899, respectively. The RARV histograms for AVFF and FRADE in Fig.~\ref{fig:classwise_masking}(a) show pronounced shifts toward higher fake probabilities, indicating that zero-filling for missing audio can cause authentic content to be classified as fake. In contrast, the adverse effects on forged clips are more limited: no detector shows an increase of at least 0.05 in either FNR or MSE on RAFV or FAFV clips.

\item \textbf{Video masking produces both false alarms and missed detections.}
The effects of video masking vary across detectors and authenticity types. On RARV clips, AVFF, Baseline, and AVH show increases of at least 0.05 in both FPR and MSE. AVFF exhibits an FPR increase of 0.842, indicating a strong tendency to classify authentic content as fake after video masking.

Video masking also impairs detection when the retained audio is forged. On FARV clips, Baseline, FRADE, and FauForensics show increases of at least 0.05 in both FNR and MSE. On FAFV clips, five detectors exceed this threshold for both metrics. All seven detectors assign lower mean fake probabilities and incur higher MSE on FAFV clips, although FNR does not increase in every case. FauForensics shows the largest deterioration, with FNR and MSE increasing by 0.863 and 0.612, respectively. The corresponding histograms show many forged clips receiving low fake probabilities after video masking. These observations indicate that retaining a forged modality does not consistently preserve detection performance under the zero-filling setting, with prediction shifts varying across detectors and authenticity types.

\end{itemize}

Overall, all seven detectors exhibit increases of at least 0.05 in both error rate and MSE under at least one masking condition. Strong performance with complete audio-visual inputs therefore does not ensure reliable predictions when a missing modality is replaced with zeros. Reliable detection under missing-modality conditions requires both sensitivity to the remaining forgeries and correct handling of authentic content. These findings highlight the limitations of zero-filling alone as a way to accommodate unavailable modalities in the evaluated detectors and motivate the development of modality-flexible detectors that remain reliable with either complete or missing-modality inputs.

Beyond modality availability, we examine whether predictions change when both modalities are retained but their temporal alignment is disrupted. We use temporal reassignment as a diagnostic intervention to study this response.

\ul{\textit{Q7: How do detector predictions respond to audio-visual temporal misalignment?}}
We evaluate the same seven Mega-MMDF-trained detectors on real clips from the Mega-MMDF test set. Restricting the evaluation to authentic content allows us to examine temporal reassignment without introducing forgery operations that may also alter modality-specific cues and cross-modal correspondence.

Using each detector's non-overlapping temporal sampling windows, we pair audio and video from temporally disjoint portions of the same clip. Specifically, for a clip with $N$ windows ($N \geq 2$), we replace each aligned pair $(V_i,A_i)$ with $(V_i,A_{(i+1)\bmod N})$, where $i=0,\ldots,N-1$. This circular reassignment preserves the original audio and video windows while changing their pairing, ensuring that each reassigned pair contains windows from non-overlapping source time intervals.

We measure the instance-level change in predicted fake probability as $\Delta p=p_{\mathrm{misaligned}}-p_{\mathrm{aligned}}$. Positive values indicate shifts toward fake predictions, and negative values indicate shifts toward real predictions. We characterize these responses using the distribution of $\Delta p$, the signed mean $\overline{\Delta p}$, and the mean absolute change $\overline{|\Delta p|}$:
\begin{equation}
    \overline{\Delta p}=\frac{1}{M}\sum_{j=1}^{M}\Delta p_j,\qquad
    \overline{|\Delta p|}=\frac{1}{M}\sum_{j=1}^{M}|\Delta p_j|,
\end{equation}
where $M$ is the number of evaluated instances with temporal reassignment applied. These two statistics summarize the net direction and average magnitude of prediction changes, respectively.

\begin{figure*}[t]
    \centering
    \includegraphics[width=\textwidth]{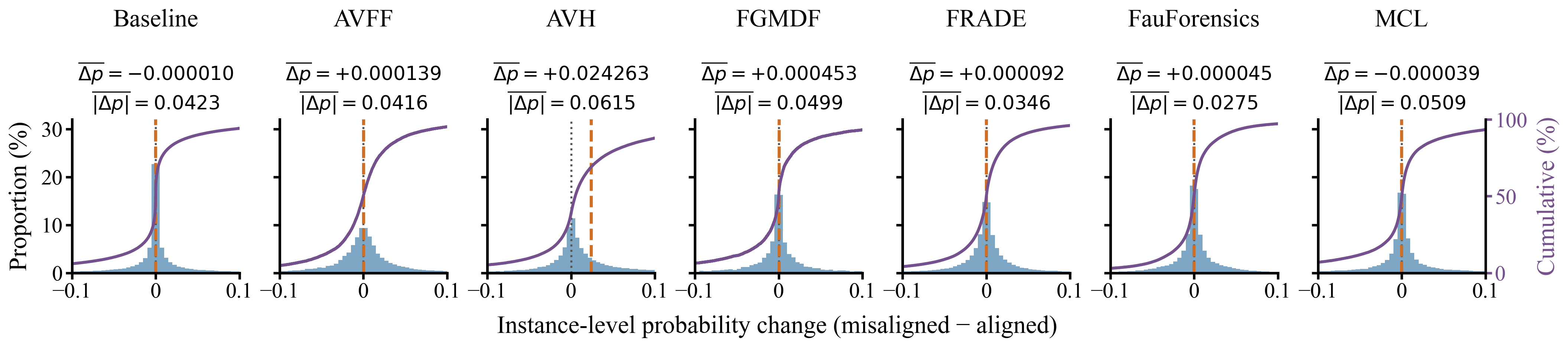}
    \caption{Instance-level prediction changes after within-clip audio-window reassignment on real Mega-MMDF clips. Blue histograms show instance proportions, purple curves show empirical cumulative distributions, and orange dashed lines indicate $\overline{\Delta p}$. Each panel reports $\overline{\Delta p}$ and $\overline{|\Delta p|}$, distinguishing net directional shifts from response magnitudes.}
    \label{fig:misalignment_response}
\end{figure*}

Across detectors, 69.7\%--85.8\% of evaluated instances have $|\Delta p|\leq 0.05$, indicating that temporal reassignment changes predicted fake probabilities by at most five percentage points for a majority of inputs. Meanwhile, $\overline{|\Delta p|}$ ranges from $0.0275$ to $0.0615$, reflecting the response magnitudes across all evaluated instances, including those with larger changes. Considering the overall direction, Fig.~\ref{fig:misalignment_response} shows that AVH exhibits a positive shift toward higher fake probabilities, with $\overline{\Delta p}=+0.0243$. The other six detectors have $\overline{\Delta p}$ between $-0.0005$ and $+0.0005$, indicating little net directional shift.

These results reveal differences in both the magnitude and direction of detector responses to disrupted audio-visual temporal correspondence. They motivate further investigation into how temporal correspondence should be defined and modeled, how detectors learn representations of it, and how the resulting cues can be combined with modality-specific forgery artifacts for deepfake detection.

\begin{figure*}[t]
    \centering
    \includegraphics[width=\textwidth]{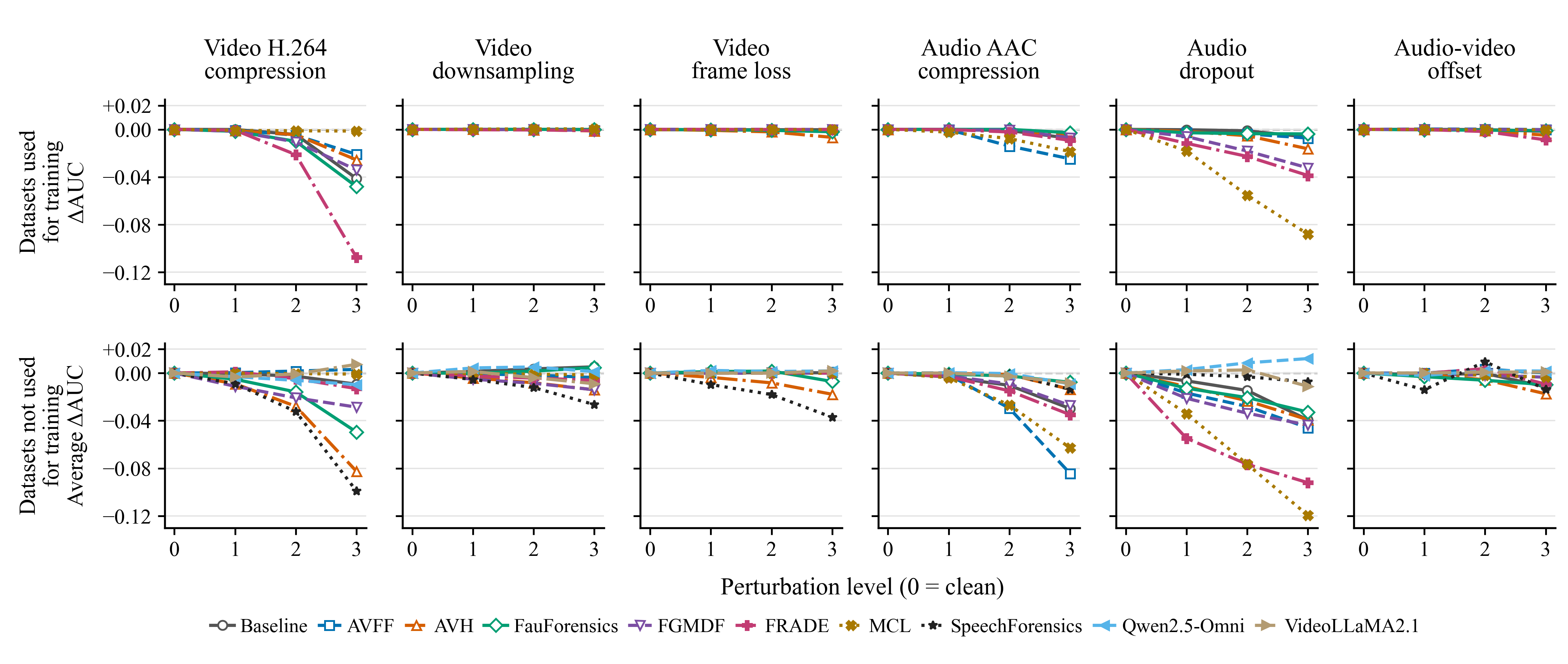}
    \caption{Changes in clip-level AUC under post-processing. Top: seven Mega-MMDF-trained detectors evaluated on Mega-MMDF. Bottom: average changes across the other seven datasets for these detectors and across all eight datasets for the three training-free methods. $\Delta\mathrm{AUC}=\mathrm{AUC}_{\mathrm{perturbed}}-\mathrm{AUC}_{\mathrm{clean}}$. Training refers to benchmark-specific training. Level 0 denotes clean inputs; levels 1--3 indicate increasing perturbation severity.}
    \label{fig:benchmark_robustness_results}
\end{figure*}

\subsection{Robustness and Fairness Evaluation}
\label{sec:analysis:robustness_and_fairness}

\ul{\textit{Q8: How robust are detectors to audio-visual post-processing?}}
The source data usually lack consistent metadata on recording devices and capture conditions. We therefore focus on post-processing of recorded audio-visual content and exclude capture-related degradations such as motion blur and defocus. We evaluate six transformations: H.264 and AAC compression simulate lossy video and audio encoding, respectively; video downsampling reduces spatial resolution; frame loss and audio dropout simulate missing video frames and audio segments; and audio-visual offsets introduce temporal misalignment. Detailed settings are provided in Section~\ref{appendix:benchmark_evaluation} and Table~\ref{tab:benchmark_robustness}.

Fig.~\ref{fig:benchmark_robustness_results} reports changes in clip-level AUC relative to the clean condition. The top row shows results for the seven Mega-MMDF-trained detectors evaluated on Mega-MMDF. The bottom row reports average AUC changes across the other seven datasets for these detectors and across all eight datasets for the three training-free methods. 
Sensitivity varies across transformations and detectors. For the trained detectors, video compression and audio dropout cause noticeable degradation on Mega-MMDF, while audio compression and dropout produce larger average losses across the other seven datasets. Downsampling, frame loss, and temporal offsets generally have smaller effects within the tested ranges. Among the training-free methods, SpeechForensics shows a pronounced decline under video compression. In contrast, Qwen2.5-Omni and VideoLLaMA2.1 exhibit smaller changes in average AUC, with absolute changes below 0.013 across all tested transformations and severity levels. These results show that robustness depends on the transformation and that robustness observed on Mega-MMDF does not necessarily extend to other datasets, highlighting the importance of evaluating cross-dataset robustness under different post-processing conditions.

\ul{\textit{Q9: How does detection performance vary across demographic groups?}}
We evaluate the seven Mega-MMDF-trained detectors and three training-free methods on the Mega-MMDF test split. Following~\cite{AIFace}, we report four fairness metrics: demographic parity (DP), equal odds (EO), overall accuracy equality (OAE), and maximum equalized odds (MEO), all evaluated at a fixed decision threshold of 0.5. We also report $\mathrm{SD}_{\mathrm{AUC}}$~\cite{Hu2024AttributeNeutral}, defined as the standard deviation of clip-level AUC across the eight demographic groups. The threshold-based metrics assess disparities in prediction rates or classification performance across groups, while $\mathrm{SD}_{\mathrm{AUC}}$ measures variation in ranking performance. Detailed definitions are provided in Section~\ref{appendix:benchmark_evaluation}.

\begin{table}[t]
    \centering
    \caption{Clip-level demographic group disparities on Mega-MMDF. Bold and underline indicate the lowest and second-lowest distinct values in each column, respectively.}
    \label{tab:fairness-megammdf}
    \setlength{\tabcolsep}{3pt}
    \renewcommand{\arraystretch}{1.1}
    \small
    \begin{tabular}{lccccc}
    \toprule
    Detector
    & DP $\downarrow$
    & EO $\downarrow$
    & OAE $\downarrow$
    & MEO $\downarrow$
    & SD$_{\mathrm{AUC}}$ $\downarrow$ \\
    \midrule
    Baseline
    & 0.0257 & 0.5387 & 0.0358 & 0.2064 & 0.0057 \\
    AVFF~\cite{AVFF}
    & 0.0634 & 0.3131 & 0.0529 & 0.0652 & 0.0102 \\
    AVH~\cite{AVH}
    & 0.0207 & 0.3051 & \underline{0.0197} & 0.0935 & \underline{0.0027} \\
    FauForensics~\cite{FauForensics}
    & 0.0735 & 0.2074 & 0.0571 & 0.0596 & 0.0064 \\
    FGMDF~\cite{FGMDF}
    & 0.0688 & 0.2605 & 0.0542 & 0.0606 & 0.0061 \\
    FRADE~\cite{FRADE}
    & \underline{0.0185} & 0.2867 & \textbf{0.0119} & 0.1229 & \textbf{0.0021} \\
    MCL~\cite{MCL}
    & 0.0393 & 0.5937 & 0.0377 & 0.2348 & 0.0155 \\
    \midrule
    SpeechForensics~\cite{SpeechForensics}
    & \textbf{0.0000} & \textbf{0.0000} & 0.0239 & \textbf{0.0000} & 0.0199 \\
    Qwen2.5-Omni~\cite{Qwen2.5-Omni}
    & 0.0777 & 0.3276 & 0.0540 & 0.0800 & 0.0346 \\
    VideoLLaMA2.1~\cite{VideoLLaMA2.1}
    & 0.0223 & \underline{0.0667} & 0.0425 & \underline{0.0241} & 0.0489 \\
    \bottomrule
    \end{tabular}
\end{table}

Table~\ref{tab:fairness-megammdf} reports demographic disparities under these five metrics. Among the training-based detectors, FRADE achieves the lowest DP, OAE, and $\mathrm{SD}_{\mathrm{AUC}}$, while FauForensics achieves the lowest EO and MEO. SpeechForensics obtains zero DP, EO, and MEO because it classifies all clips as real at the threshold of 0.5. Its AUC nevertheless varies across demographic groups, yielding an $\mathrm{SD}_{\mathrm{AUC}}$ of 0.0199. This example shows that low threshold-based disparities can coexist with a failure to detect forgeries at the chosen threshold. These metrics should therefore be interpreted jointly with detection performance rather than as a standalone ranking of detector fairness. Together, these results highlight the importance of evaluating demographic fairness in multimodal deepfake detection.

\section{Conclusion}
\label{sec: conclusion}

In this work, we introduce Mega-MMDF, a million-scale multimodal deepfake dataset covering 30 forgery methods and 22 forgery pipelines, and develop DeepfakeBench-MM to evaluate its training value and systematically investigate detector behavior under unified experimental settings. Future work will expand Mega-MMDF's forgery coverage and extend DeepfakeBench-MM with additional detectors and evaluation protocols.

Our experiments demonstrate the advantages of Mega-MMDF as a training source under the benchmark protocol and support the value of data scale and pipeline diversity. Using Mega-MMDF as a common experimental foundation, we compare detectors and reveal bidirectional and asymmetric cross-pipeline generalization, showing that training on composite pipelines does not necessarily enable detection of their constituent operations in isolation. False alarms and missed forgeries under the evaluated missing-modality conditions motivate modality-flexible detection, while responses to temporal misalignment motivate further study of how audio-visual correspondence is defined, learned, and used for detection. Post-processing robustness and demographic disparity evaluations provide protocols and reference results for subsequent research.

\flushbottom

\bibliographystyle{IEEEtran}
\bibliography{main}

\clearpage
\raggedbottom
\twocolumn[{
    \centering
    \Large\bfseries Appendix\par
    \vspace{1em}
}]

% Appendix headings use A, A.1, A.1.1; float and equation counters stay global.
\setcounter{section}{0}
\setcounter{subsection}{0}
\setcounter{subsubsection}{0}
\setcounter{paragraph}{0}
\renewcommand{\thesection}{\Alph{section}}
\renewcommand{\thesubsection}{\thesection.\arabic{subsection}}
\renewcommand{\thesubsubsection}{\thesubsection.\arabic{subsubsection}}
\renewcommand{\thesectiondis}{\thesection.}
\renewcommand{\thesubsectiondis}{\thesubsection}
\renewcommand{\thesubsubsectiondis}{\thesubsubsection}
% Keep PDF destinations distinct from the main-text sections after the reset.
\renewcommand{\theHsection}{supp.\Alph{section}}
\renewcommand{\theHsubsection}{\theHsection.\arabic{subsection}}
\renewcommand{\theHsubsubsection}{\theHsubsection.\arabic{subsubsection}}
% \clearpage
% \section*{Appendix}
This appendix provides additional details on Mega-MMDF and DeepfakeBench-MM. Appendix~\ref{appendix:dataset} details the dataset's ethics and release policy, licenses and usage terms of the source and benchmark datasets, forgery methods and pipelines, and dataset statistics. Appendix~\ref{appendix:benchmark} documents the benchmark's modular implementation, experimental configurations, and additional evaluation and analysis results.

\section{Additional Details on Mega-MMDF}
\label{appendix:dataset}
This section documents the construction, release, and assessment of Mega-MMDF. We first discuss ethical considerations and responsible release policies, followed by the licenses and usage terms of the datasets used for construction and benchmark evaluation. We then summarize the generation methods within each forgery category and illustrate how they are combined into individual pipelines, including the provenance of their inputs. Finally, we present dataset statistics covering its composition and training, validation, and test partitions. 

% ===== Appendix A.1 Mega-MMDF Ethics and Responsible Release ===== %
\subsection{Mega-MMDF Ethics and Responsible Release}
\label{appendix:megammdf_ethics}

Mega-MMDF is intended for research on multimodal deepfake detection. To support responsible use and mitigate potential misuse, we adopt the access and dataset maintenance policies described below.

\paragraph{Data provenance and access policy}
We will provide source indices and reconstruction scripts instead of redistributing real clips. Users must obtain the corresponding real source clips from the respective providers, and reconstruct the complete dataset locally. The applicable source-dataset licenses and usage agreements are summarized in Section~\ref{appendix:license}. Access to the media distributed through Mega-MMDF will require email verification and a data-use agreement signed by all researchers requesting access. Applications will record their names, institutional affiliations, and contact information to support access management, documentation of agreement acceptance, and subsequent update notifications.

\paragraph{Privacy requests and dataset maintenance} We will maintain a dedicated contact channel for privacy and rights-related removal requests from individuals depicted in the dataset, their authorized representatives, and relevant rights holders. Upon verification, we will remove the affected clips and associated derivatives identified through our provenance records from the distribution channels under our control. Each removal update will be assigned a new release version, accompanied by a record of affected clip identifiers and a cleanup utility for users to update local copies. When the update is issued, all registered users who have signed the Mega-MMDF data-use agreement will be notified by email and required to remove the affected clips. Users will be required to report the release version of our Mega-MMDF used in their experiments, together with any subsequent removal updates applied, so that the documented dataset composition can be checked for inclusion of affected clips. Once removed, affected clips will no longer be accessible through our Mega-MMDF data release platform.

% ===== Appendix A.2 Licenses of Utilized Datasets ===== %
\subsection{Licenses of Utilized Datasets}
\label{appendix:license}

The datasets used in this work fall into two categories: real-data sources for constructing Mega-MMDF and existing deepfake datasets incorporated into DeepfakeBench-MM. For each category, we first summarize the licenses, usage agreements, and access conditions published by the dataset providers, and then describe the corresponding usage and distribution practices adopted in this work. Official license and policy documents are referenced in the footnotes.

\subsubsection{Real-Data Sources for Mega-MMDF}
\label{appendix:license_real_sources}

VoxCeleb2~\cite{VoxCeleb2}, MEAD~\cite{MEAD}, and CelebV-HQ~\cite{CelebV-HQ} provide the real audio-visual material used to construct Mega-MMDF. Their respective licensing and usage conditions are summarized below.

\paragraph{VoxCeleb2~\cite{VoxCeleb2}}
The official VoxCeleb license file specifies CC BY 4.0, whereas the VoxCeleb2 webpage separately specifies CC BY-SA 4.0 for the provided metadata. The license file also states that copyright in both the original and cropped videos remains with the original owners. We report both statements because the official sources do not explicitly clarify their relationship.\footnote{\url{https://www.robots.ox.ac.uk/~vgg/data/voxceleb/files/license.txt}} \footnote{ \url{https://www.robots.ox.ac.uk/~vgg/data/voxceleb/vox2.html}}

\paragraph{MEAD~\cite{MEAD}}
MEAD is governed by a dataset-specific Terms of Use agreement permitting non-commercial, non-profit research. The agreement restricts third-party copying, distribution, hosting, and transfer of dataset content, subject to a provision for research associates and colleagues who accept the same terms. It also requires preservation of proprietary notices and citation of the original work. These dataset terms are distinct from the MIT software license in the repository.\footnote{\url{https://github.com/uniBruce/Mead/blob/master/Terms\%20of\%20Use_\%20Mead.pdf}} \footnote{ \url{https://github.com/uniBruce/Mead/blob/master/LICENSE}}

\paragraph{CelebV-HQ~\cite{CelebV-HQ}}
CelebV-HQ is governed by an agreement restricting use to non-commercial research. The agreement prohibits commercial use of the videos and derived data and restricts further copying, publication, and distribution of dataset content, with an exception for internal use at a single site within the same organization. It also states that the videos originate from the Internet and are not owned by the participating institutions.\footnote{\url{https://github.com/CelebV-HQ/CelebV-HQ\#agreement}}

\paragraph{Use in Mega-MMDF}
These datasets are used as source material for the non-commercial research underlying Mega-MMDF. The release documentation will preserve source attribution and applicable license information and identify the transformations applied during dataset construction. As described in Section~\ref{appendix:megammdf_ethics}, real source clips will be represented by indices rather than redistributed media files; users must obtain the corresponding samples through the respective providers or other provider-authorized channels, where available, and observe the applicable licenses or usage agreements. This arrangement avoids direct redistribution of the original source recordings and does not confer additional rights over source content retained in generated samples. Access to the media distributed by Mega-MMDF and the handling of verified removal requests follow the policies specified in Section~\ref{appendix:megammdf_ethics}.

\subsubsection{Deepfake Datasets for DeepfakeBench-MM}
\label{appendix:license_evaluation}

Seven existing deepfake datasets are used for benchmark evaluation: FakeAVCeleb~\cite{FakeAVCeleb}, LAV-DF~\cite{LAV-DF}, PolyGlotFake~\cite{PolyGlotFake}, IDForge~\cite{IDForge}, TalkingHeadBench~\cite{TalkingHeadBench}, DigiFakeAV~\cite{DigiFakeAV&DigiShield}, and MMDF~\cite{MMDF&X-AVDT}. Their published terms include dataset-specific requirements and, in some cases, explicit obligations concerning the underlying source datasets.

\paragraph{FakeAVCeleb~\cite{FakeAVCeleb}}
The official FakeAVCeleb website identifies CC BY 4.0 while also describing the dataset as available for non-commercial research and education.\footnote{\url{https://sites.google.com/view/fakeavcelebdash-lab/license}} Separately, the official access-request form, labelled as last updated in September 2024, requires approval and acceptance of dataset-specific terms restricting further copying, publication, and distribution of the dataset and derived data. It also provides for access by research associates and colleagues who first accept the same terms. These access terms are reported separately from the website's CC BY designation.\footnote{\href{https://docs.google.com/forms/d/e/1FAIpQLSfPDd3oV0auqmmWEgCSaTEQ6CGpFeB-ozQJ35x-B_0Xjd93bw/viewform}{FakeAVCeleb access-request form (Google Forms)}.}

\paragraph{LAV-DF~\cite{LAV-DF}}
The official LAV-DF repository specifies CC BY-NC 4.0. Its accompanying dataset terms limit use to non-commercial research and education and require users to accept the applicable agreements of VoxCeleb2, from which LAV-DF is derived.\footnote{\url{https://github.com/ControlNet/LAV-DF/blob/master/LICENSE}} \footnote{\url{https://github.com/ControlNet/LAV-DF/blob/master/TERMS_AND_CONDITIONS.md}} The VoxCeleb2 licensing statements are summarized in Section~\ref{appendix:license_real_sources}.

\paragraph{PolyGlotFake~\cite{PolyGlotFake}}
PolyGlotFake is made available under a published usage policy restricting access to academic institutions and use for research purposes. The official ethics statement also provides for the removal of contested material in response to requests from rights holders, but does not specify a standard dataset license.\footnote{\url{https://github.com/tobuta/PolyGlotFake\#ethics-statement}}

\paragraph{IDForge~\cite{IDForge}}
The official IDForge repository provides links to the original video version (v1) and a preprocessed version (v2), but does not identify a standard dataset license. Its linked application page nevertheless sets explicit conditions: academic research use only, acceptance of responsibility and indemnification obligations by users and their affiliated institutions, and the authors' right to terminate access.\footnote{\url{https://github.com/xyyandxyy/IDForge}} \footnote{\url{https://request.idforge.cfd/}} The paper's ethics statement also describes access as limited to academic purposes and subject to application review.\footnote{\url{https://arxiv.org/html/2401.11764v2\#S3.SS4}}

\paragraph{TalkingHeadBench~\cite{TalkingHeadBench}}
The official TalkingHeadBench dataset card labels the dataset as MIT-licensed and separately requires compliance with the original licenses and attribution requirements of FFHQ~\cite{FFHQ}, CelebV-HQ~\cite{CelebV-HQ}, and FaceForensics++~\cite{FaceForensics++}. The MIT label does not replace those source-specific conditions.\footnote{\url{https://huggingface.co/datasets/luchaoqi/TalkingHeadBench}} The CelebV-HQ conditions are summarized in Section~\ref{appendix:license_real_sources}; the FFHQ licensing statement and FaceForensics++ Terms of Use are available from their respective providers.\footnote{\url{https://github.com/NVlabs/ffhq-dataset\#licenses}} \footnote{\url{https://kaldir.vc.in.tum.de/faceforensics_tos.pdf}}

\paragraph{DigiFakeAV~\cite{DigiFakeAV&DigiShield}}
The official DigiFakeAV dataset card labels the dataset as MIT-licensed. The associated code repository also contains an MIT license, requiring preservation of its copyright and permission notices in copies or substantial portions of the software. We cite the dataset card for the dataset's license designation.\footnote{\url{https://huggingface.co/datasets/cambrain/DigiFakeAV}} \footnote{\url{https://github.com/DigiFakeAV/DigiFakeAV-v1/blob/main/LICENSE}}

\paragraph{MMDF~\cite{MMDF&X-AVDT}}
The official MMDF dataset card specifies CC BY-NC 4.0. Its public access notice states that users must share their contact information and accept access conditions before obtaining the dataset content; the detailed conditions are presented through the Hugging Face access process.\footnote{\url{https://huggingface.co/datasets/zaqxsw0526/MMDF}}

\paragraph{Use in DeepfakeBench-MM}
These seven datasets are used for non-commercial academic research, including training, validation, and testing as specified in the main experimental protocol, with attribution to the original works. For dataset preparation, DeepfakeBench-MM will provide preprocessing code rather than redistribute the original media or preprocessed data. This policy respects redistribution restrictions, including those specified for FakeAVCeleb~\cite{FakeAVCeleb}, and preserves the independent access procedures established by the respective providers. Researchers reproducing the experiments must obtain each dataset through its official channels, satisfy the applicable access conditions, and execute the preprocessing code locally. Dataset-specific usage agreements and any applicable source-dataset licensing and attribution requirements remain in effect.

% ===== Appendix A.3 Forgery Method ===== %
\subsection{Forgery Methods}
\label{appendix:method}

Table~\ref{tab:dataset_method} summarizes the forgery methods used to construct Mega-MMDF across eight forgery types. These include methods from academic publications as well as publicly available models and tools. Their combinations form the forgery pipelines described in Section~\ref{appendix:pipeline_illustration}.

\begin{table*}[ht]
    \centering
    \caption{Forgery methods used to construct Mega-MMDF, together with their publication venues and public engagement statistics.}
    \label{tab:dataset_method}
    \resizebox{\textwidth}{!}{
    \begin{tabular}{c c c c c c}
        \toprule
        Forgery Type
        & Forgery Method
        & Year
        & Publication Venue
        & Public Engagement
        & Output Data Format \\
        \midrule
        
        \multirow{5}{*}{\shortstack{TTS\\(Text-to-Speech)}}
        & YourTTS~\cite{YourTTS}
        & 2022 & ICML 2022
        & \href{https://github.com/Edresson/YourTTS}{GitHub}: 1,053 stars
        & Audio \\
        
        & SpeechT5~\cite{SpeechT5}
        & 2022 & ACL 2022
        & \href{https://github.com/microsoft/SpeechT5}{GitHub}: 1,450 stars
        & Audio \\
        
        & OpenVoice V1~\cite{OpenVoiceV1}
        & 2023 & arXiv 2023
        & \href{https://github.com/myshell-ai/OpenVoice}{GitHub}: 37,532 stars
        & Audio \\
        
        & XTTS~\cite{XTTS}
        & 2024 & Interspeech 2024
        & \href{https://github.com/coqui-ai/TTS}{GitHub}: 46,016 stars
        & Audio \\
        
        & Vevo~\cite{Vevo}
        & 2025 & ICLR 2025
        & \href{https://github.com/open-mmlab/Amphion}{GitHub}: 10,295 stars
        & Audio \\
        
        \cmidrule(lr){1-6}
        
        \multirow{3}{*}{\shortstack{VC\\(Voice Conversion)}}
        & FreeVC~\cite{FreeVC}
        & 2023 & ICASSP 2023
        & \href{https://github.com/OlaWod/FreeVC}{GitHub}: 717 stars
        & Audio \\
        
        & Diff-HierVC~\cite{Diff-HierVC}
        & 2023 & Interspeech 2023
        & \href{https://github.com/hayeong0/Diff-HierVC}{GitHub}: 237 stars
        & Audio \\
        
        & Vevo~\cite{Vevo}
        & 2025 & ICLR 2025
        & \href{https://github.com/open-mmlab/Amphion}{GitHub}: 10,295 stars
        & Audio \\
        
        \cmidrule(lr){1-6}
        
        \multirow{2}{*}{\shortstack{PF\\(Partial Fake)}}
        & FluentSpeech~\cite{FluentSpeech}
        & 2023 & Findings of ACL 2023
        & \href{https://github.com/Zain-Jiang/Speech-Editing-Toolkit}{GitHub}: 207 stars
        & Audio \\
        
        & VoiceCraft~\cite{VoiceCraft}
        & 2024 & ACL 2024
        & \href{https://github.com/jasonppy/VoiceCraft}{GitHub}: 8,575 stars
        & Audio \\
        
        \midrule
        
        \multirow{4}{*}{\shortstack{FS\\(Face Swapping)}}
        & UniFace~\cite{Uniface}
        & 2022 & ECCV 2022
        & \href{https://github.com/xc-csc101/UniFace}{GitHub}: 58 stars
        & Video \\
        
        & BlendFace~\cite{BlendFace}
        & 2023 & ICCV 2023
        & \href{https://github.com/mapooon/BlendFace}{GitHub}: 207 stars
        & Video \\
        
        & InSwapper~\cite{inswapper}
        & 2023 & ---
        & \href{https://github.com/haofanwang/inswapper}{GitHub}: 749 stars
        & Video \\
        
        & CSCS~\cite{CSCS}
        & 2024 & ACM TOG 2024
        & \href{https://github.com/ICTMCG/CSCS}{GitHub}: 73 stars
        & Video \\
        
        \cmidrule(lr){1-6}
        
        \multirow{5}{*}{\shortstack{EFS\\(Entire Face Synthesis)}}
        & ThisPersonDoesNotExist~\cite{ThisPersonDoesNotExist}
        & 2019 & ---
        & ---
        & Image \\
        
        & Stable Diffusion XL Base~\cite{SDXL}
        & 2024 & ICLR 2024
        & \href{https://huggingface.co/stabilityai/stable-diffusion-xl-base-1.0}{Hugging Face}: 2,570,790 downloads
        & Image \\
        
        & RealVisXL V3.0~\cite{sg1612222023realvisxl3}
        & 2023 & ---
        & \href{https://huggingface.co/SG161222/RealVisXL_V3.0}{Hugging Face}: 14,098 downloads
        & Image \\
        
        & Realistic Vision V4.0 (noVAE)~\cite{sg1612222023realisticvision4}
        & 2023 & ---
        & \href{https://huggingface.co/SG161222/Realistic_Vision_V4.0_noVAE}{Hugging Face}: 11,896 downloads
        & Image \\
        
        & RealVisXL V5.0~\cite{sg1612222024realvisxl5}
        & 2024 & ---
        & \href{https://huggingface.co/SG161222/RealVisXL_V5.0}{Hugging Face}: 326,003 downloads
        & Image \\
        
        \cmidrule(lr){1-6}
        
        \multirow{3}{*}{\shortstack{FE\\(Face Editing)}}
        & IP-Adapter~\cite{IP-Adapter}
        & 2023 & arXiv 2023
        & \href{https://github.com/tencent-ailab/IP-Adapter}{GitHub}: 6,688 stars
        & Image \\
        
        & PhotoMaker~\cite{PhotoMaker}
        & 2024 & CVPR 2024
        & \href{https://github.com/TencentARC/PhotoMaker}{GitHub}: 10,086 stars
        & Image \\
        
        & PuLID~\cite{PuLID}
        & 2024 & NeurIPS 2024
        & \href{https://github.com/ToTheBeginning/PuLID}{GitHub}: 3,550 stars
        & Image \\
        
        \midrule
        
        \multirow{6}{*}{\shortstack{FR\\(Audio-Driven Face Reenactment)}}
        & Wav2Lip~\cite{Wav2Lip}
        & 2020 & ACM MM 2020
        & \href{https://github.com/Rudrabha/Wav2Lip}{GitHub}: 13,203 stars
        & Video \\
        
        & SadTalker~\cite{SadTalker}
        & 2023 & CVPR 2023
        & \href{https://github.com/OpenTalker/SadTalker}{GitHub}: 14,085 stars
        & Video \\
        
        & MuseTalk~\cite{MuseTalk}
        & 2024 & arXiv 2024
        & \href{https://github.com/TMElyralab/MuseTalk}{GitHub}: 6,565 stars
        & Video \\
        
        & Diff2Lip~\cite{Diff2Lip}
        & 2024 & WACV 2024
        & \href{https://github.com/soumik-kanad/diff2lip}{GitHub}: 379 stars
        & Video \\
        
        & EDTalk~\cite{EDTalk}
        & 2024 & ECCV 2024
        & \href{https://github.com/tanshuai0219/EDTalk}{GitHub}: 467 stars
        & Video \\
        
        & LatentSync~\cite{LatentSync}
        & 2024 & arXiv 2024
        & \href{https://github.com/bytedance/LatentSync}{GitHub}: 6,073 stars
        & Video \\
        
        \cmidrule(lr){1-6}
        
        \multirow{2}{*}{\shortstack{T2AV\\(Joint Audio-Video Generation)}}
        & Ovi~\cite{Ovi}
        & 2025 & arXiv 2025
        & \href{https://github.com/character-ai/Ovi}{GitHub}: 1,760 stars
        & Audio-Video \\
        
        & LTX-2~\cite{LTX-2}
        & 2026 & arXiv 2026
        & \href{https://github.com/Lightricks/LTX-2}{GitHub}: 9,430 stars
        & Audio-Video \\
        
        \bottomrule
    \end{tabular}
    }
    \par\vspace{2pt}
    \begin{minipage}{\textwidth}
    \footnotesize
    \textit{Notes:} Public engagement statistics were retrieved on
    September 16, 2026. GitHub stars refer to the linked repositories,
    which may cover multiple methods or versions. Hugging Face downloads
    refer to the last month as reported by the platform. These indicators
    are not directly comparable across platforms.
    \end{minipage}
\end{table*}

% ===== Appendix A.4 Forgery Pipeline ===== %
\subsection{Pipeline Illustration}
\label{appendix:pipeline_illustration}

We organize Mega-MMDF into 23 pipelines, comprising one real pipeline (P0) and 22 forgery pipelines (P1--P22). These pipelines cover four combinations of modality authenticity: RARV (Real Audio, Real Video), RAFV (Real Audio, Fake Video), FARV (Fake Audio, Real Video), and FAFV (Fake Audio, Fake Video). Depending on the pipeline, audio forgery, visual forgery, and audio-driven face reenactment are applied individually or in combination, while T2AV jointly generates audio and video from text prompts. Paired real clips provide the inputs required by the corresponding operations. EFS images and T2AV clips are generated independently of these paired clips, associated with pair IDs through their assigned indices, and allocated to the corresponding data splits.

\begin{figure*}[ht]
    \centering
    \includegraphics[width=0.9\linewidth]{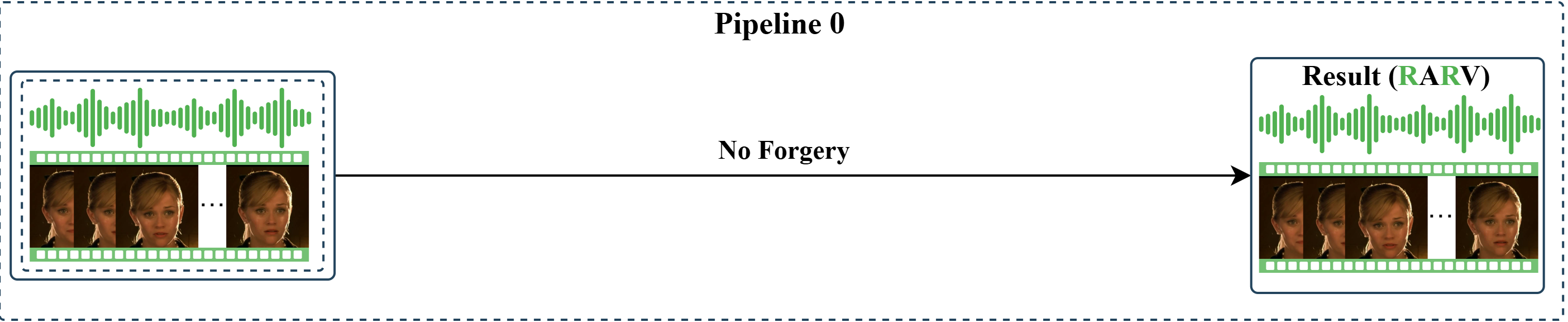}
    
    \vspace{0.2cm}  
    \includegraphics[width=0.9\linewidth]{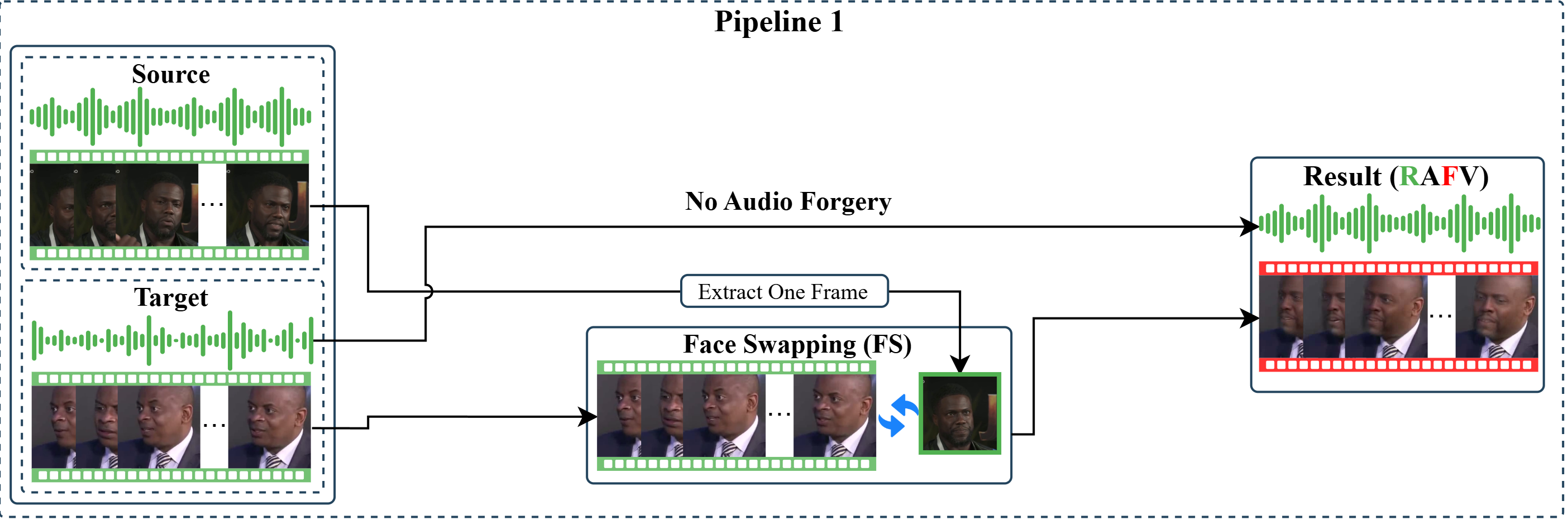}

    \vspace{0.2cm}  
    \includegraphics[width=0.9\linewidth]{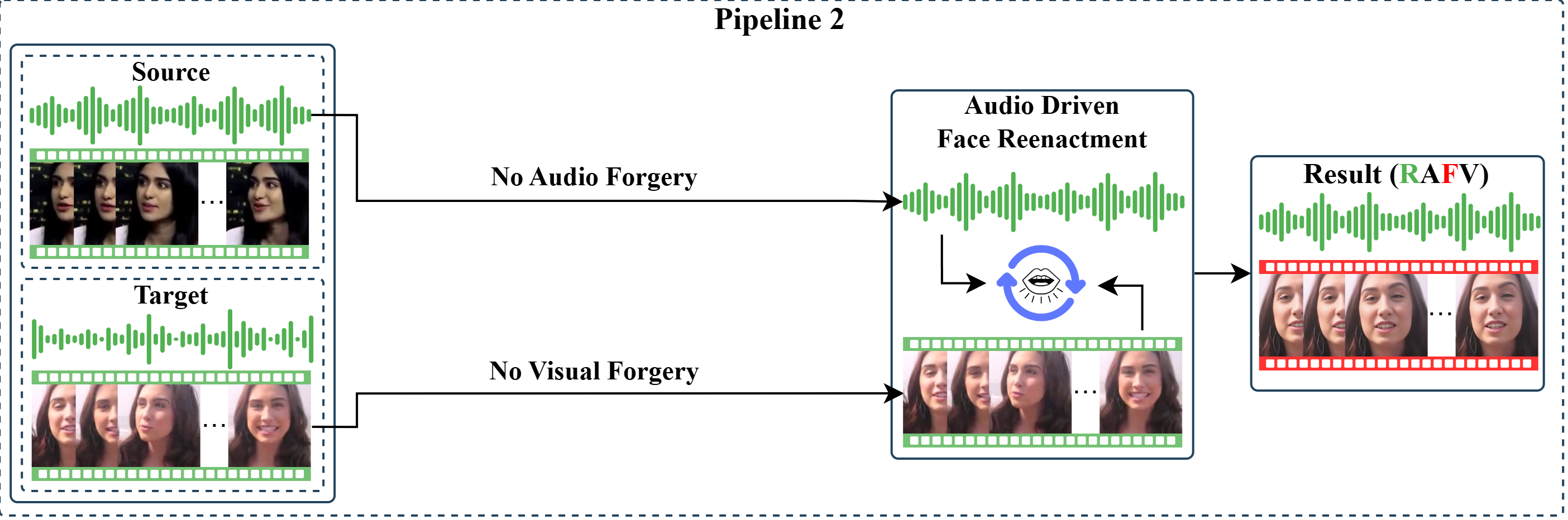}

    \vspace{0.2cm}  
    \includegraphics[width=0.9\linewidth]{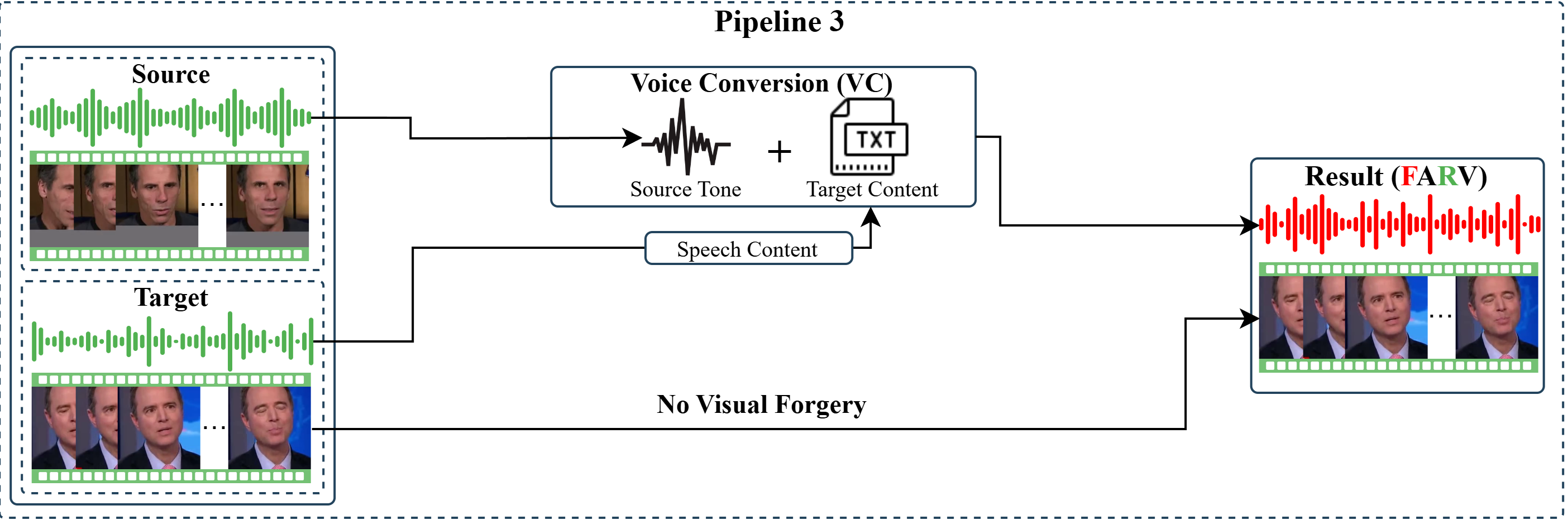}
    
    \caption{Illustration of Pipelines 0 to 3.}
    \label{fig:pipeline0-3}
\end{figure*}

\begin{figure*}[ht]
    \centering
    \includegraphics[width=0.9\linewidth]{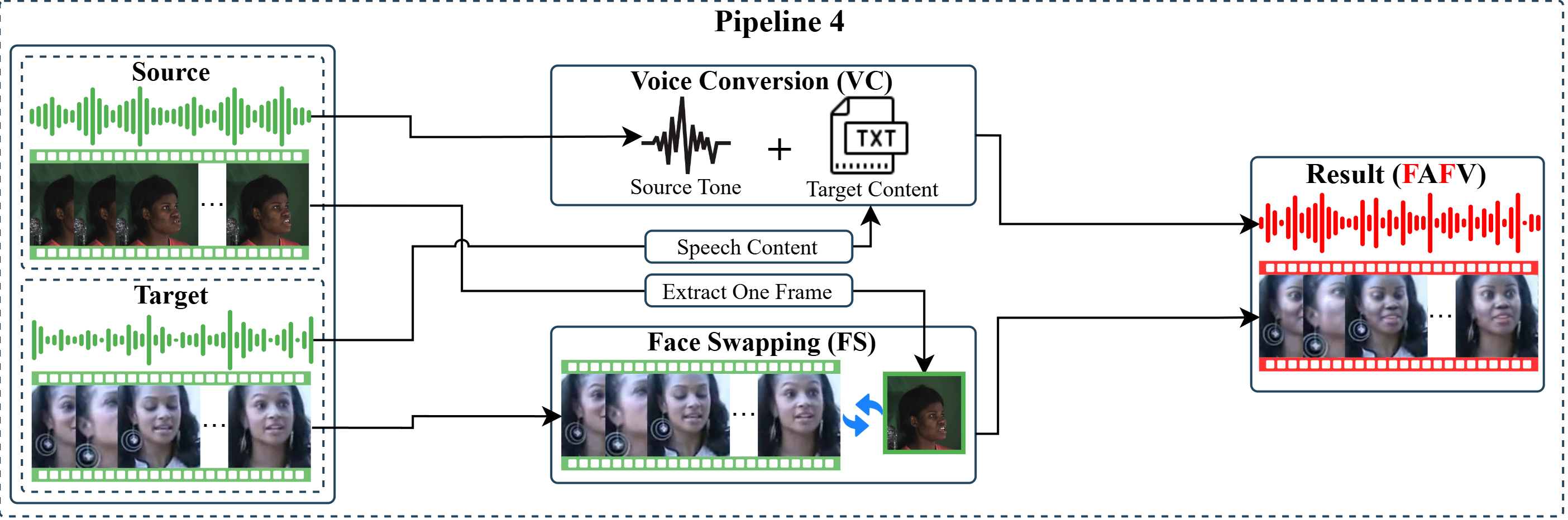}
    
    \vspace{0.2cm}  
    \includegraphics[width=0.9\linewidth]{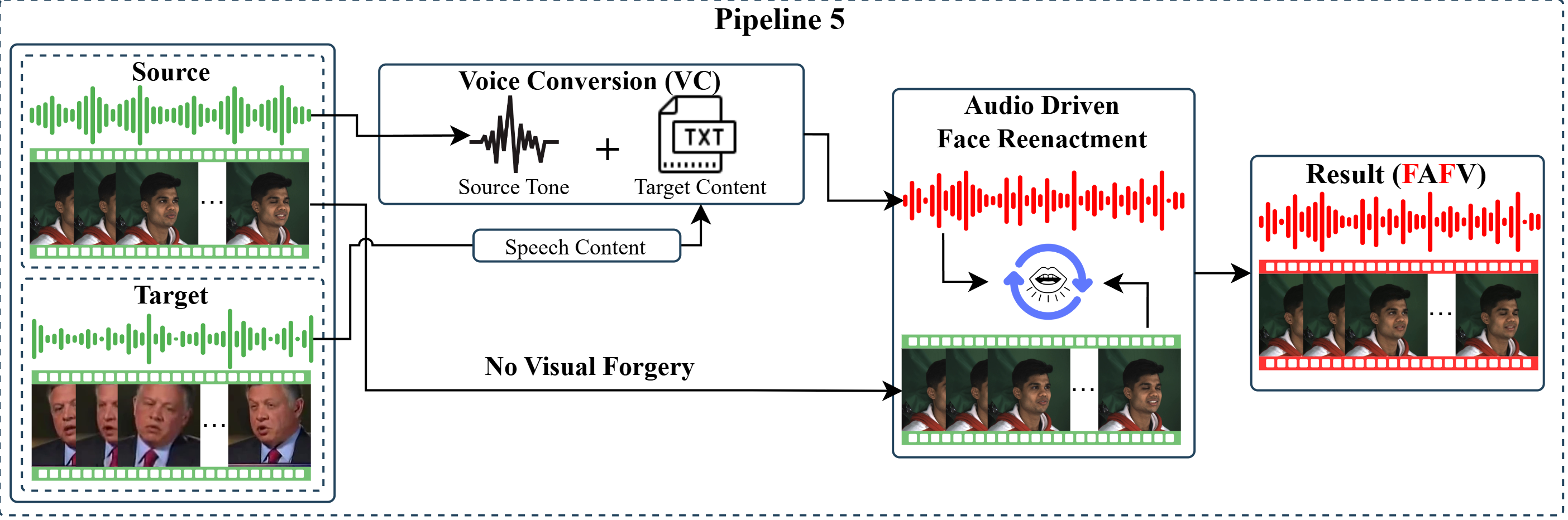}

    \vspace{0.2cm}  
    \includegraphics[width=0.9\linewidth]{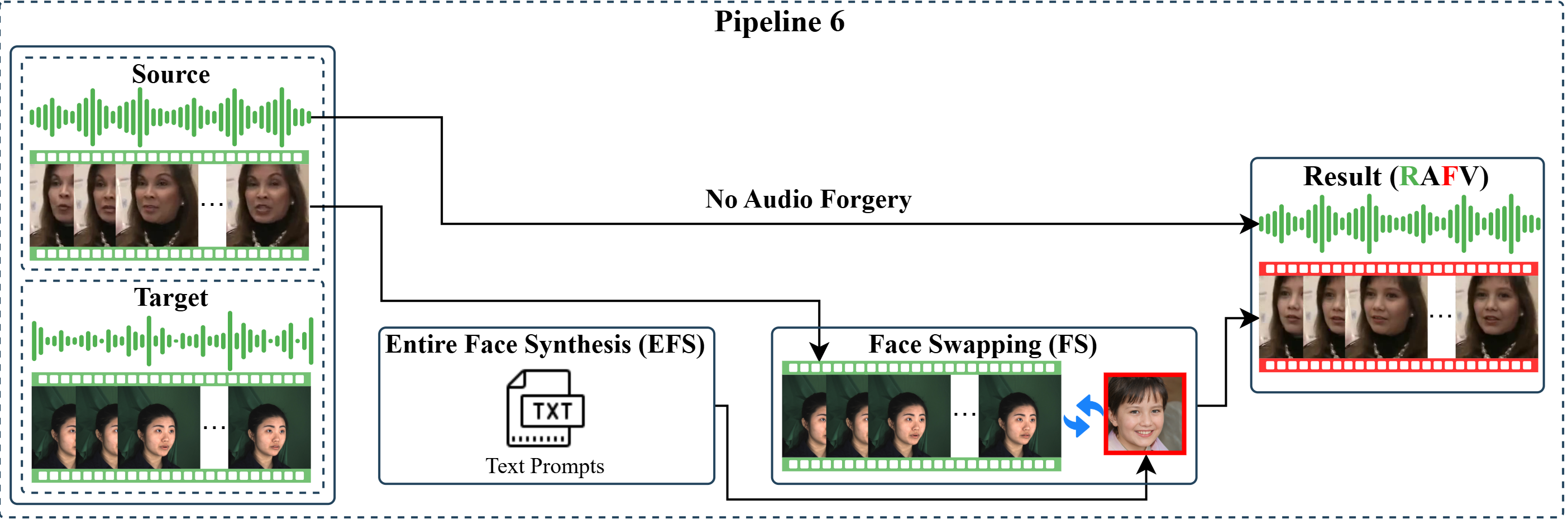}

    \vspace{0.2cm}  
    \includegraphics[width=0.9\linewidth]{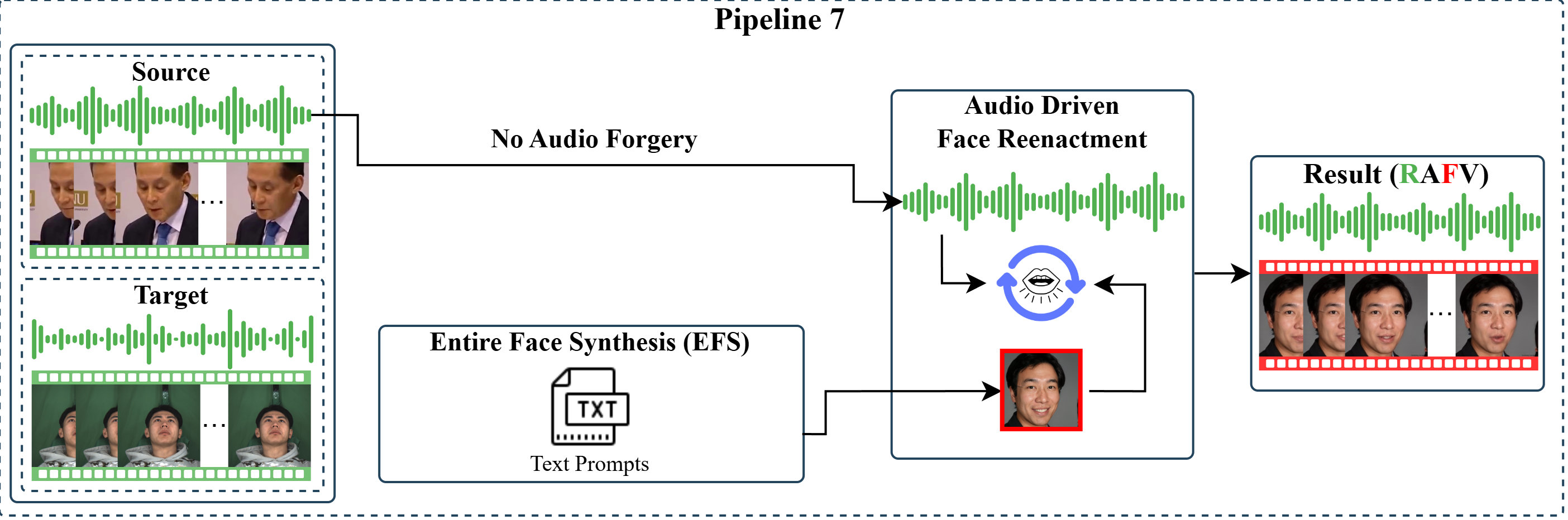}
    
    \caption{Illustration of Pipelines 4 to 7.}
    \label{fig:pipeline4-7}
\end{figure*}

\begin{figure*}[ht]
    \centering
    \includegraphics[width=0.9\linewidth]{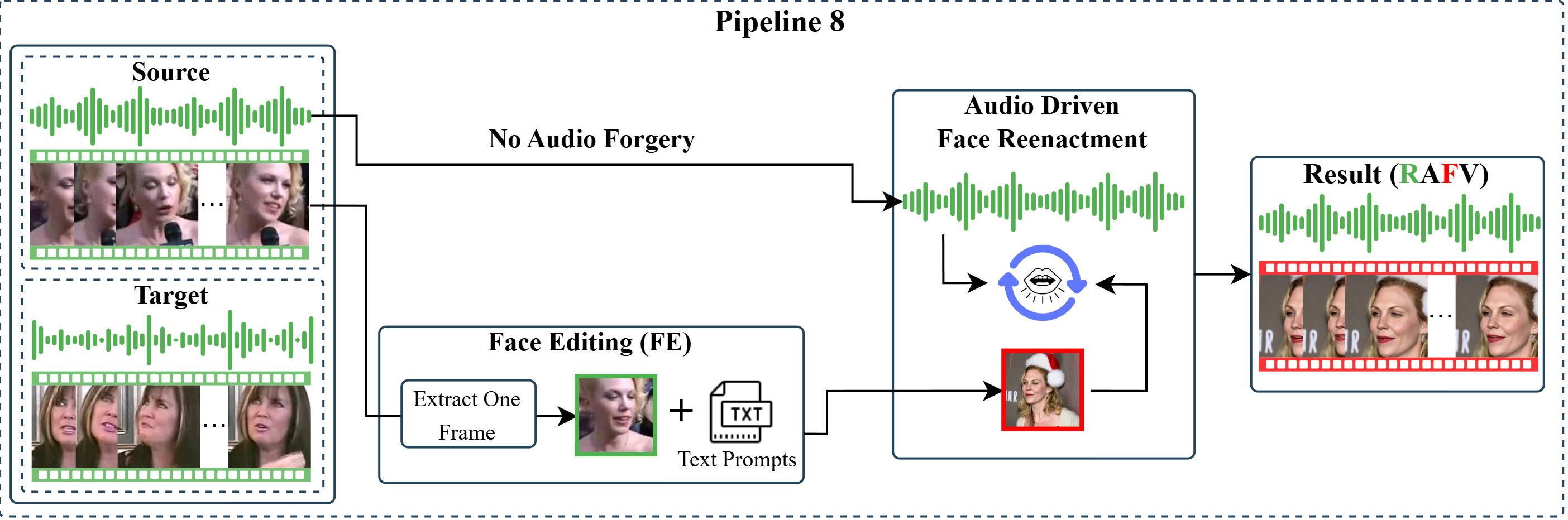}
    
    \vspace{0.2cm}  
    \includegraphics[width=0.9\linewidth]{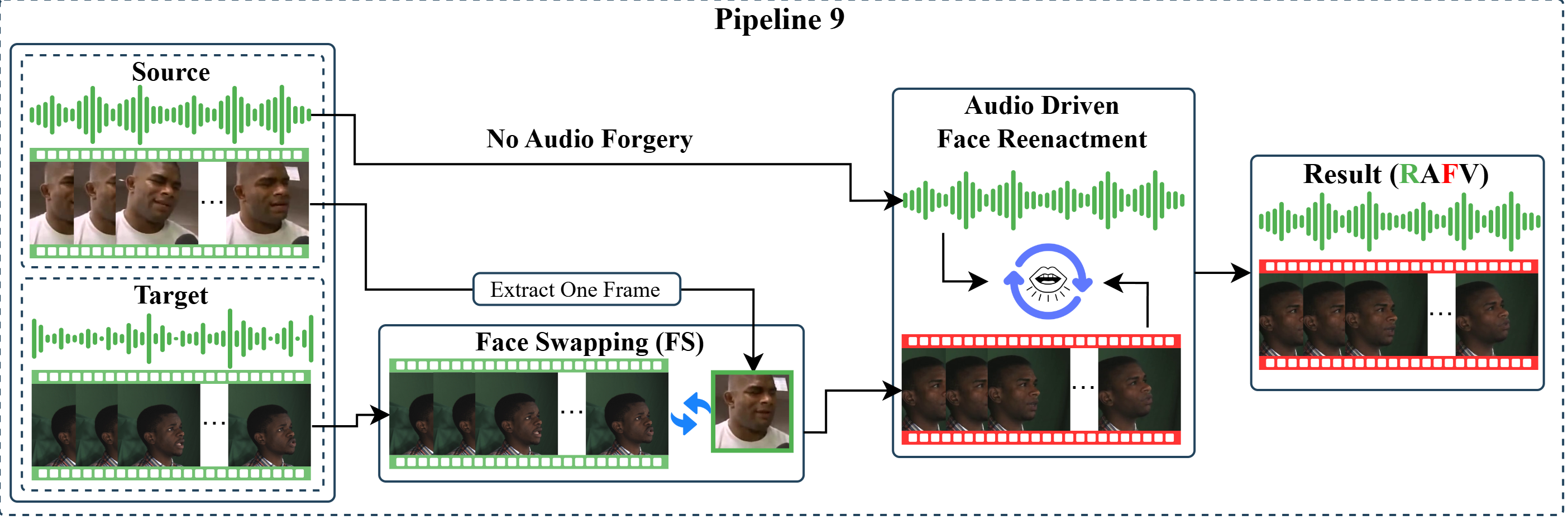}

    \vspace{0.2cm}  
    \includegraphics[width=0.9\linewidth]{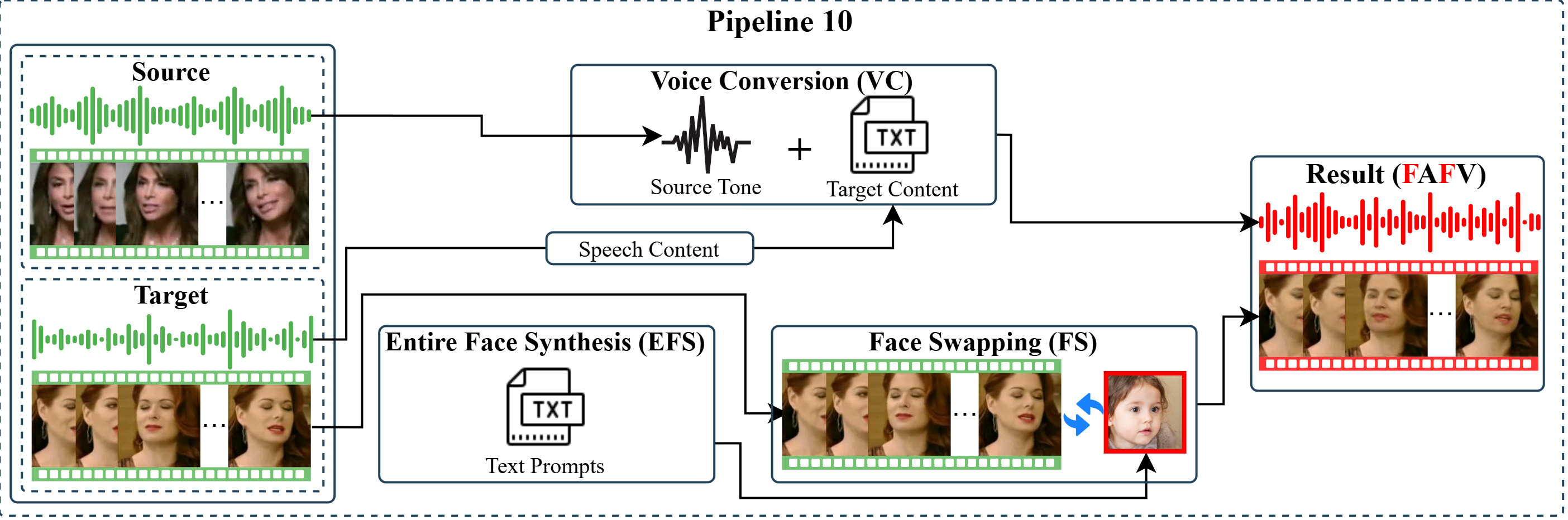}

    \vspace{0.2cm}  
    \includegraphics[width=0.9\linewidth]{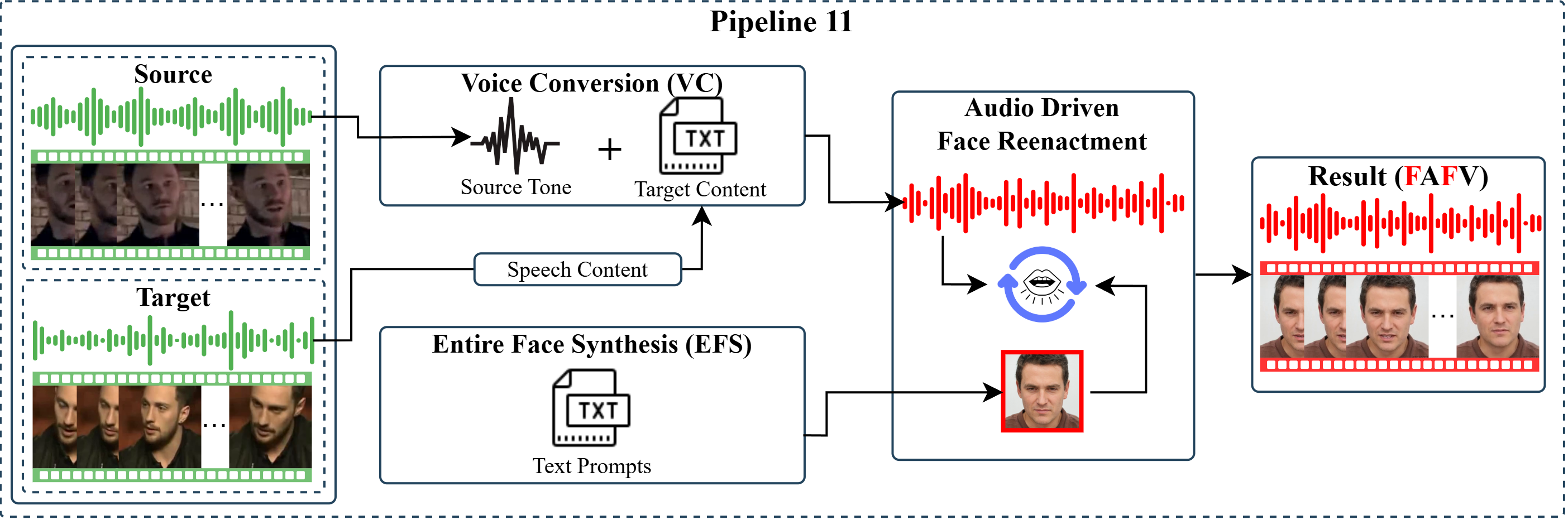}
    
    \caption{Illustration of Pipelines 8 to 11.}
    \label{fig:pipeline8-11}
\end{figure*}

\begin{figure*}[ht]
    \centering
    \includegraphics[width=0.9\linewidth]{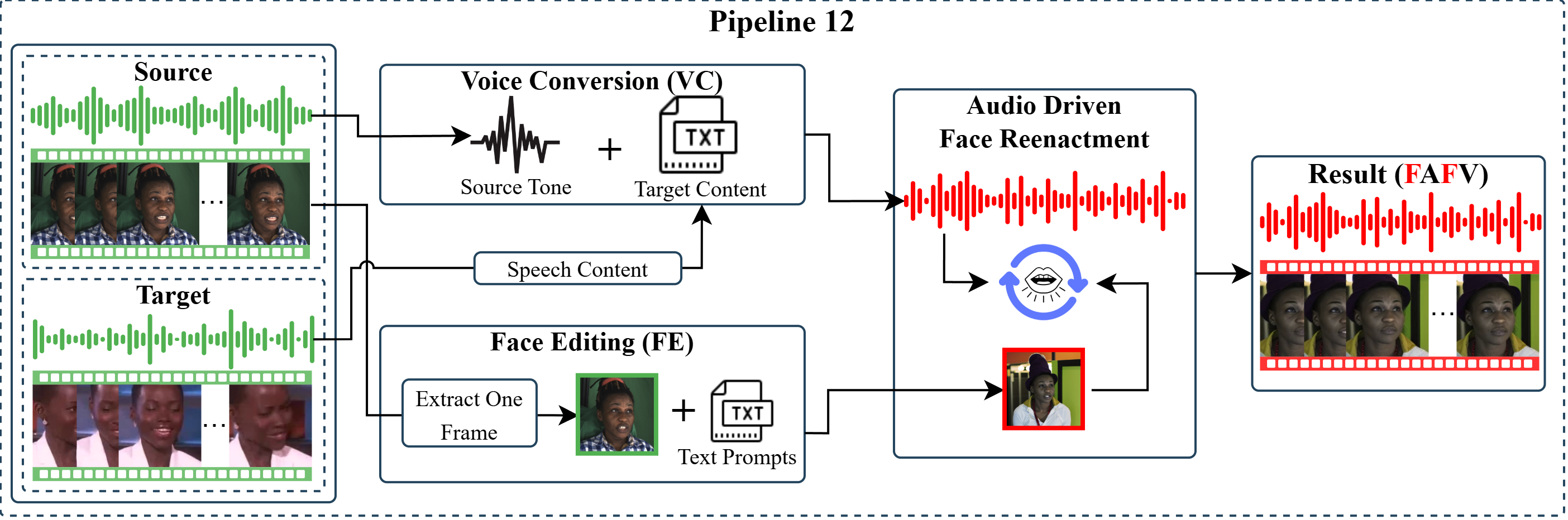}
    
    \vspace{0.2cm}  
    \includegraphics[width=0.9\linewidth]{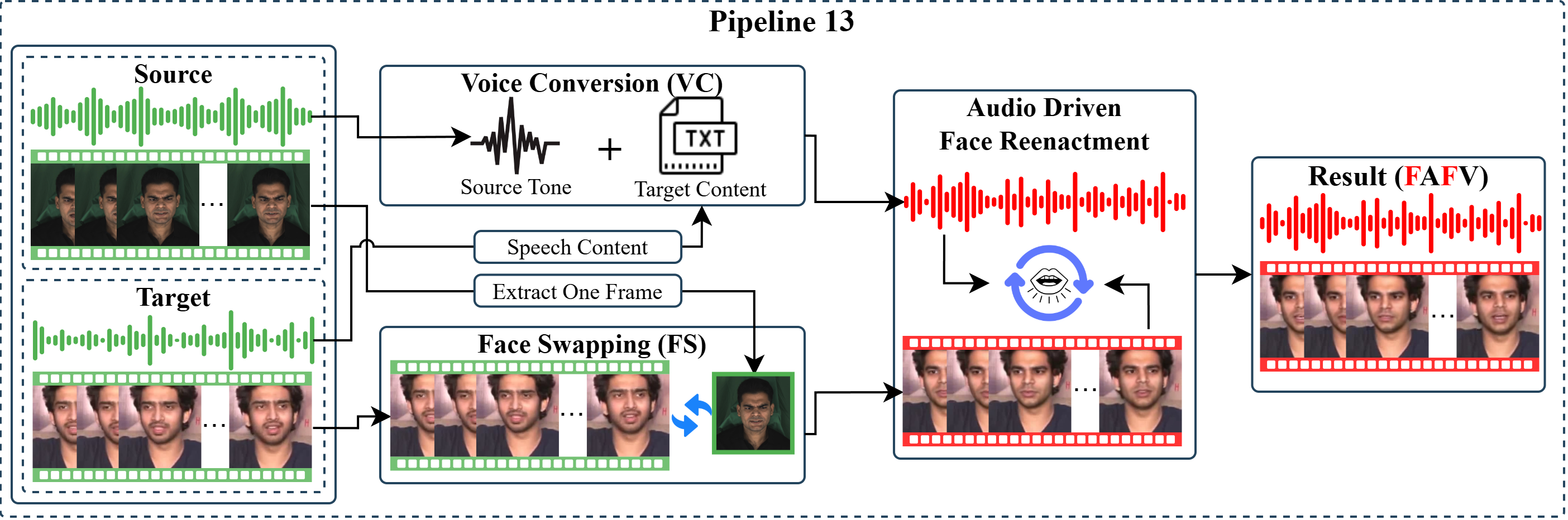}

    \vspace{0.2cm}  
    \includegraphics[width=0.9\linewidth]{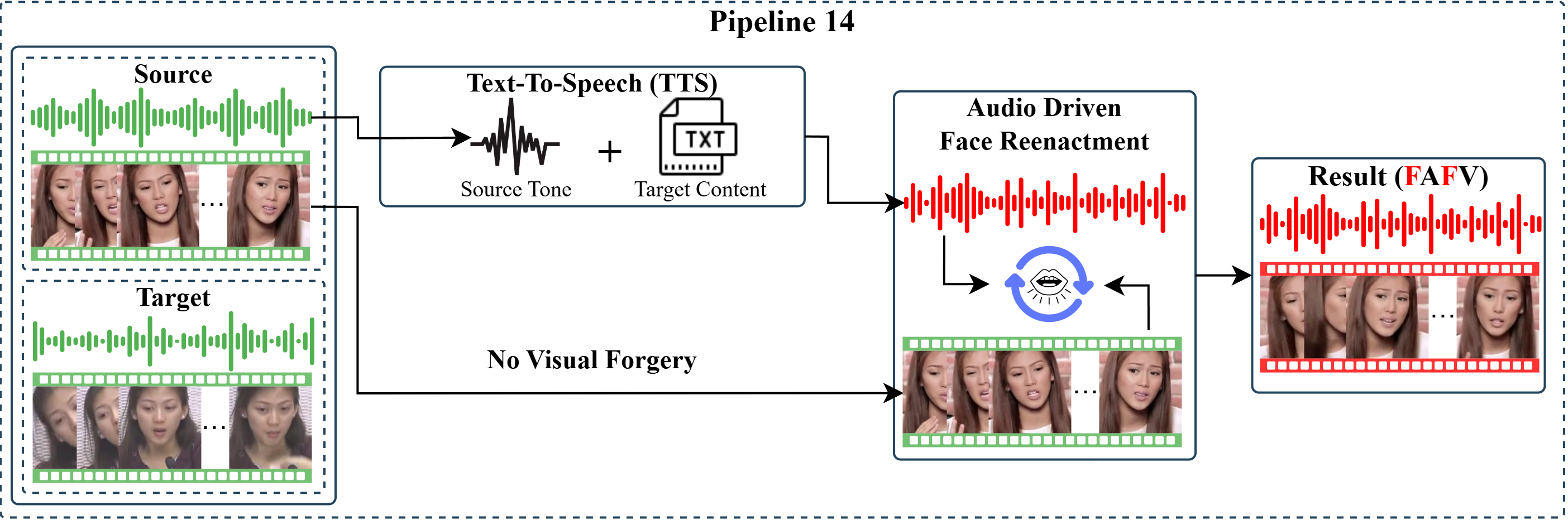}

    \vspace{0.2cm}  
    \includegraphics[width=0.9\linewidth]{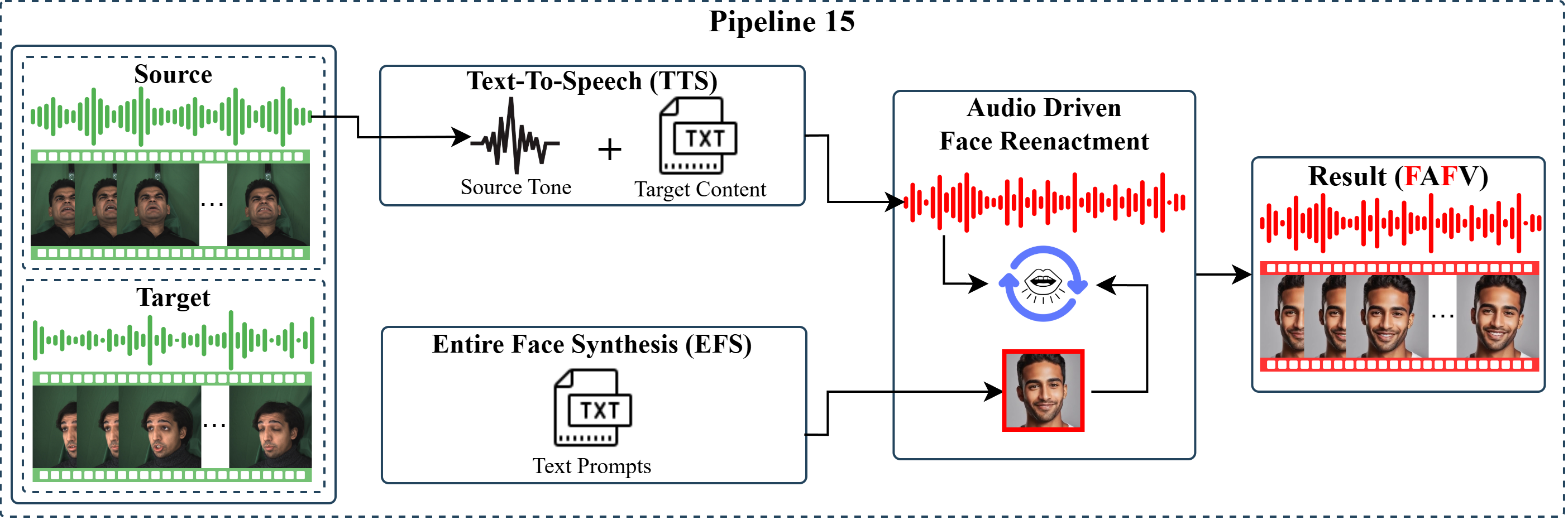}
    
    \caption{Illustration of Pipelines 12 to 15.}
    \label{fig:pipeline12-15}
\end{figure*}

\begin{figure*}[ht]
    \centering
    \includegraphics[width=0.9\linewidth]{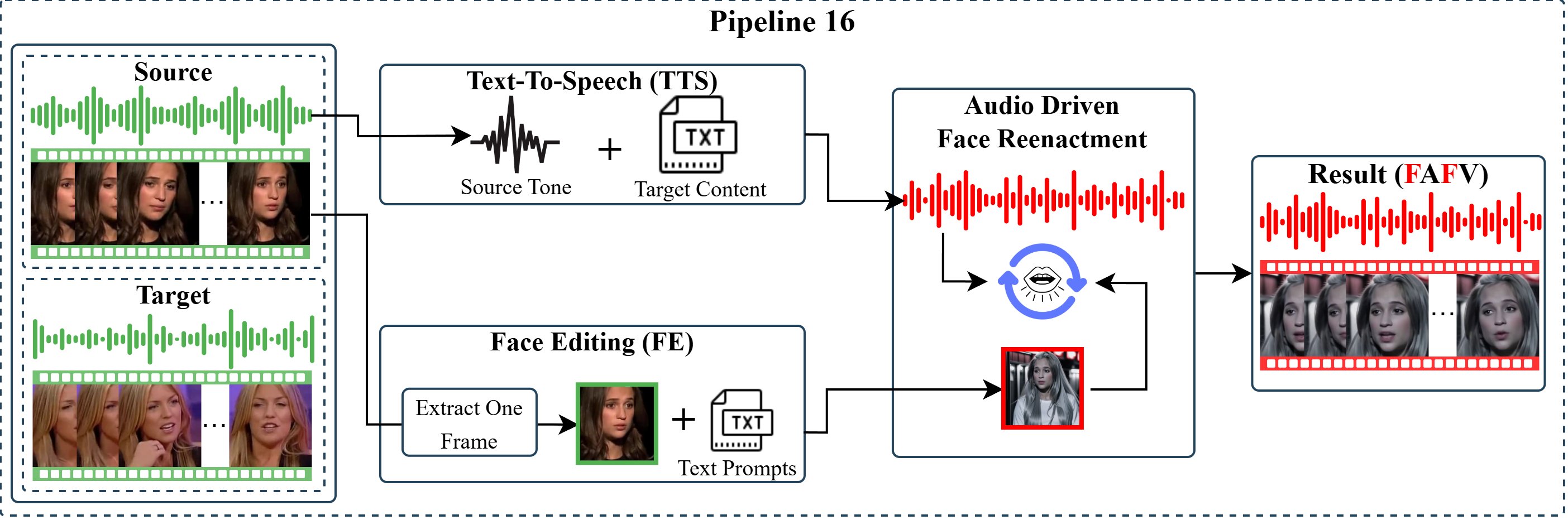}
    
    \vspace{0.2cm}  
    \includegraphics[width=0.9\linewidth]{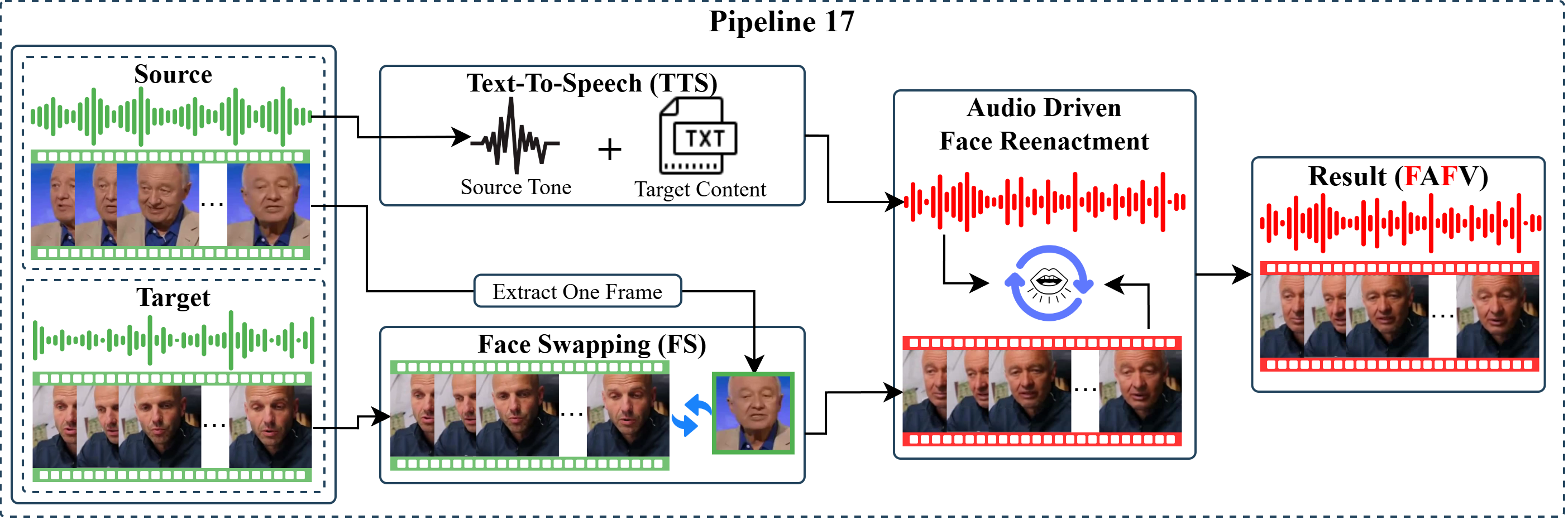}

    \vspace{0.2cm}  
    \includegraphics[width=0.9\linewidth]{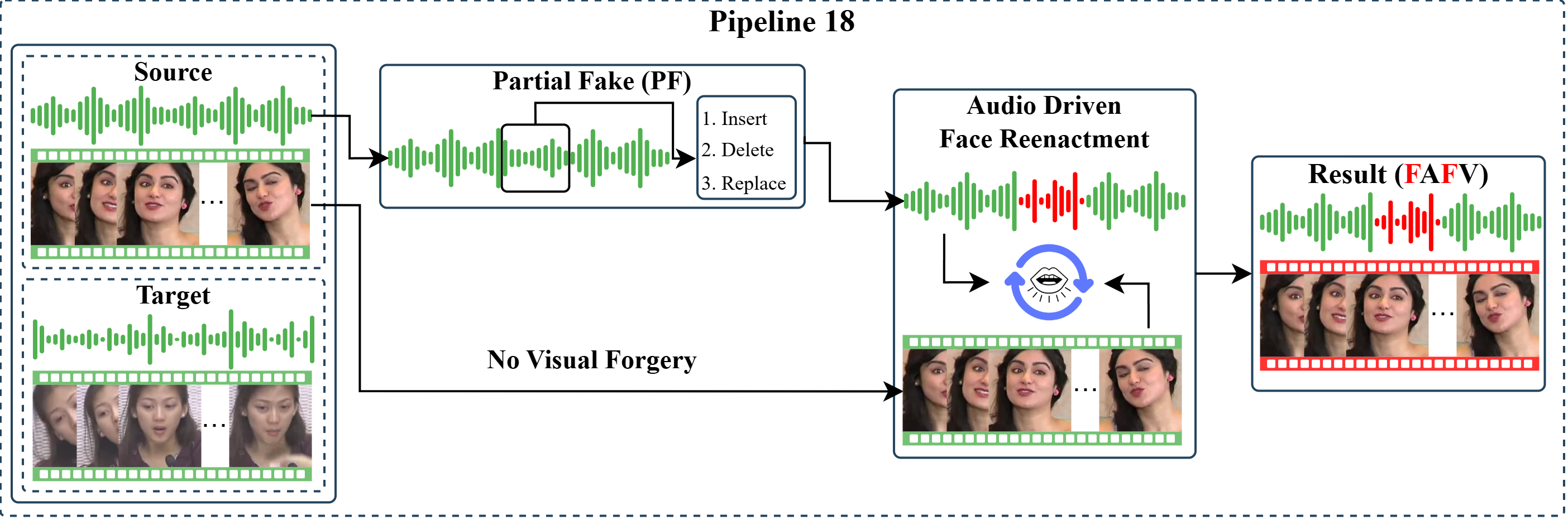}

    \vspace{0.2cm}  
    \includegraphics[width=0.9\linewidth]{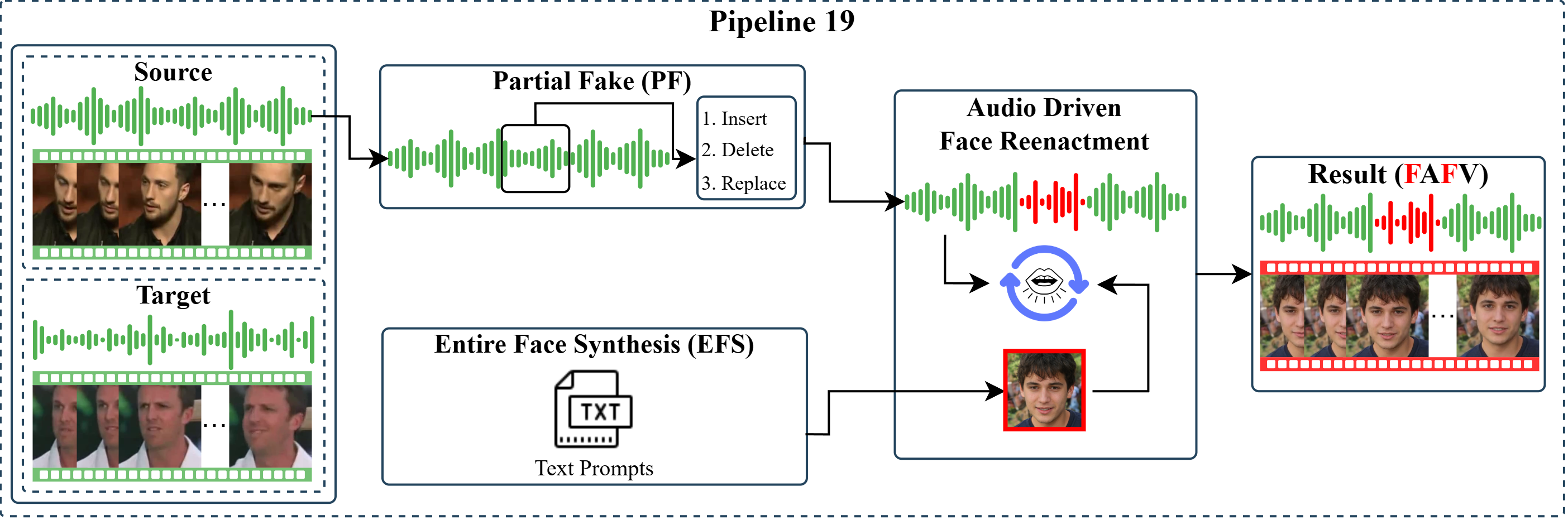}
    
    \caption{Illustration of Pipelines 16 to 19.}
    \label{fig:pipeline16-19}
\end{figure*}

\begin{figure*}[ht]
    \centering
    \includegraphics[width=0.9\linewidth]{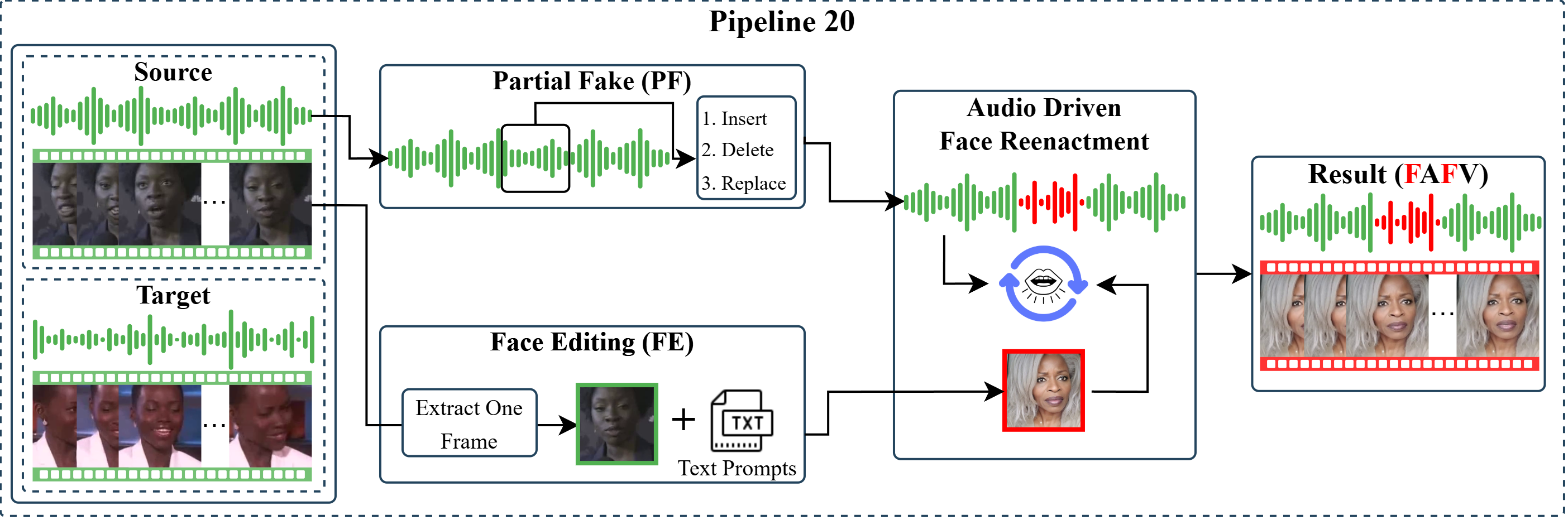}
    
    \vspace{0.2cm}  
    \includegraphics[width=0.9\linewidth]{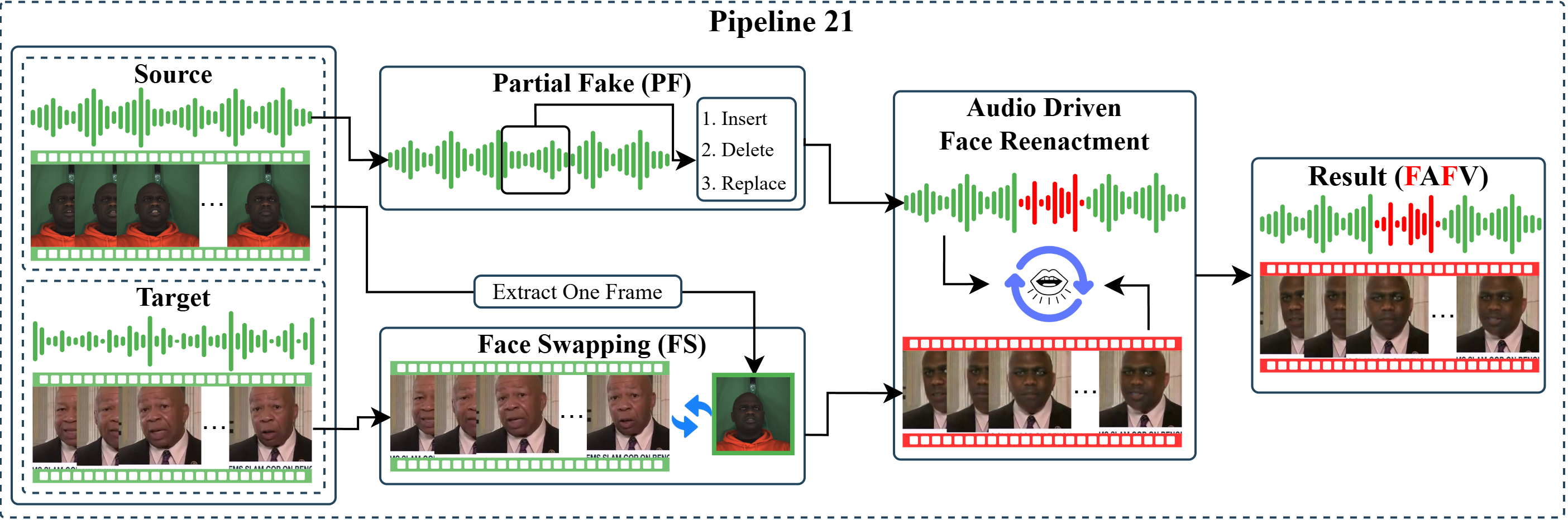}
    
    \vspace{0.2cm}
    \includegraphics[width=0.9\linewidth]{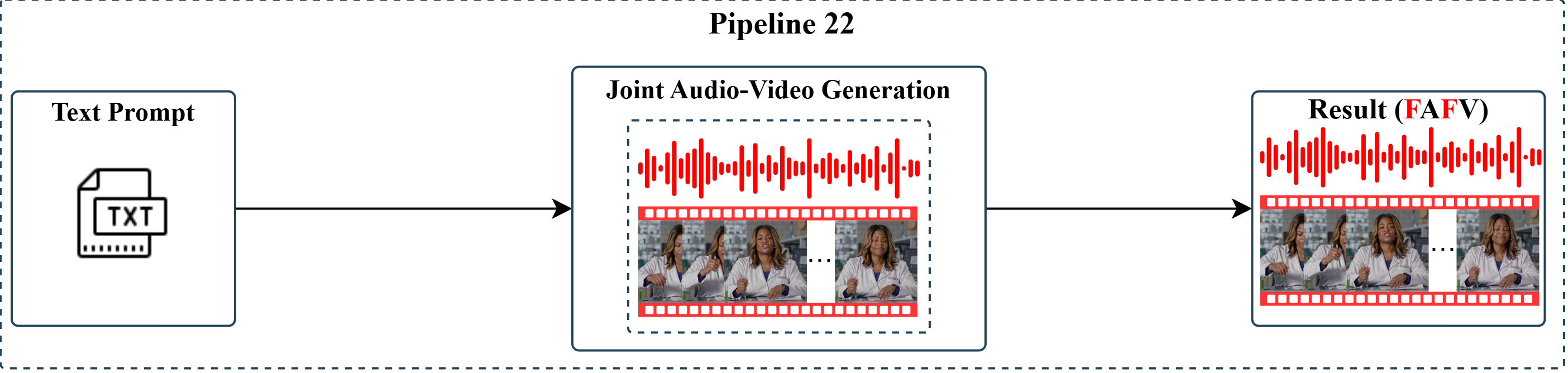}

    \caption{Illustration of Pipelines 20 to 22.}
    \label{fig:pipeline20-22}
\end{figure*}

Figs.~\ref{fig:pipeline0-3}, \ref{fig:pipeline4-7}, \ref{fig:pipeline8-11}, \ref{fig:pipeline12-15}, \ref{fig:pipeline16-19}, and~\ref{fig:pipeline20-22} illustrate the individual pipelines. Randomly selected clips from Mega-MMDF are shown in Figs.~\ref{fig:pipeline_samples_0_9}, \ref{fig:pipeline_samples_10_19}, and~\ref{fig:pipeline_samples_20_22}.

% ===== Appendix A.5 Data Statistics ===== %
% annotations
\subsection{Dataset Statistics}
\label{appendix:dataset_statistics}

\begin{figure}[t]
    \centering
    \includegraphics[width=\columnwidth]{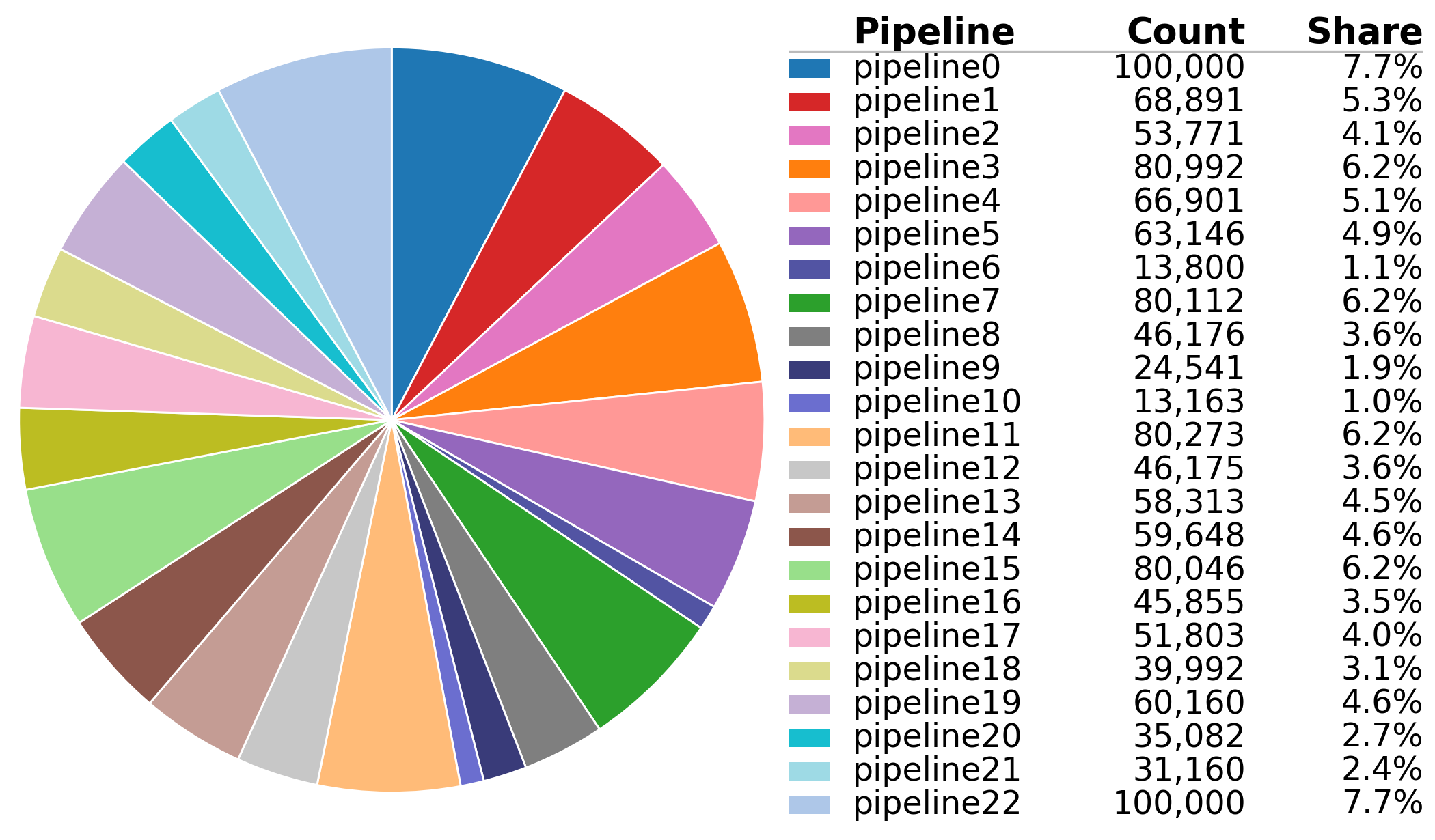}
    \caption{Clip distribution by pipeline ID in Mega-MMDF}
    \label{fig:pipeline_distribution}
\end{figure}

\paragraph{Dataset Composition} Mega-MMDF contains 1.3 million audio-visual clips, including 100,000 authentic clips and 1.2 million forged clips. The forged clips comprise 287,291 clips with manipulated video only, 80,992 with manipulated audio only, and 831,717 with both modalities manipulated. Fig.~\ref{fig:pipeline_distribution} presents the clip
counts by pipeline ID. The training, validation, and testing sets contain $780,866$, $259,163$, and $259,971$ clips, respectively, approximately following a 6:2:2 split.

\begin{figure}[t]
    \centering
    \includegraphics[width=\columnwidth]{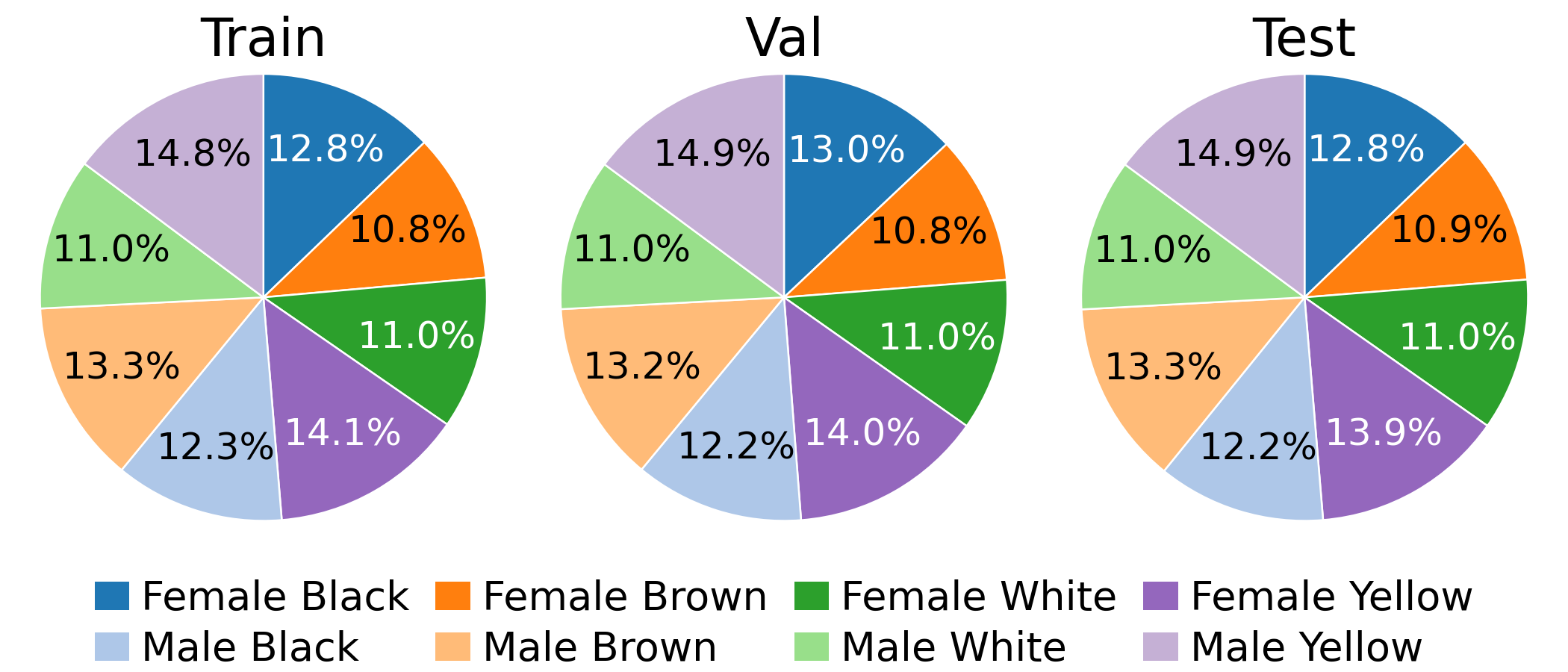}
    \caption{Demographic distributions in the training, validation, and test sets.}
    \label{fig:demographic_distribution}
\end{figure}

\paragraph{Demographic Distribution} 
Fig.~\ref{fig:demographic_distribution} summarizes the clip distribution across the eight annotated demographic groups. Each group accounts for 10.81\%--14.81\% of the dataset. The group proportions remain closely aligned across the training, validation, and test sets. For each demographic group, its proportions in the training, validation, and test sets differ by less than 0.18 percentage points.

\begin{figure}[t]
    \centering
    \includegraphics[width=\columnwidth]{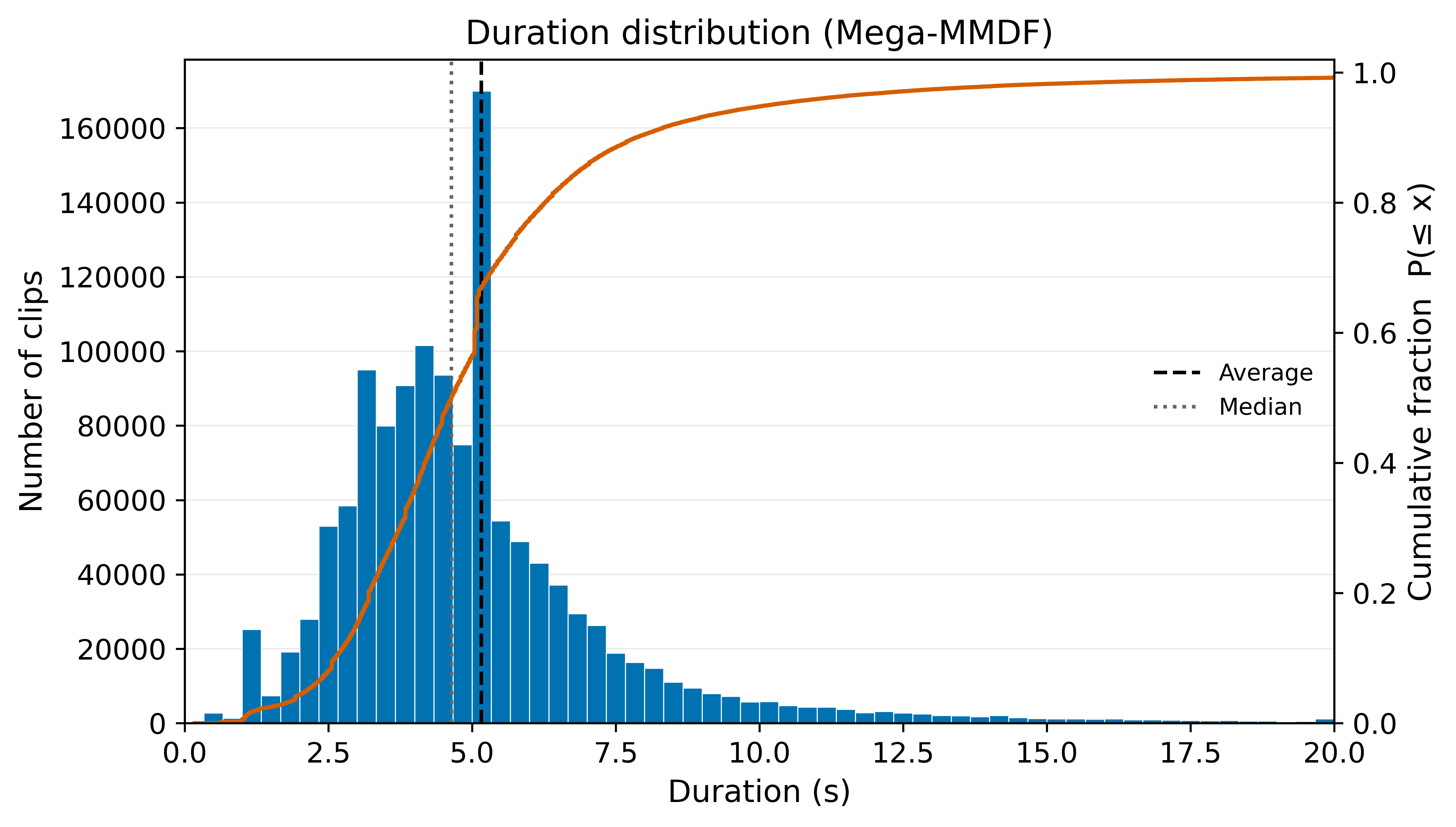}
    \caption{Distribution of clip durations in Mega-MMDF. The dashed and dotted lines indicate the mean and median, respectively.}
    \label{fig:duration_distribution}
\end{figure}

\paragraph{Clip Duration} Fig.~\ref{fig:duration_distribution} presents the distribution of clip durations in Mega-MMDF. The mean and median durations are 5.16~s and 4.64~s, respectively, with an interquartile range of 3.47--5.76~s. Most clips span several seconds: 95.70\% are at least 2~s long, while only 0.32\% are shorter than 1~s.

\begin{figure}[t]
    \centering
    \includegraphics[width=\columnwidth]{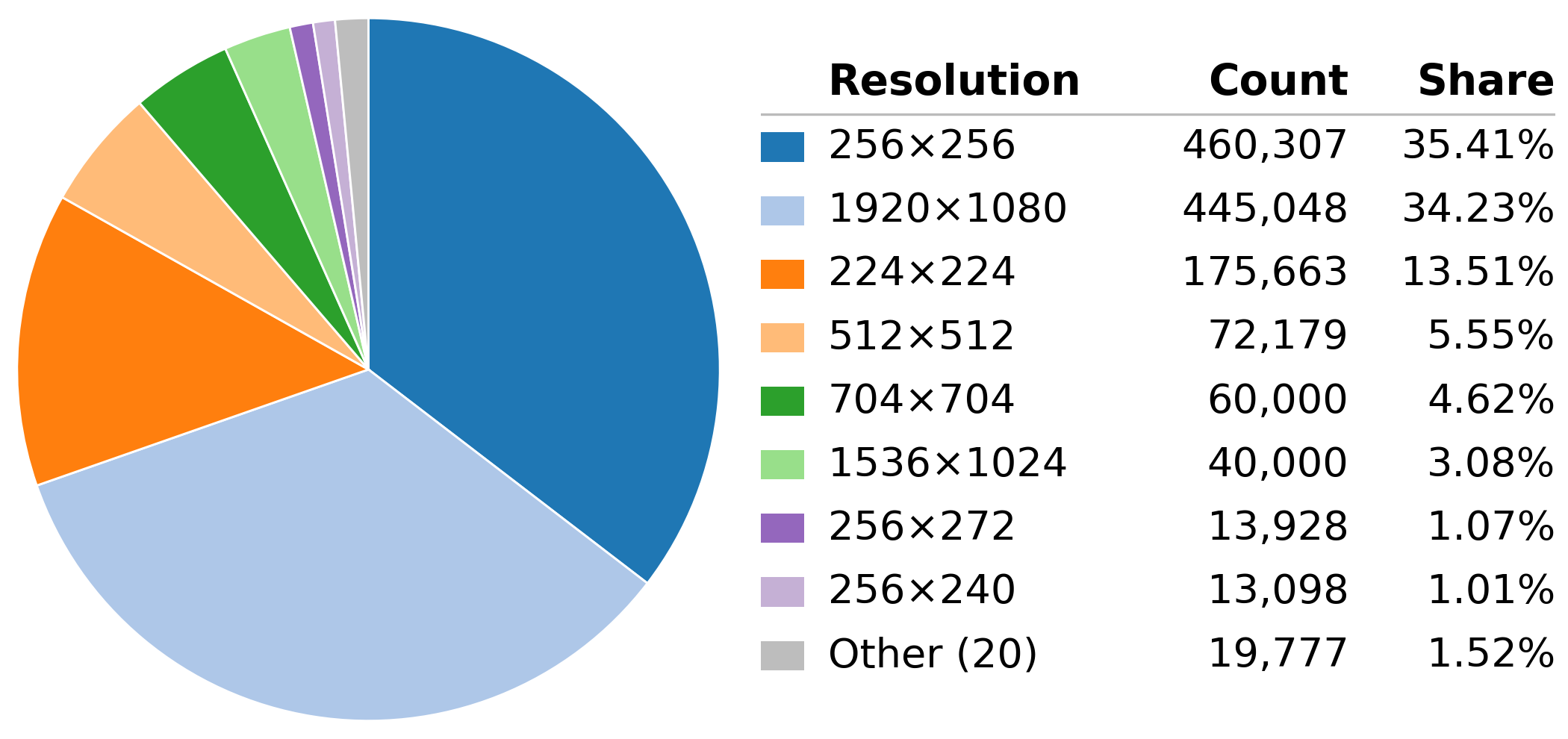}
    \caption{Distribution of video resolution in Mega-MMDF.}
    \label{fig:resolution_distribution}
\end{figure}

\paragraph{Spatial Resolution} Fig.~\ref{fig:resolution_distribution} summarizes the distribution of stored video resolutions in Mega-MMDF. The dataset contains 28 distinct frame resolutions. The three most common resolutions are $256 \times 256$, $1920 \times 1080$, and $224 \times 224$, accounting for 35.41\%, 34.23\%, and 13.51\% of the videos, respectively.

\subsection{Carbon Emissions}
We estimate that the construction of Mega-MMDF resulted in approximately 3,100~kg of CO$_2$-equivalent emissions, using the ML CO$_2$ Impact tool \footnote{\url{https://mlco2.github.io/impact/}}.

\section{Additional Details on DeepfakeBench-MM}
\label{appendix:benchmark}

DeepfakeBench-MM provides a modular framework for standardized comparison and systematic analysis of audio--visual deepfake detectors. This section describes its six modules: data preprocessing, data loading, detector, training, evaluation, and analysis. We detail their organization, extension interfaces, and the procedures and configurations used in this work.

% ===== 01_preprocessing =====
\subsection{Preprocessing Module}
\label{appendix:benchmark_preprocessing}

Audio--visual datasets differ in media formats, frame rates, audio sampling rates, directory structures, and annotation formats. These differences require dataset-specific handling before clips can be accessed through a common interface. The preprocessing module converts source audio--visual clips and their annotations into aligned face videos, resampled audio, facial landmarks, and metadata in a common format.

\paragraph{Framework and Extensibility}
The module combines a shared clip-processing procedure with dataset-specific adapters. The shared procedure performs face detection, continuous-sequence selection, and face alignment. Each adapter reads the dataset's files and annotations, prepares the required inputs, and organizes the outputs. We describe the shared procedure first, followed by the dataset-level organization and extension mechanism.

\begin{figure}[t]
    \centering
    \includegraphics[width=\columnwidth]{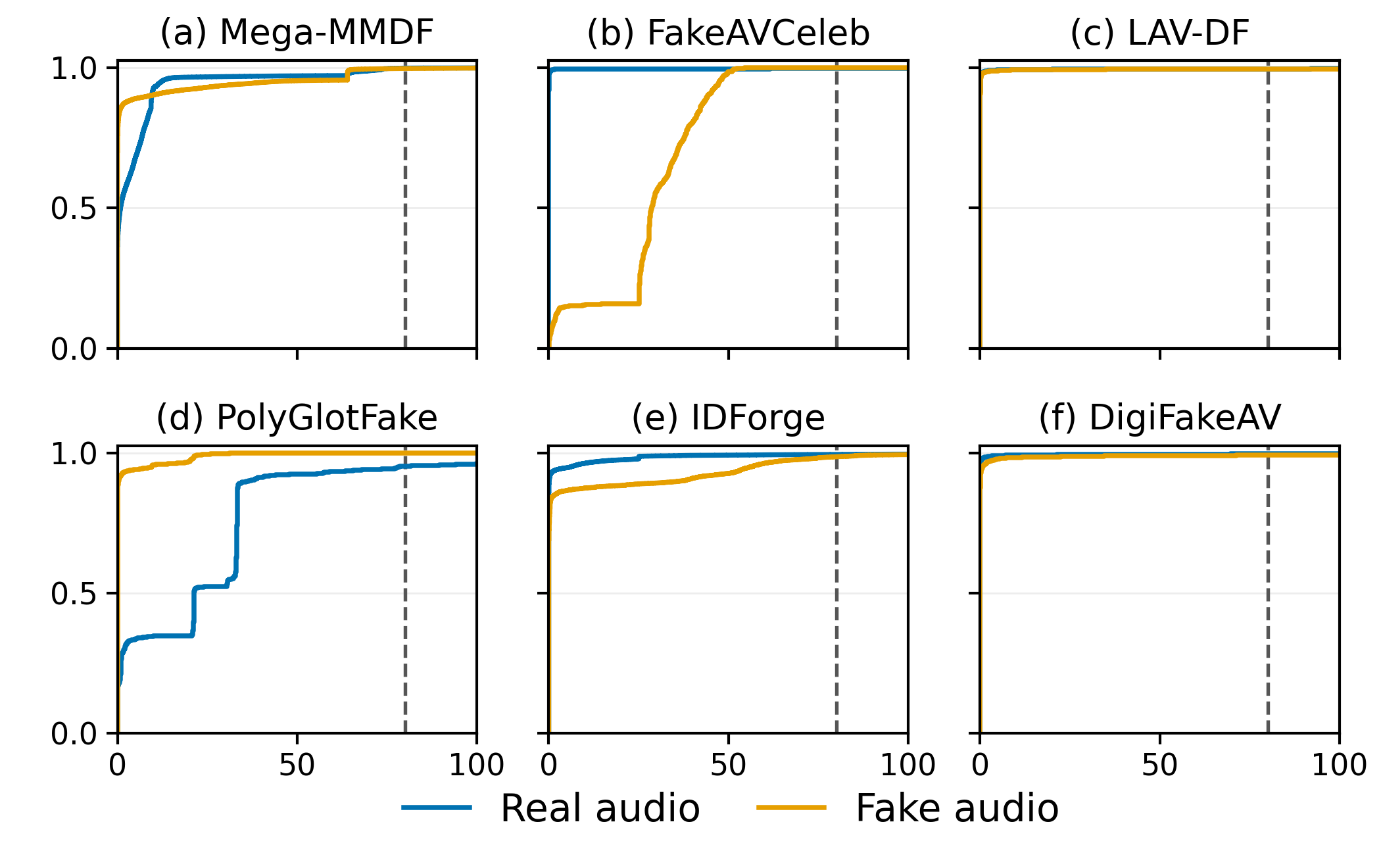}
    \caption{Empirical cumulative distributions of leading silence for real and fake audio across six datasets. Horizontal axes show leading-silence duration in milliseconds; vertical axes show cumulative proportions. Dashed lines mark 80~ms, corresponding to two frames at 25~fps. Following~\cite{AVH}, leading silence is measured on 16~kHz mono audio as the time before the first sample whose absolute waveform amplitude exceeds $5\times10^{-4}$.}
    \label{fig:benchmark_leading_silence}
\end{figure}

\paragraph{Processing an Individual Recording}
Algorithm~\ref{alg:benchmark_preprocessing} summarizes the procedure. The input can be a single audio--video file or separate audio and video files. We first convert the video to 25 frames per second and the audio to 16\,kHz mono PCM. Motivated by AVH~\cite{AVH}, we examine leading-silence durations across the six benchmark datasets shown in Fig.~\ref{fig:benchmark_leading_silence}. For the vast majority of recordings, leading silence does not exceed 80\,ms, corresponding to two video frames. Therefore, we skip the first two frames to reduce the risk that detectors exploit leading silence as a shortcut for predicting authenticity. This operation excludes the corresponding initial audio interval from subsequent sampling. We then detect faces in every remaining video frame.

% Replace this placeholder with the measured leading-silence distribution.

A dlib frontal-face detector is used first. If it detects no face, we apply a CNN face detector to the same frame. For each frame containing exactly one detected face, we extract 68 facial landmarks.

A frame with no detected face or more than one detected face ends the current candidate sequence. We retain the longest sequence of consecutive frames that each contain exactly one detected face. This sequence must contain at least 25 frames, corresponding to one second. If no sequence meets this requirement, preprocessing is marked as unsuccessful and the recording is excluded from the processed split. Multiple-face frames are excluded from the retained sequence. Their presence does not cause the entire recording to be rejected if another sufficiently long sequence is available.

\begin{algorithm}[t]
\caption{Preprocessing an audio--visual clip}
\label{alg:benchmark_preprocessing}
\begin{algorithmic}[1]
\REQUIRE Data $x$, labels, and available manipulation intervals.
\ENSURE Aligned face sequence, audio, landmarks; or failure.
\STATE Resample video to 25 fps and audio to 16 kHz mono.
\STATE Skip the first 2 video frames.
\FOR{each remaining frame $I_t$}
    \STATE Detect faces using the frontal-face detector.
    \IF{no face is detected}
        \STATE Apply the CNN face detector.
    \ENDIF
    \IF{exactly one face is detected}
        \STATE Extract its 68 landmarks $L_t$; mark frame $t$ as valid.
    \ELSE
        \STATE Mark frame $t$ as invalid.
    \ENDIF
\ENDFOR
\STATE Find the longest consecutive run of valid frames, $[s,e)$.
\IF{no run exists or $e-s<25$}
    \STATE Mark preprocessing as unsuccessful and \RETURN failure.
\ENDIF
\FOR{$t=s,\ldots,e-1$}
    \STATE Average landmarks from $t$ to $\min(t+11,e-1)$.
    \STATE Estimate a similarity transform to the reference face.
    \STATE Align $I_t$ to $256\times256$ and transform $L_t$.
\ENDFOR
\STATE Save the aligned sequence, landmarks, and resampled audio.
\STATE Preserve source indices $[s,e)$, labels, and manipulation intervals.
\STATE \RETURN processed recording
\end{algorithmic}
\end{algorithm}

Following~\cite{AVHuBERT}, for each retained frame, we average its landmarks with those of the following frames, using a window of at most 12 frames in total, \ie, a full window contains the current frame and up to the next 11 frames. Five averaged landmarks around the eyes and nose determine a similarity transform to a reference face. We apply this transform to the current frame to obtain a $256\times256$ aligned face image. We also apply it to the current frame's original landmarks and save the resulting coordinates.

The output consists of the aligned video, its landmark sequence, and the resampled audio. The audio file retains the full recording, while the video filename records the retained sequence's start and end frame indices. Both indices refer to the normalized source timeline. The data-loading module uses them to extract audio from the same interval as the selected video frames.

If a recording is partially manipulated, we also preserve its annotated manipulation intervals on this timeline. We convert an interval $(t_s,t_e)$ in seconds to the half-open frame interval $[\operatorname{round}(25t_s),\operatorname{round}(25t_e)+1)$ to determine the labels of sampled frames and audio segments. Keeping these intervals allows the data-loading module to determine whether a sampled window overlaps a manipulated region.

\paragraph{Relation to Existing Preprocessing Procedures}
DeepfakeBench~\cite{DeepfakeBench} estimates the face-alignment transform separately for each selected frame using facial landmarks detected in this frame.\footnote{\url{https://github.com/SCLBD/DeepfakeBench/blob/main/preprocessing/preprocess.py}} Although this places faces in a common spatial layout, frame-to-frame changes in landmark positions can also change the crop's translation, scale, and rotation. Such crop jitter can introduce apparent motion into the processed video. We follow the temporal landmark-smoothing approach used in AVHuBERT~\cite{AVHuBERT}, with a 12-frame window to estimate the alignment transform.\footnote{\url{https://github.com/facebookresearch/av_hubert/blob/main/avhubert/preparation/align_mouth.py}}

However, the AVHuBERT preprocessing script fills in missing landmarks by linear interpolation between detected frames. Interpolation across a long gap may not reflect the actual face motion during that interval. Its face-selection procedure also does not explicitly associate the same face across frames. When multiple faces are visible, the selected face may therefore change over time.\footnote{\url{https://github.com/facebookresearch/av_hubert/blob/main/avhubert/preparation/detect_landmark.py}} Moreover, a clip-level forgery label may not identify which visible face was manipulated. Therefore, we exclude multiple-face frames to avoid this ambiguity and retain a continuous sequence with one detected face per frame. 

\paragraph{Dataset-Level Organization and Extensibility}
This study includes FakeAVCeleb~\cite{FakeAVCeleb}, LAV-DF~\cite{LAV-DF}, PolyGlotFake~\cite{PolyGlotFake}, IDForge~\cite{IDForge}, TalkingHeadBench~\cite{TalkingHeadBench}, MMDF~\cite{MMDF&X-AVDT}, DigiFakeAV~\cite{DigiFakeAV&DigiShield}, and our Mega-MMDF. For each dataset, dataset-specific code reads its directory structure and annotations to construct a list of clips. Each entry specifies the source file path, the output location, the split assignment, and the available labels and manipulation intervals. The shared clip-level procedure processes these entries, and successfully processed recordings are collected into split-specific metadata files.

Support for a new dataset requires extracting these inputs from its original files and annotations. The shared face-processing procedure and output format remain the same. This organization allows additional datasets with different file layouts or annotation formats to use the existing preprocessing and data-loading code.

In addition, different train--validation--test partitions can change the difficulty of an evaluation and make results from separate studies difficult to compare. To reduce this source of variation, DeepfakeBench-MM uses fixed partitions across detectors for datasets such as FakeAVCeleb~\cite{FakeAVCeleb}, PolyGlotFake~\cite{PolyGlotFake}, and DigiFakeAV~\cite{DigiFakeAV&DigiShield}, where official dataset splits are unavailable. The corresponding split information will accompany the benchmark release to support reproducible comparisons. 

\paragraph{Dataset Availability and Future Extensions}
To the best of our knowledge, test-set annotations specifying manipulation intervals for AVDeepfake1M~\cite{AVDeepfake1M} and AVDeepfake1M++~\cite{AVDeepfake1M++} have not been publicly released at the time of writing. The official challenge documentation describes interval annotations for training and validation, with test predictions evaluated through the challenge protocol.\footnote{\url{https://deepfakes1m.github.io/2024/details}; \url{https://deepfakes1m.github.io/2025/details}} For partially manipulated clips, a clip-level label does not determine whether a shorter sampled window contains a manipulation. Without temporal annotations, we cannot reliably assign authenticity labels to the windows selected after preprocessing and during data loading. Therefore, these datasets are not included in our DeepfakeBench-MM.

At the time of writing, access requests, downloading, and preprocessing for DeepSpeak~\cite{DeepSpeak}, KoDF~\cite{KoDF}, and ArEnAV~\cite{ArEnAV} are still in progress. We plan to extend the benchmark to these datasets as the necessary access, annotations, and processing become available, and to additional audio--visual face deepfake datasets in the future. If the source dataset's license and any required permissions explicitly allow redistribution of processed data, we also plan to provide preprocessed versions to users whose authorization to access the source dataset has been verified. Otherwise, users can run the scripts on their own authorized copies. This release strategy aims to reduce repeated preprocessing effort and facilitate subsequent research.

% ===== 02_data_loading =====
\subsection{Data Loading Module}
\label{appendix:benchmark_data_loading}

The data-loading module reads the standardized outputs of the preprocessing module and constructs detector-specific input instances from preprocessed audio--visual clips. Each instance consists of video frames and the corresponding audio segment sampled from a temporal window, together with its labels and metadata. A shared loading interface supports detector-specific input formats through configuration files.

\paragraph{Framework and Extensibility}
Temporal sampling is defined separately for training, validation, and testing. Let $N$ denote the number of frames in a preprocessed clip, $T$ the number of frames loaded per input instance, and $s$ the frame stride. Each instance occupies a window of $T\times s$ frames, corresponding to $T\times s/25$ seconds at the preprocessing frame rate of 25 frames per second.

For training and validation, the preprocessed audio--visual clip is divided into $K=\min(M,\lfloor N/(Ts)\rfloor)$ approximately equal temporal partitions, where $M$ is a configurable maximum number of instances per clip for training and validation, and is set to two in our detector configurations. Training randomly selects a complete window within each partition. This allows the detector to receive different temporal windows of the same clip across epochs. Validation instead selects a fixed window starting at each partition boundary. Using a small number of fixed windows reduces validation time and resource consumption while keeping temporal sampling deterministic.

Testing places complete windows consecutively along the preprocessed clips without overlap. It produces $\lfloor N/(Ts)\rfloor$ input instances and discards remaining frames that cannot form a complete instance window. This strategy covers as much of the clip as possible using complete, non-overlapping windows, without padding the final video window with zeros or repeating inference over overlapping windows. A clip shorter than one window contributes no input instances in any split.

The reading configuration specifies the visual resolution, RGB or grayscale format, frame stride, and pixel normalization mean and standard deviation. For audio, it specifies the target sampling rate and whether to return a waveform or extracted features. Supported feature types include STFT, Mel, Log-Mel, and MFCC. Feature extraction settings and optional spectral normalization depend on the selected backend, \eg, \texttt{torchaudio}, \texttt{librosa}, \texttt{python\_speech\_features}. The audio segment is read from the same temporal window as the video window.

Dataset manifests provide paths, split membership, and authenticity labels, together with manipulation intervals and demographic annotations when available. Datasets following this format can use the same loading procedure. Additional input representations or sampling policies can be implemented in the corresponding loading utilities without changing the detector's forward computation. The module also provides processing hooks for training augmentation, robustness evaluation, and modality masking. Their settings are described in Sections~\ref{appendix:benchmark_training} and~\ref{appendix:benchmark_evaluation}.

\paragraph{Input Configurations}
Table~\ref{tab:benchmark_input} lists the input configurations used by the evaluated detectors. Spatial sizes and frame counts refer to the loader outputs. Additional frame selection, cropping, and other model-specific input operations are described below the table.

\begin{table*}[t]
    \centering
    \caption{Detector input configurations.}
    \label{tab:benchmark_input}
    \small
    \setlength{\tabcolsep}{4pt}
    \renewcommand{\arraystretch}{1.15}

    \resizebox{\textwidth}{!}{%
    \begin{tabular}{ccccccccccc}
        \toprule
        \multirow{2}{*}{Detector}
        & \multicolumn{5}{c}{Video}
        & \multicolumn{4}{c}{Audio}
        & \multirow{2}{*}{Duration} \\
        \cmidrule(lr){2-6}
        \cmidrule(lr){7-10}
        & Color mode
        & Size
        & Mean / Std
        & \# Frames
        & Stride
        & SR (Hz)
        & \makecell[c]{Remove\\DC offset}
        & Spectrogram
        & Mean / Std
        & \\
        \midrule
        Baseline & RGB & 224 & \makecell[c]{$[0.485, 0.456, 0.406]$\\$[0.229, 0.224, 0.225]$} & 16 & 1 & 16,000 & No & MFCC & -- & 0.64~s \\
        \cmidrule[0.2pt](lr){1-11}
        AVFF~\cite{AVFF} & RGB & 224 & \makecell[c]{$[0.485, 0.456, 0.406]$\\$[0.229, 0.224, 0.225]$} & 16 & 5 & 16,000 & Yes & Log-Mel & \makecell[c]{$-4.2677393$\\$9.1379948$} & 3.20~s \\
        \cmidrule[0.2pt](lr){1-11}
        AVH~\cite{AVH} & Gray & $256^{*}$ & \makecell[c]{$0.421$\\$0.165$} & 25 & 1 & 16,000 & No & Log-Mel & -- & 1.00~s \\
        \cmidrule[0.2pt](lr){1-11}
        FGMDF~\cite{FGMDF} & RGB & 128 & -- & $100^{*}$ & 1 & 16,000 & No & -- & -- & 4.00~s \\
        \cmidrule[0.2pt](lr){1-11}
        FRADE~\cite{FRADE} & RGB & 224 & \makecell[c]{$[0.485, 0.456, 0.406]$\\$[0.229, 0.224, 0.225]$} & 16 & 1 & 16,000 & No & Log-Mel & -- & 0.64~s \\
        \cmidrule[0.2pt](lr){1-11}
        FauForensics~\cite{FauForensics} & RGB & 112 & \makecell[c]{$[0.485, 0.456, 0.406]$\\$[0.229, 0.224, 0.225]$} & 25 & 1 & 16,000 & No & Log-Mel & \makecell[c]{$-40$\\$40$} & 1.00~s \\
        \cmidrule[0.2pt](lr){1-11}
        MCL~\cite{MCL} & RGB & 224 & \makecell[c]{$[0.485, 0.456, 0.406]$\\$[0.229, 0.224, 0.225]$} & 30 & 1 & 32,000 & No & -- & -- & 1.20~s \\
        \cmidrule[0.2pt](lr){1-11}
        SpeechForensics~\cite{SpeechForensics} & Gray & $256^{*}$ & \makecell[c]{$0.421$\\$0.165$} & 25 & 1 & 16,000 & No & Log-Mel & -- & 1.00~s \\
        \cmidrule[0.2pt](lr){1-11}
        Qwen2.5-Omni~\cite{Qwen2.5-Omni} & RGB & $256$ & -- & 8 & Uniform$^{\dagger}$ & 16,000 & No & -- & -- & 4.00~s \\
        \cmidrule[0.2pt](lr){1-11}
        VideoLLaMA2.1~\cite{VideoLLaMA2.1} & RGB & $256$ & -- & 8 & Uniform$^{\dagger}$ & 16,000 & No & -- & -- & 4.00~s \\
        \bottomrule
    \end{tabular}%
    }

    \par\vspace{6pt}
    \begin{minipage}{\textwidth}
        \footnotesize
        $^{*}$ AVH and SpeechForensics crop an $88\times88$ mouth region from each loaded frame. FGMDF selects 40 of the 100 loaded frames: training uses four random offsets shared across ten consecutive ten-frame blocks, whereas inference uses approximately uniform selection. $^{\dagger}$For both large multimodal models, the loader returns eight $256\times256$ RGB frames uniformly selected from each 100-frame, 4-second window at 25 FPS, together with the complete synchronized 16 kHz mono waveform (64,000 samples). The zero-based frame indices are $[0,14,28,42,57,71,85,99]$, corresponding to approximately 2 FPS rather than a fixed integer frame stride. Test windows advance by 4 seconds without overlap.
    \end{minipage}
\end{table*}

\paragraph{Labels and Metadata}
Each instance retains its source path and temporal coordinates, joint and modality-specific authenticity labels, and available demographic annotations. Optional landmarks are loaded using the same indices as the video frames. For partially manipulated sequences, the manifest preserves manipulation intervals so that labels can account for the selected window. Frame-level visual labels indicate whether each sampled frame lies within a manipulated interval. The binary instance label is positive when either modality is labeled as manipulated.

% ===== 03_models =====
\subsection{Detector Module}
\label{appendix:benchmark_models}

Audio--visual deepfake detectors use different feature extractors, fusion methods, and training objectives. The detector module provides a common interface for running these detectors within the benchmark. It receives the inputs prepared by the data-loading module, computes features and authenticity predictions, and provides the losses required for training.

\paragraph{Framework and Extensibility}
Each detector implements feature extraction, classification, and loss computation through the same interface. During a forward pass, it reads the audio and video inputs and returns prediction scores together with any features needed for its losses or subsequent analysis. During training, its loss function uses these outputs and the corresponding labels to compute an overall loss and individual loss terms. The training and evaluation modules can therefore call different detectors through the same interface while each detector retains its own architecture and objective.

The codebase maintains registries for detectors, backbones, and loss functions, allowing registered components to be selected by name in the configuration files. To add a detector, users implement its forward computation and loss calculation, register the detector, and specify its input and training configurations. Existing backbone and loss components can be reused, and additional components can be added when needed. This structure supports different audio and visual encoders, fusion operations, and combinations of joint and modality-specific supervision. Separate detector configurations can also support different stages of a training procedure, such as pretraining and subsequent classification.

\paragraph{Detectors Included in This Study}
We include a vanilla Baseline and six existing detectors: AVFF~\cite{AVFF}, AVH~\cite{AVH}, FGMDF~\cite{FGMDF}, FRADE~\cite{FRADE}, FauForensics~\cite{FauForensics}, and MCL~\cite{MCL}. In the following, we summarize their backbones, key modules, and training objectives. The input configurations and training settings are provided in Table~\ref{tab:benchmark_input} and Section~\ref{appendix:benchmark_training}, respectively. We denote commonly used cross-entropy losses for joint, audio, and video authenticity classification by $\mathcal{L}_{\mathrm{CE}}^{av}$, $\mathcal{L}_{\mathrm{CE}}^{a}$, and $\mathcal{L}_{\mathrm{CE}}^{v}$.

\paragraph{Baseline}
Our baseline uses a 3D ResNet-18 video encoder and an SE-ResNet audio encoder. Temporal averaging on features produces one 512-dimensional vector in either modality. Their concatenation is passed to an MLP consisting of a 512-dimensional hidden layer, batch normalization, ReLU activation, dropout with probability 0.5, and a two-class output layer. The model is trained with joint cross-entropy, $\mathcal{L}_{\mathrm{CE}}^{av}$.

\paragraph{AVFF}
AVFF~\cite{AVFF} follows a two-stage learning procedure with MARLIN~\cite{MARLIN} and AudioMAE~\cite{AudioMAE} as the video and audio backbones. The first stage learns audio--visual correspondence from real data. Complementary Masking hides different temporal portions of the two modalities, so that a masked portion in one modality remains visible in the other. Audio-to-video and video-to-audio mappings provide cross-modal features to reconstruct the masked content. The pretraining objective combines an audio--visual contrastive loss, $\mathcal{L}_{c}$, with reconstruction and adversarial losses, $\mathcal{L}_{\mathrm{rec}}$ and $\mathcal{L}_{\mathrm{adv}}$. The contrastive loss brings paired audio and video representations closer, while reconstruction and adversarial training encourage the features to preserve the information needed to recover the missing content. Generator and discriminator updates alternate during adversarial pretraining.

The second stage uses the pretrained encoders and cross-modal mappings without complementary masking. Original and cross-modal features are combined and passed to an MLP classification head. Training with $\mathcal{L}_{\mathrm{CE}}^{av}$ adapts the learned representations to final decisions.

\paragraph{AVH}
We adopt the supervised variant of AVH, termed AVH-sup in the original paper~\cite{AVH}, and refer to it as AVH in this benchmark. A frozen AVHuBERT~\cite{AVHuBERT} backbone extracts audio and visual features separately. An MLP scores the paired features at each time step, and log-sum-exp pooling aggregates the temporal scores into a prediction for the input window. Our implementation uses a two-class output trained with $\mathcal{L}_{\mathrm{CE}}^{av}$.

\paragraph{FGMDF}
FGMDF~\cite{FGMDF} represents audio and video features as a heterogeneous graph. A convolutional visual encoder and a waveform encoder based on sinc filtering and residual blocks extract modality-specific features. Cross-modal graph interaction exchanges information between audio and visual nodes to capture their temporal relationships. Graph attention aggregates the node features, while De-homogenization Graph Pooling removes redundant nodes to preserve diverse graph-level information. We retain the audio and video classification heads and use a binary joint head for this benchmark. The training objective is
\begin{equation}
    \mathcal{L}_{\mathrm{FGMDF}}=
    \mathcal{L}_{\mathrm{CE}}^{v}+
    \mathcal{L}_{\mathrm{CE}}^{a}+
    \mathcal{L}_{\mathrm{CE}}^{av},
\end{equation}
which encourages the model to identify both modality-specific forgeries and the overall authenticity of an audio--video pair.

\paragraph{FRADE}
FRADE~\cite{FRADE} uses a shared ViT-B/16 backbone to process visual frames and audio spectrograms. Its Adaptive Forgery-aware Injection (AFI) module extracts high-frequency cues and injects them into attention computations, allowing modality-specific forgery information to influence the pretrained features. Its Audio-distilled Cross-modal Interaction (ACI) module uses learnable latent tokens to extract information from audio before interacting with visual features. This intermediate representation is intended to reduce interference from modality-specific information when learning audio--visual inconsistencies. The resulting audio and visual features are fused for classification. The original method freezes the pretrained ViT and learns the added modules.

The training objective combines three terms:
\begin{equation}
    \mathcal{L}_{\mathrm{FRADE}}=
    \mathcal{L}_{\mathrm{CE}}^{av}
    +0.3\mathcal{L}_{\mathrm{CMC}}
    +0.4\mathcal{L}_{\mathrm{CC}}.
\end{equation}
The Cross-modal Contrastive Loss, $\mathcal{L}_{\mathrm{CMC}}$, brings audio and visual features closer for authentic pairs and penalizes their similarity above a margin for manipulated pairs. The Audio-Visual Center Cluster Loss, $\mathcal{L}_{\mathrm{CC}}$, brings authentic fused features toward a common center and encourages manipulated features to lie farther away. Together with authenticity classification, these objectives encourage compact authentic representations and separation from manipulated content.

\paragraph{FauForensics}
FauForensics~\cite{FauForensics} augments audio--visual detection with Facial Action Unit (FAU) features. Our implementation uses a CSN-101 video encoder, a Whisper~\cite{Whisper} audio encoder, and a frozen MEFARG~\cite{MEFARG} FAU extractor with a ResNet-50 backbone. FAU-enhanced Feature Learning combines facial-action features with visual features to capture subtle facial dynamics. Implicit Feature Alignment uses shared learnable queries to align the audio and visual representations. A Temporal Attentional Pooler then captures relationships within and between the two modalities, and separate classifiers predict joint and modality-specific authenticity. The objective is
\begin{equation}
    \mathcal{L}_{\mathrm{FauForensics}}=
    0.8\mathcal{L}_{\mathrm{CE}}^{av}
    +0.1\mathcal{L}_{\mathrm{CE}}^{a}
    +0.1\mathcal{L}_{\mathrm{CE}}^{v}.
\end{equation}
The auxiliary audio and video losses provide direct supervision for each modality along with the joint detection objective.

\paragraph{MCL}
MCL~\cite{MCL} combines frame, video, and audio information through two training stages. The first stage trains an EfficientNet-B7 frame teacher with frame authenticity labels. The second freezes this teacher and learns the video and audio branches. Following the implementation provided by the authors, we use ViViT and Cnn14, respectively. Frame and video features are combined into a frame--video representation, while an attention-based fusion module forms an audio--video representation. These compositional features allow the model to use spatial forgery cues from the teacher together with temporal and cross-modal information.

Multimodal contrastive learning uses authenticity labels to bring compatible representations closer and separate incompatible ones. The audio-related contrastive loss, $\mathcal{L}_{A}$, relates video features to audio and audio--video features; the frame-related contrastive loss, $\mathcal{L}_{I}$, relates video features to frame and frame--video features. A prediction-consistency term, $\mathcal{L}_{\mathrm{cons}}$, transfers information from the frame--video prediction to the video branch. The original paper defines this term as a Jensen--Shannon divergence loss, $\mathcal{L}_{\mathrm{jsd}}$. Noise-based Feature Augmentation (NFA) additionally perturbs fused features with learned multiplicative and additive noise during training to improve robustness to feature variation.

Using the weights in our training configurations, the second-stage objective is
\begin{equation}
    \mathcal{L}_{\mathrm{MCL}}=
    2\mathcal{L}_{\mathrm{CE}}^{v}
    +\mathcal{L}_{\mathrm{CE}}^{a}
    +\mathcal{L}_{A}+\mathcal{L}_{I}
    +\mathcal{L}_{\mathrm{cons}}
    +\mathcal{L}_{\mathrm{dis}}
    +\mathcal{L}_{\mathrm{NFA}}.
\end{equation}
Here, $\mathcal{L}_{\mathrm{dis}}$ classifies the audio--video compositional features, and $\mathcal{L}_{\mathrm{NFA}}$ classifies the features after noise augmentation. The modality-specific classification losses retain supervision for audio and video, while the contrastive and consistency terms encourage information transfer across representations. Noise augmentation is disabled during inference.

\paragraph{SpeechForensics}
SpeechForensics~\cite{SpeechForensics} detects inconsistencies between the speech content conveyed by lip movements and the corresponding audio. We use the pretrained AVHuBERT-Large~\cite{AVHuBERT} checkpoint used by the official implementation. Its visual frontend is a ResNet-18 with an initial 3D convolution, while a linear projection processes stacked Log-Mel audio features. A shared encoder with 24 Transformer blocks produces contextual speech representations. The original representation learning procedure uses authentic speech videos, with iteratively refined cluster assignments providing frame-level targets and masked prediction encouraging temporal context modeling. Modality dropout enables the encoder to process either modality independently. In our benchmark, the backbone remains frozen, with no benchmark-specific training or additional classification head. At inference, four consecutive Log-Mel feature frames are concatenated and layer-normalized before audio encoding. The model processes the mouth sequence and audio separately to obtain final-layer visual and audio embeddings. Following the official fixed time-offset alignment, we compute frame-wise cosine similarities for each integer offset from $-15$ to $15$, with out-of-range audio embeddings set to zero. The matching score $S$ is the maximum of the temporally averaged similarities across these offsets, selecting one constant offset for the entire input window. Lower matching scores indicate greater audio--visual inconsistency. We convert this score to $p_{\mathrm{fake}}=(1-S)/2$ on the benchmark's $[0,1]$ scale without learning a probability calibration. This procedure is applied to the non-overlapping 25-frame (1-second) windows specified in Table~\ref{tab:benchmark_input}; the original evaluation instead uses whole-video inputs.

\paragraph{Qwen2.5-Omni}
Qwen2.5-Omni~\cite{Qwen2.5-Omni} is a pretrained multimodal model with a Thinker--Talker architecture. The Thinker integrates visual and audio representations with text in an autoregressive language model, while the Talker supports speech generation. Time-aligned Multimodal Rotary Position Embedding (TMRoPE) aligns the temporal positions of video and audio inputs. We use the Qwen2.5-Omni-7B checkpoint and only its Thinker for authenticity scoring; the speech-generation branch is not used. All parameters remain frozen, with no benchmark-specific fine-tuning or additional classification head. Each input consists of eight uniformly sampled RGB frames and the synchronized 16 kHz mono audio from a non-overlapping 4-second window, as specified in Table~\ref{tab:benchmark_input}. The checkpoint processor prepares the multimodal inputs and applies the model's chat template.

We use the prompt: ``\texttt{Carefully analyze the video and audio to determine whether this audiovisual content is real or fake. Provide only a single-word response: Real or Fake.}'' Instead of sampling a textual answer, we obtain the next-token logits for the single-token labels ``\texttt{Real}'' and ``\texttt{Fake}'' at the final attended prompt position. Denoting these logits by $z_{\mathrm{Real}}$ and $z_{\mathrm{Fake}}$, respectively, we compute the detection score as
\begin{equation}
    p_{\mathrm{fake}} =
    \frac{\exp(z_{\mathrm{Fake}})}
    {\exp(z_{\mathrm{Real}})+\exp(z_{\mathrm{Fake}})}.
    \label{eq:mllm_authenticity_score}
\end{equation}
This two-label normalization produces a continuous score for benchmark evaluation without fitting a probability calibration to the test data.

\paragraph{VideoLLaMA2.1}
VideoLLaMA2.1~\cite{VideoLLaMA2.1} combines modality-specific encoders with a language model for audio--visual understanding. We use the VideoLLaMA2.1-7B-AV checkpoint, which comprises a SigLIP visual encoder, a BEATs audio encoder, and a Qwen2 language backbone. A spatial--temporal convolutional connector aggregates visual features across frames and maps them into the language model's embedding space. A two-layer MLP with GELU activation projects the audio features into the same space, allowing the language backbone to jointly process the audio--visual representations and the textual instruction. The audio branch uses 128-bin log-Mel filterbank features.

As with Qwen2.5-Omni, we keep the pretrained model frozen and evaluate it in a zero-shot setting without a learned deepfake classification head. We use the same synchronized 4-second windows, eight uniformly sampled frames, 16 kHz mono audio, and detection prompt, with the checkpoint's own visual preprocessing and chat template. The model processes both modalities in a single forward pass, and we compute $p_{\mathrm{fake}}$ from the next-token logits using Eq.~\ref{eq:mllm_authenticity_score}. Our implementation applies the vocabulary projection only to the final attended hidden state, yielding the same next-token scores as projecting every sequence position while reducing intermediate memory usage. Neither model is optimized on the benchmark training split.

% ===== 04_training =====
\subsection{Training Module}
\label{appendix:benchmark_training}

The training module uses the input instances provided by the data-loading module and the losses returned by each detector to optimize model parameters. It supports different training procedures through configuration files and manages validation and checkpoint storage.

\paragraph{Framework and Extensibility}
The module provides two training procedures. The basic trainer updates model parameters using the overall loss returned by the detector. The alternating trainer updates different parameter groups in turn, each with its own optimizer and loss, supporting procedures such as adversarial pretraining. Separate configurations define pretraining and subsequent classification stages. Both trainers support distributed training and the resumption from saved checkpoints.

The optimizer interface supports Adam, AdamW, AdamP, SGD, and RMSprop. The scheduler interface provides step decay through StepLR and cosine annealing through CosineAnnealingLR. These components and their parameters are selected through configuration files. Additional optimizers and schedulers can be integrated through the corresponding interfaces without changing the detector's forward computation.

\paragraph{Training Datasets and Configurations}
We train detectors separately on FakeAVCeleb~\cite{FakeAVCeleb}, LAV-DF~\cite{LAV-DF}, PolyGlotFake~\cite{PolyGlotFake}, IDForge~\cite{IDForge}, and our Mega-MMDF. All training runs are conducted on a single NVIDIA H100 GPU. Table~\ref{tab:training_optimizers} summarizes the training configurations on Mega-MMDF. AVFF and MCL have separate pretraining and fine-tuning stages. AVFF pretraining uses authentic clips and alternates generator and discriminator updates. MCL pretraining learns the frame teacher, which is subsequently frozen during multimodal training.

\begin{table}[t]
    \centering
    \caption{Configured training settings on Mega-MMDF. }
    \label{tab:training_optimizers}
    \small
    \setlength{\tabcolsep}{4pt}
    \renewcommand{\arraystretch}{1.15}
    \resizebox{\columnwidth}{!}{%
    \begin{tabular}{lcccc}
        \toprule
        Detector / Stage & Optimizer & Learning Rate & Epochs & Batch size \\
        \midrule

        Baseline
        & Adam & $1 \times 10^{-3}$ & 10 & 64 \\

        \multirow{2}{*}{AVFF (Pretrain)}
        & \multirow{2}{*}{AdamW}
        & G: $1.5\times10^{-4}$
        & \multirow{2}{*}{80}
        & \multirow{2}{*}{16} \\
        & & D: $1 \times 10^{-3}$ & & \\

        AVFF (Finetune)
        & AdamW & $1 \times 10^{-4}$ & 10 & 16 \\

        AVH
        & Adam & $1 \times 10^{-3}$ & 10 & 64 \\

        FGMDF
        & Adam & $1 \times 10^{-3}$ & 5 & 32 \\

        FRADE
        & AdamW & $1 \times 10^{-5}$ & 3 & 16 \\

        FauForensics
        & AdamP & $5\times10^{-4}$ & 5 & 64 \\

        MCL (Pretrain)
        & Adam & $5\times10^{-4}$ & 5 & 64 \\

        MCL (Finetune)
        & Adam & $2\times10^{-4}$ & 5 & 32 \\

        \bottomrule
    \end{tabular}%
    }
\end{table}

\paragraph{Default Data Augmentation}
The default training configuration applies augmentation to both video and audio. Visual augmentation includes horizontal flipping, rotation within $\pm10^{\circ}$, Gaussian blur with kernel sizes from 3 to 7, and image compression with quality values from 40 to 100. The same visual transformation parameters are applied to all frames within an instance to maintain temporal consistency.

Audio augmentation includes Gaussian noise with amplitudes from 0.001 to 0.01, pitch shifts within $\pm2$ semitones, and gain changes within $\pm6$ dB. An additional temporal augmentation shifts the audio within $\pm0.1$ s relative to the video. Each augmentation is applied with probability 0.5. These augmentations are disabled during validation and testing.

The module also supports optional modality masking during training. Dedicated configurations replace audio with silence or video with black frames, set the masked modality's label to real, and recompute the joint authenticity label from the remaining modality. Modality masking is disabled in the standard training configurations.

\paragraph{Validation and Checkpoint Selection}
For each training source, we validate the detector on the validation splits of the same five datasets: FakeAVCeleb, LAV-DF, PolyGlotFake, IDForge, and Mega-MMDF. At each validation round, we compute instance-level AUC on each validation split and take the average of the five scores. We save the checkpoint whenever this average improves and use the checkpoint with the highest average validation AUC for the reported evaluations. This rule provides a common checkpoint-selection criterion across detectors and training sources. All 8 datasets are involved in testing to compute the instance-level and clip-level metrics. Clip-level aggregation and test metrics are described in Section~\ref{appendix:benchmark_evaluation}.

% ===== 05_evaluation =====
\subsection{Evaluation Module}
\label{appendix:benchmark_evaluation}

The evaluation module measures detection performance, demographic fairness, and robustness to post-processing through three evaluation protocols. It receives detector predictions together with their labels and metadata, and computes metrics for comparison across models and datasets.

\paragraph{Framework and Extensibility}
Detector inference and metric computation are handled separately. Prediction records contain source paths, sampled temporal intervals, authenticity scores, labels, and demographic annotations when available. These records can be reused to compute different metrics without repeating inference. A common metric interface supports instance-level and clip-level evaluation. Additional metrics can be implemented through this interface, while additional test-time perturbations can be incorporated into the data-loading module.

\paragraph{Detection Performance Evaluation}
The benchmark currently implements four performance metrics: area under the receiver operating characteristic curve (AUC), average precision (AP), accuracy (ACC), and equal error rate (EER). AUC is the primary metric in this study, while the other three provide complementary measurements. The metric interface can be extended to include balanced accuracy, precision, recall, and other evaluation criteria. ACC uses a decision threshold of 0.5 by default, with scores greater than 0.5 classified as manipulated. EER is estimated at the empirical ROC operating point with the smallest absolute difference between the false-positive and false-negative rates.

Within-dataset evaluation measures performance on the held-out test split of the training dataset, whereas cross-dataset evaluation measures generalization to a different dataset. We consider five training sources: FakeAVCeleb~\cite{FakeAVCeleb}, LAV-DF~\cite{LAV-DF}, PolyGlotFake~\cite{PolyGlotFake}, IDForge~\cite{IDForge}, and Mega-MMDF. Evaluation covers these five datasets together with TalkingHeadBench~\cite{TalkingHeadBench}, DigiFakeAV~\cite{DigiFakeAV&DigiShield}, and MMDF~\cite{MMDF&X-AVDT}. 

Both instance-level and clip-level evaluation are supported. Instance-level evaluation directly uses the prediction and label of each instance. For clip-level evaluation, predictions from the same source clip are aggregated. Let $\mathcal{S}_c$ denote the set of evaluated windows from clip $c$. Its manipulation score and authenticity label are
\begin{equation}
    P_c = \max_{s\in\mathcal{S}_c} p_s,
    \qquad
    Y_c = \max_{s\in\mathcal{S}_c} y_s,
\end{equation}
where $p_s$ is the predicted manipulation probability and $y_s\in\{0,1\}$ is the label of instance $s$. Thus, a clip is labeled as manipulated when at least one evaluated instance contains a manipulation. This aggregation applies to the instances retained and evaluated by the benchmark. Our primary results use clip-level metrics.

The complete AUC results are reported in Table~\ref{tab:test-publication-full} in the main text. They include seven training-based detectors evaluated after training on each of the five training datasets. Three additional methods, SpeechForensics, Qwen2.5-Omni, and VideoLLaMA2.1, are evaluated without task-specific training.

\paragraph{Fairness Evaluation}
We adopt the four group-fairness criteria considered in~\cite{AIFace} and additionally report the standard deviation of group-wise AUC~\cite{Hu2024AttributeNeutral}: demographic parity (DP), equal odds (EO), maximum equalized odds (MEO), overall accuracy equality (OAE), and standard deviation of AUC (SD$_{\mathrm{AUC}}$). These metrics support both instance-level and clip-level evaluation; the fairness results in this study use clip-level predictions. The same decision threshold of 0.5 is applied to all demographic groups.

Let $D\in\mathcal{G}$ denote an annotated demographic group, and let $\hat{y}=\mathbf{1}[p>0.5]$ denote the predicted authenticity label. We define
\begin{equation}
    \begin{aligned}
        r_g &= P(\hat{y}=1\mid D=g),\\
        r_{g,q} &= P(\hat{y}=1\mid D=g,y=q),\\
        r_q &= P(\hat{y}=1\mid y=q),\\
        a_g &= P(\hat{y}=y\mid D=g).
    \end{aligned}
\end{equation}
The five fairness disparities are computed as
\begin{equation}
    \begin{aligned}
        F_{\mathrm{DP}}
        &= \max_{g\in\mathcal{G}} r_g
        - \min_{g\in\mathcal{G}} r_g,\\
        F_{\mathrm{EO}}
        &= \sum_{g\in\mathcal{G}}\sum_{q\in\{0,1\}}
        |r_{g,q}-r_q|,\\
        F_{\mathrm{MEO}}
        &= \max_{q\in\{0,1\}}
        \left(
        \max_{g\in\mathcal{G}} r_{g,q}
        - \min_{g\in\mathcal{G}} r_{g,q}
        \right),\\
        % F_{\mathrm{OAE}}
        % &= \max_{g\in\mathcal{G}} a_g
        % - \min_{g\in\mathcal{G}} a_g.
        F_{\mathrm{OAE}}
        &= \max_{g\in\mathcal{G}} a_g
        - \min_{g\in\mathcal{G}} a_g,\\
        \mathrm{SD}_{\mathrm{AUC}}
        &= \sqrt{
        \frac{1}{|\mathcal{G}|}
        \sum_{g\in\mathcal{G}}
        \left(\mathrm{AUC}_g-\overline{\mathrm{AUC}}\right)^2
        },
        \overline{\mathrm{AUC}}
        = \frac{1}{|\mathcal{G}|}
        \sum_{g\in\mathcal{G}}\mathrm{AUC}_g.
    \end{aligned}
\end{equation}

DP measures differences in predicted manipulation rates across demographic groups. EO measures deviations of group-specific false-positive and true-positive rates from the corresponding population-level rates. MEO captures the largest between-group difference in either rate, while OAE measures the largest accuracy difference. SD$_{\mathrm{AUC}}$ quantifies variation in AUC across groups. Lower values indicate smaller demographic disparities.

All detectors are evaluated using the same demographic group definitions, label conventions, and metric implementations. Each instance inherits the demographic annotation of its source clip. A common decision threshold of 0.5 is used for DP, EO, MEO, and OAE, while SD$_{\mathrm{AUC}}$ is computed from group-wise AUCs using continuous prediction scores. These shared settings support consistent comparisons across detectors.

\paragraph{Robustness Evaluation}
We focus on post-processing operations that can occur when audio--visual content is transcoded, resized, or transmitted online. Capture-related effects, such as motion blur, defocus, and overexposure, are outside the scope of this evaluation.

Six operations cover video compression, spatial downsampling, frame loss, audio compression, audio dropout, and audio--video temporal offset. Each operation is applied separately at three severity levels, as summarized in Table~\ref{tab:benchmark_robustness}. These settings represent increasing degradation inspired by practical distribution conditions.

\begin{figure*}
    \centering
    \includegraphics[width=\linewidth]{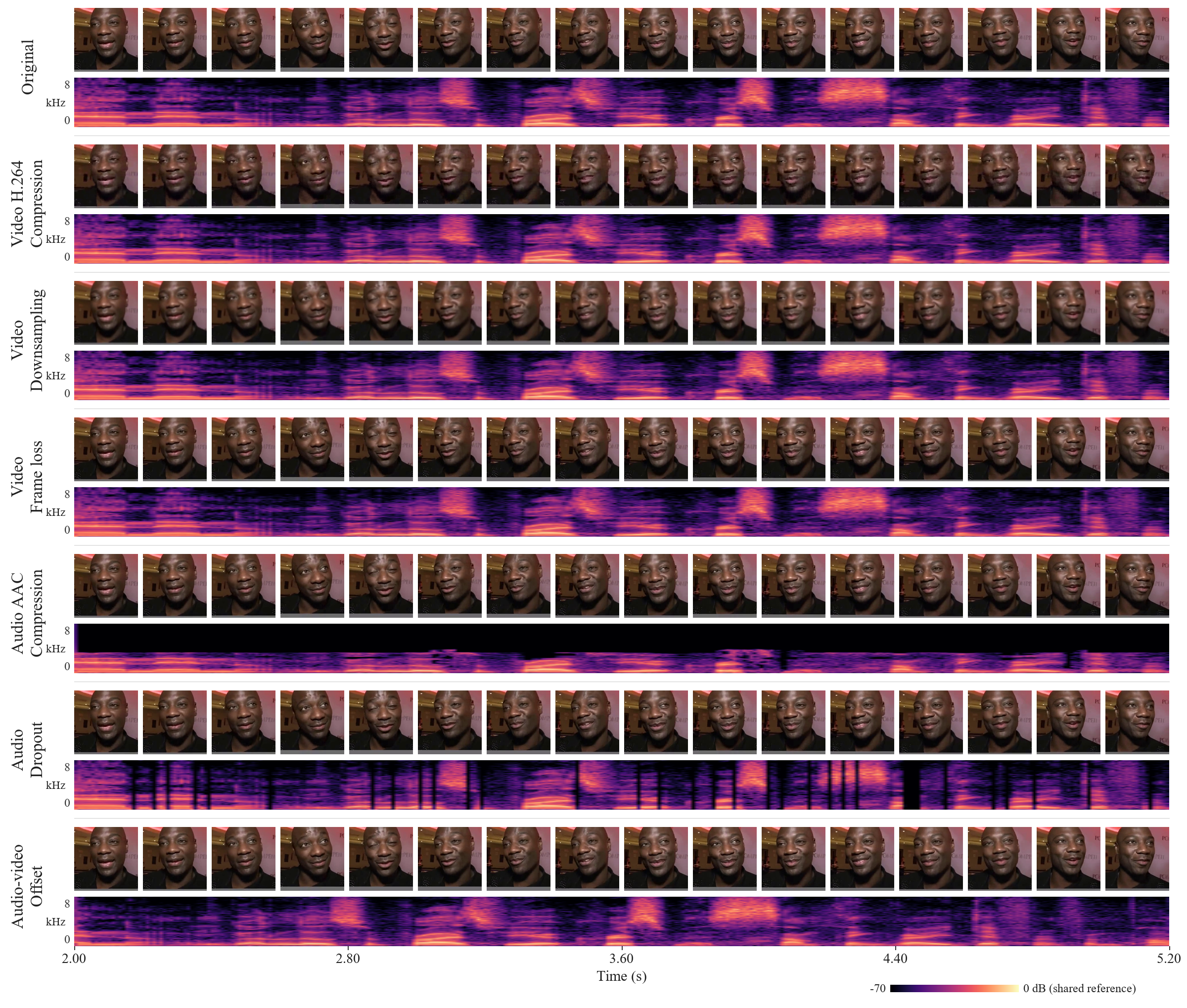}
    \caption{Examples of the six post-processing operations.}
    \label{fig:post_processing_sample}
\end{figure*}

\begin{table*}[t]
    \centering
    \caption{Post-processing operations and severity levels
    for robustness evaluation. Spatial ratios apply to both
    frame dimensions. Frame-loss and audio-dropout values
    are probabilities.}
    \label{tab:benchmark_robustness}
    \small
    \setlength{\tabcolsep}{4pt}
    \renewcommand{\arraystretch}{1.15}
    \begin{tabular}{lcccl}
        \toprule
        Operation / Parameter
        & Level 1
        & Level 2
        & Level 3
        & Intended degradation \\
        \midrule

        H.264 compression / CRF
        & 23 & 30 & 37
        & Increasing video compression during transcoding. \\

        Spatial downsampling / Ratio
        & $2/3$ & $4/9$ & $1/3$
        & Progressively reduced spatial detail during
        resolution adaptation. \\

        Frame loss / Probability
        & 0.05 & 0.10 & 0.20
        & Increasing frequency of frozen frames during
        interrupted delivery. \\

        AAC compression / Bitrate (kb/s)
        & 64 & 32 & 16
        & Increasing audio compression under reduced
        bitrate budgets. \\

        Audio dropout / Probability
        & 0.05 & 0.10 & 0.20
        & Increasing frequency of short audio
        interruptions. \\

        Video delay / Offset (ms)
        & 80 & 160 & 320
        & Increasing audio--video offset from
        unequal buffering or processing delays. \\

        \bottomrule
    \end{tabular}
\end{table*}

\textit{Video compression.}
We re-encode preprocessed videos using H.264 with the medium preset, High profile, and YUV420p pixel format. CRF values of 23, 30, and 37 introduce progressively stronger compression, representing increasing loss of visual detail during transcoding.

\textit{Spatial downsampling.}
We reduce both frame dimensions by factors of $2/3$, $4/9$, and $1/3$ using area interpolation, then restore the original dimensions using bilinear interpolation. These factors match the relative reductions from 1080p to 720p, 480p, and 360p, respectively. They are applied to the preprocessed face frames rather than to full-resolution source videos.

\textit{Frame loss.}
Each sampled frame is dropped with probability 0.05, 0.10, or 0.20 and replaced by the most recent retained frame. The first frame is always retained. This operation preserves sequence length while introducing increasingly frequent temporal freezes.

\textit{Audio compression.}
We apply a mono AAC-LC encode--decode round trip at 64, 32, or 16 kb/s. Lower bitrates introduce stronger compression under a reduced audio bitrate budget. The original sampling rate and sample count are preserved.

\textit{Audio dropout.}
% We replace randomly selected 20-ms waveform intervals with silence, using dropout probabilities of 0.05, 0.10, and 0.20. Two-millisecond fades are applied to adjacent available samples. Increasing the dropout probability produces more frequent interruptions in the audio signal.
We divide each waveform into non-overlapping 20-ms blocks and independently replace each block with silence with probability 0.05, 0.10, or 0.20. Two-millisecond fades are applied to adjacent available waveform samples.

\textit{Audio--video temporal offset.}
We introduce video delays of 80, 160, and 320 ms relative to audio. The implementation advances the audio content sampled for each visual window by the specified offset and uses silence padding when the requested interval exceeds the available waveform. These settings introduce progressively larger synchronization errors.

Video and audio compression are prepared offline, while the other operations are applied during data loading. For stochastic perturbations, the random seed is the configured base seed plus the sample index. Robustness comparisons use the same detector checkpoint and sampled visual windows across the clean and perturbed conditions, with audio intervals adjusted only for the temporal-offset test.

% ===== 06_analysis =====
\subsection{Analysis}
\label{appendix:benchmark_analysis}

\paragraph{Framework and Extensibility}
The analysis module uses prediction records saved during evaluation and intermediate features returned by the detector module. These records contain clip paths, instance time ranges, authenticity labels, prediction scores, and available demographic annotations. Features are associated with the corresponding instances through their clip paths and time ranges. The module provides ROC and precision--recall curves, confusion matrices, prediction-score distributions, demographic-group comparisons, pipeline-wise evaluation, and t-SNE visualization. Prediction-based tools support both instance-level and clip-level analysis, following the shared evaluation protocol for clip-level aggregation. Additional tools can reuse the saved records and exported features through the same interfaces.

\begin{figure*}[t]
    \centering
    \includegraphics[width=\textwidth]{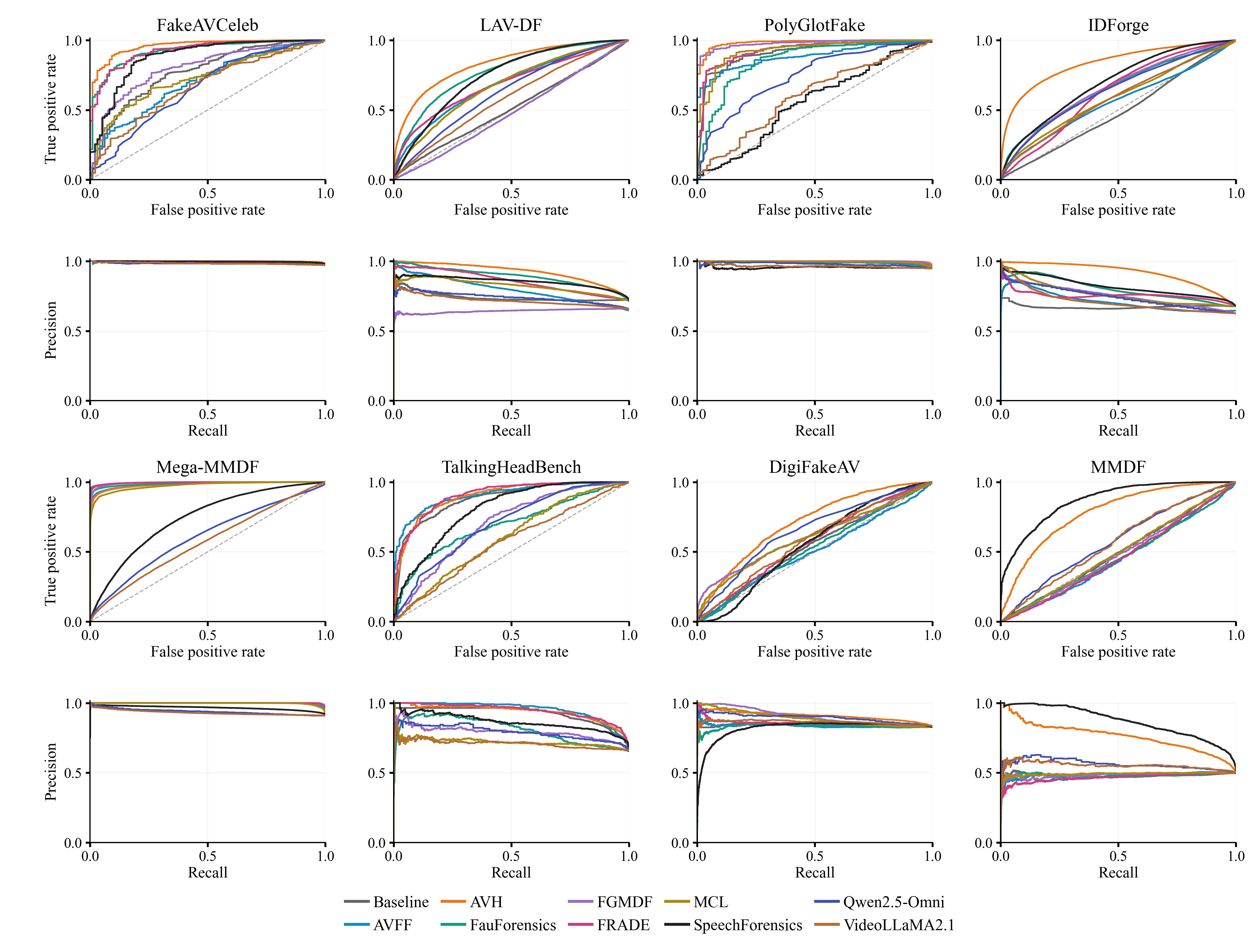}
    \caption{Clip-level ROC and precision--recall curves on the eight test datasets. Rows alternate between ROC and precision--recall curves; each dataset label applies to the corresponding pair of rows. The seven training-based detectors are trained on Mega-MMDF.}
    \label{fig:appendix-roc-pr}
\end{figure*}

\begin{figure*}[t]
    \centering
    \includegraphics[width=\textwidth]{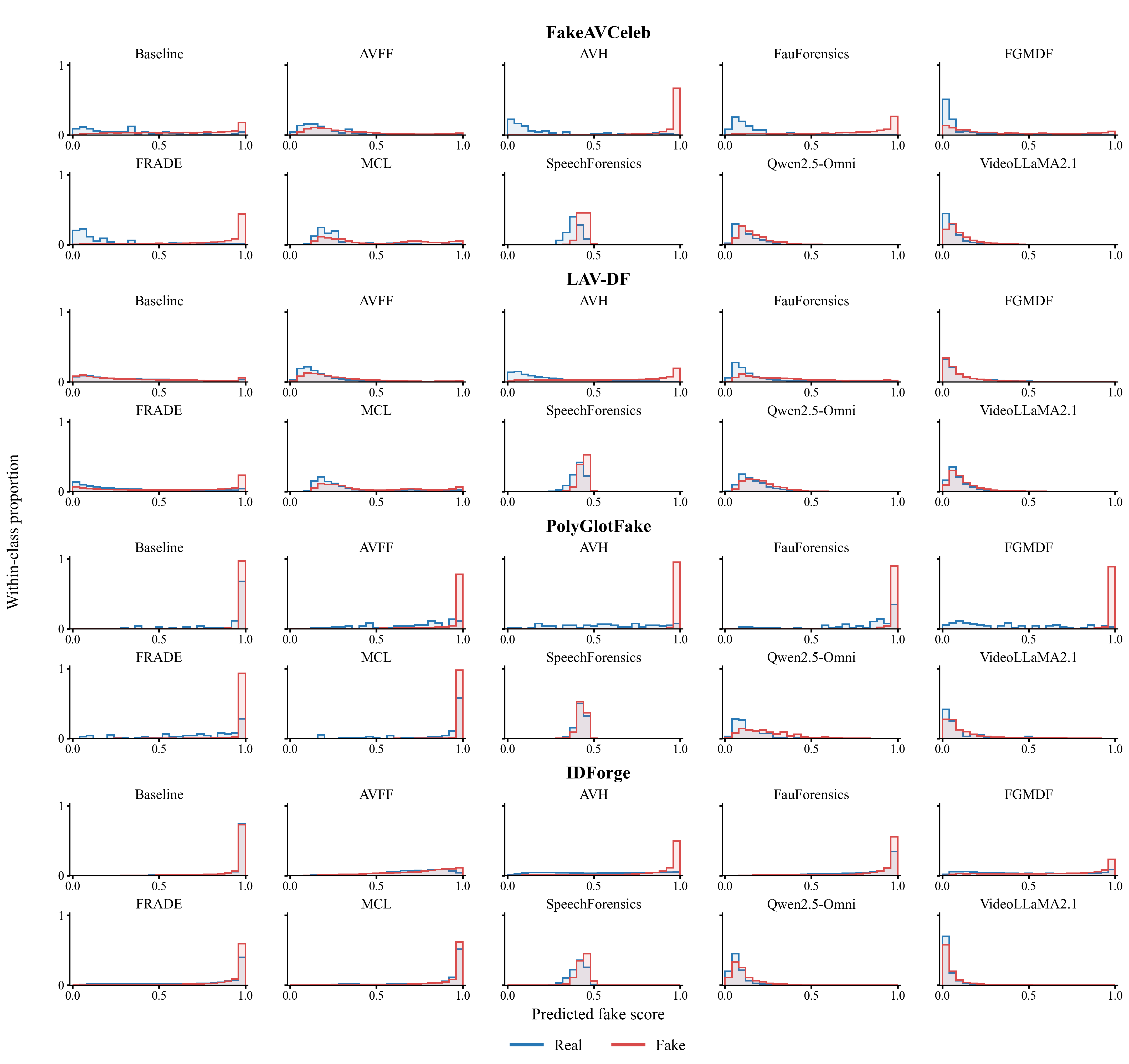}
    \caption{Clip-level prediction-score distributions on FakeAVCeleb, LAV-DF, PolyGlotFake, and IDForge. Each dataset occupies two rows. Blue and red denote real and fake clips, respectively; each class is normalized separately. The seven training-based detectors are trained on Mega-MMDF.}
    \label{fig:appendix-scores-01}
\end{figure*}

\begin{figure*}[t]
    \centering
    \includegraphics[width=\textwidth]{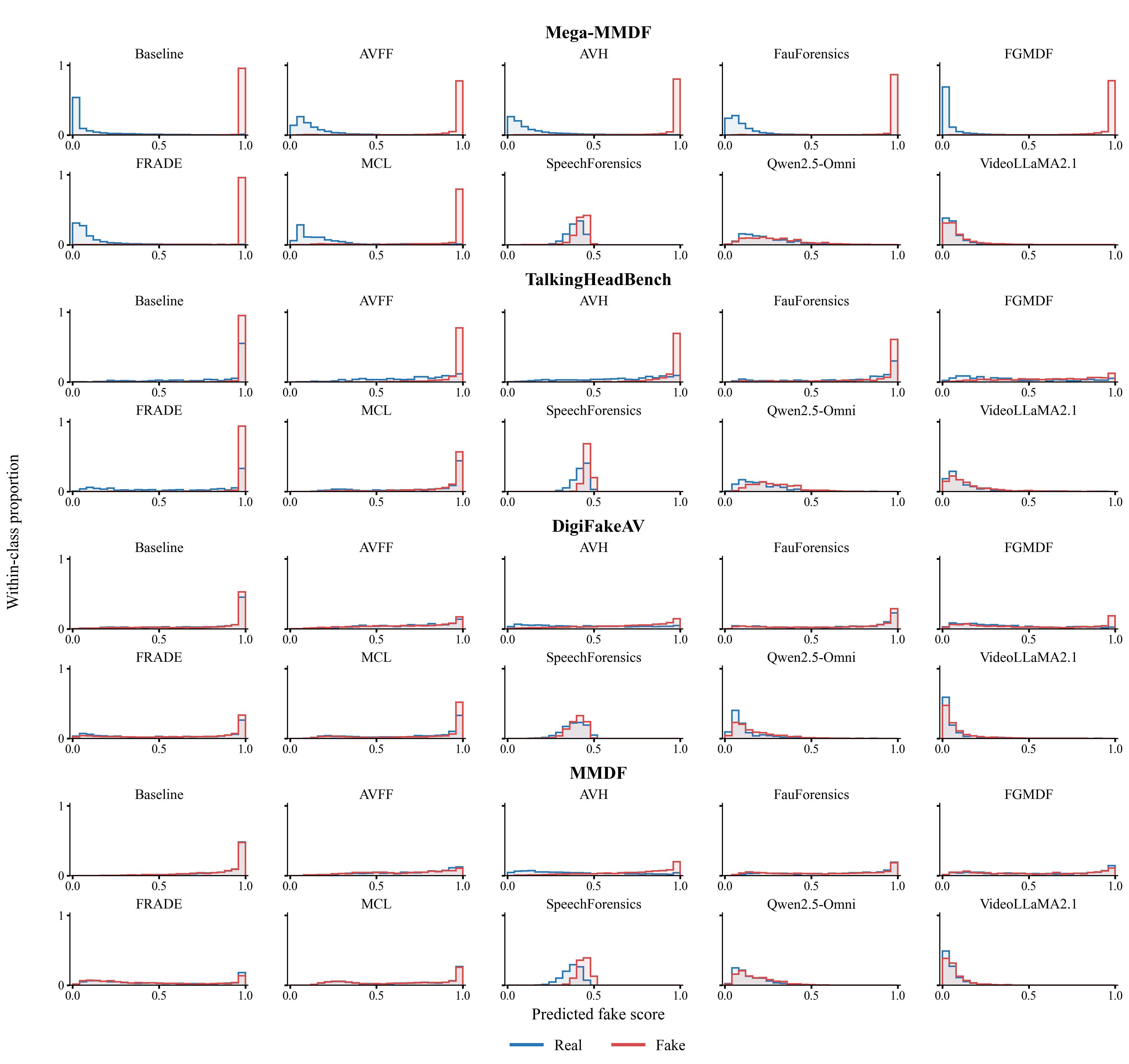}
    \caption{Clip-level prediction-score distributions on Mega-MMDF, TalkingHeadBench, DigiFakeAV, and MMDF. Each dataset occupies two rows. Blue and red denote real and fake clips, respectively; each class is normalized separately. The seven training-based detectors are trained on Mega-MMDF.}
    \label{fig:appendix-scores-02}
\end{figure*}

\begin{figure*}[t]
    \centering
    \includegraphics[width=\textwidth]{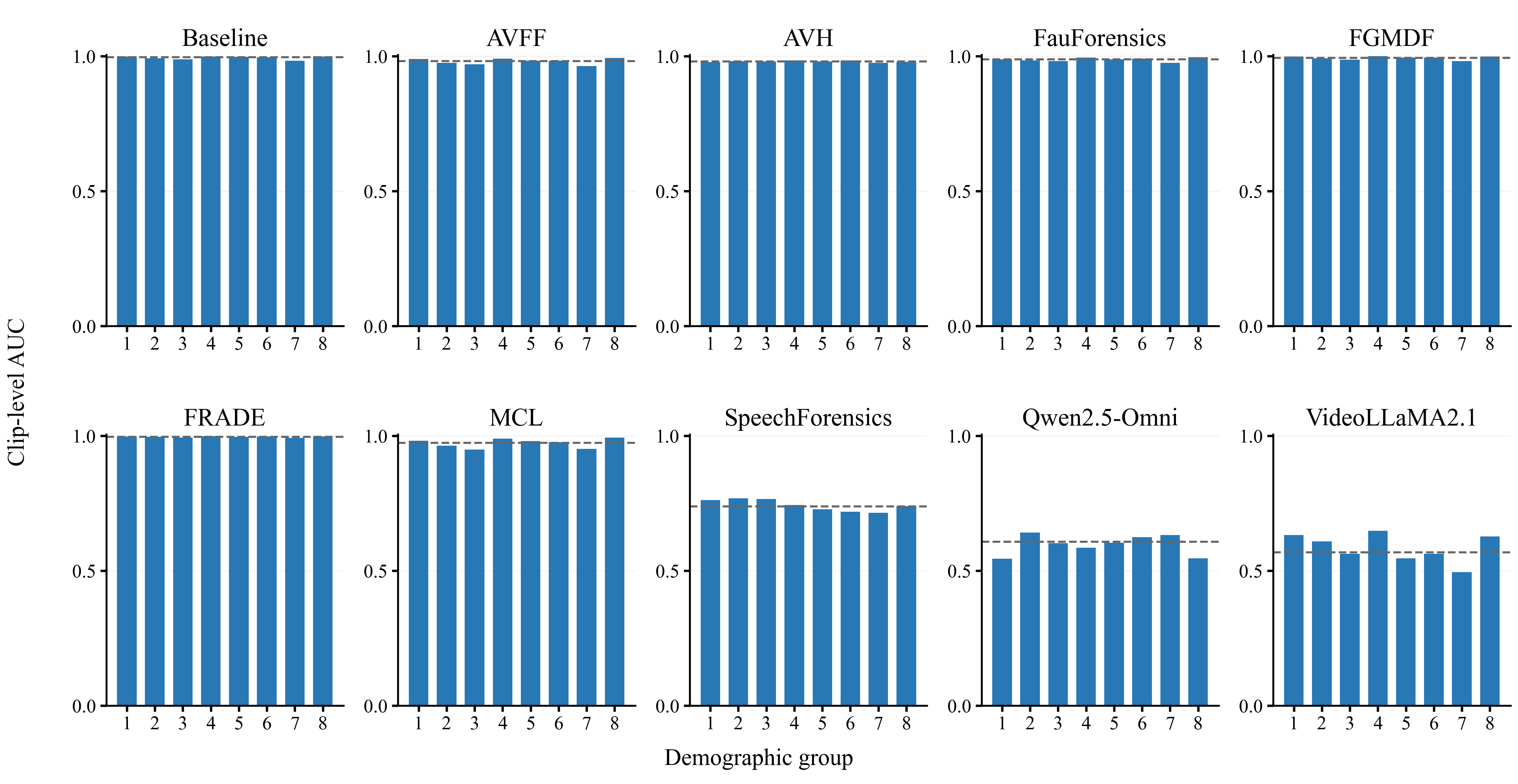}
    \caption{Clip-level AUC across demographic groups in Mega-MMDF. Dashed lines indicate overall AUC. Groups 1--8 are Female/Black, Female/Brown, Female/White, Female/Yellow, Male/Black, Male/Brown, Male/White, and Male/Yellow, respectively. The seven training-based detectors are trained on Mega-MMDF.}
    \label{fig:appendix-demographic-auc-Mega-MMDF}
\end{figure*}

\paragraph{Additional Prediction Results}
Figure~\ref{fig:appendix-roc-pr} presents clip-level ROC and precision--recall curves on the eight test datasets. Figures~\ref{fig:appendix-scores-01} and~\ref{fig:appendix-scores-02} show prediction-score distributions. Figure~\ref{fig:appendix-demographic-auc-Mega-MMDF} reports AUC across the eight demographic groups in Mega-MMDF. These figures compare the seven Mega-MMDF-trained detectors and three training-free methods.

\begin{figure*}[t]
    \centering
    \includegraphics[width=\textwidth]{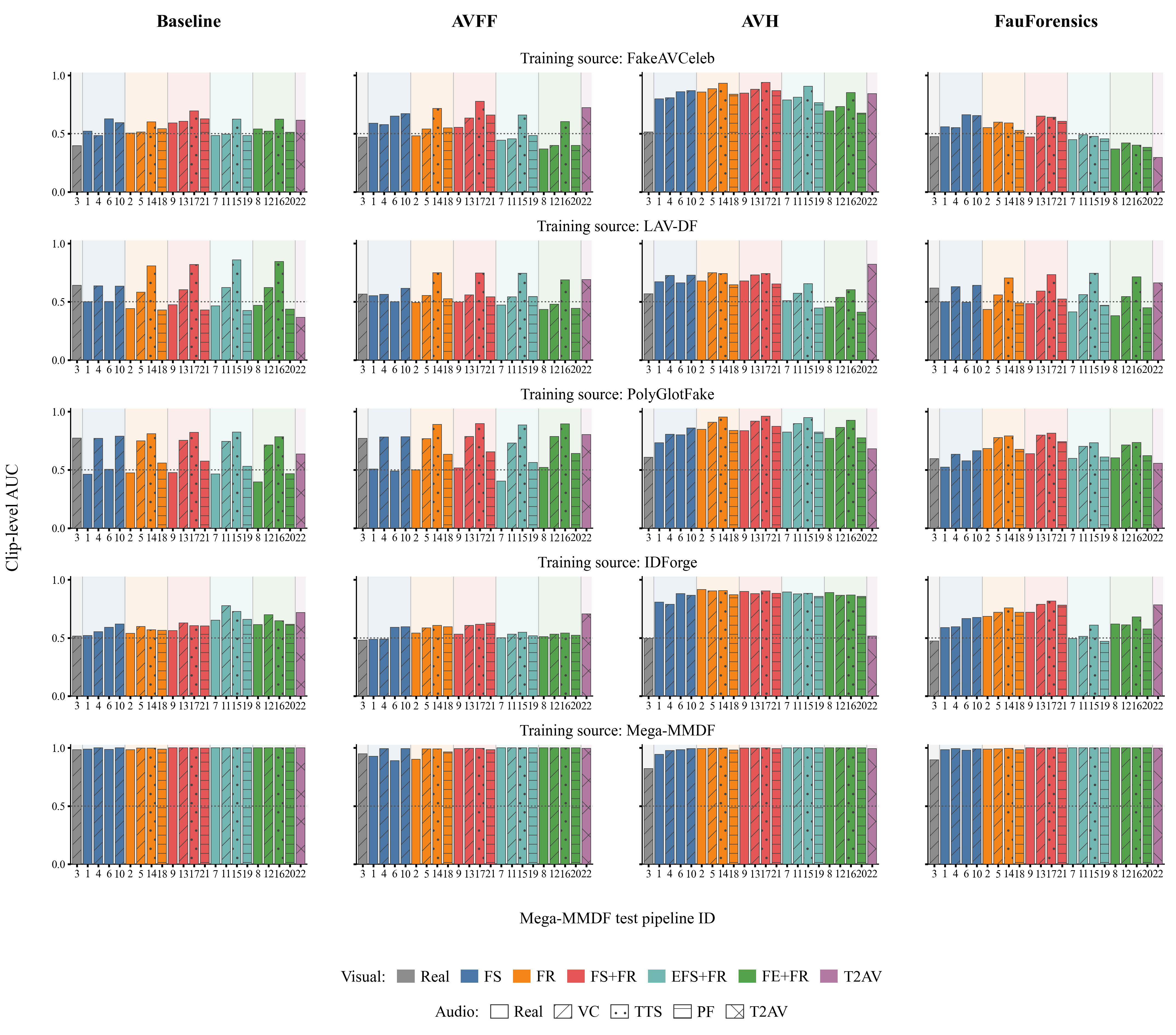}
    \caption{Pipeline-wise clip-level AUC for Baseline, AVFF, AVH, and FauForensics. Rows indicate training datasets and columns indicate detectors. Colors encode visual forgery groups and hatches encode audio forgery groups. The FS color includes EFS+FS pipelines P6 and P10. Dotted lines indicate AUC = 0.5.}
    \label{fig:appendix-pipelines-01}
\end{figure*}

\begin{figure*}[t]
    \centering
    \includegraphics[width=\textwidth]{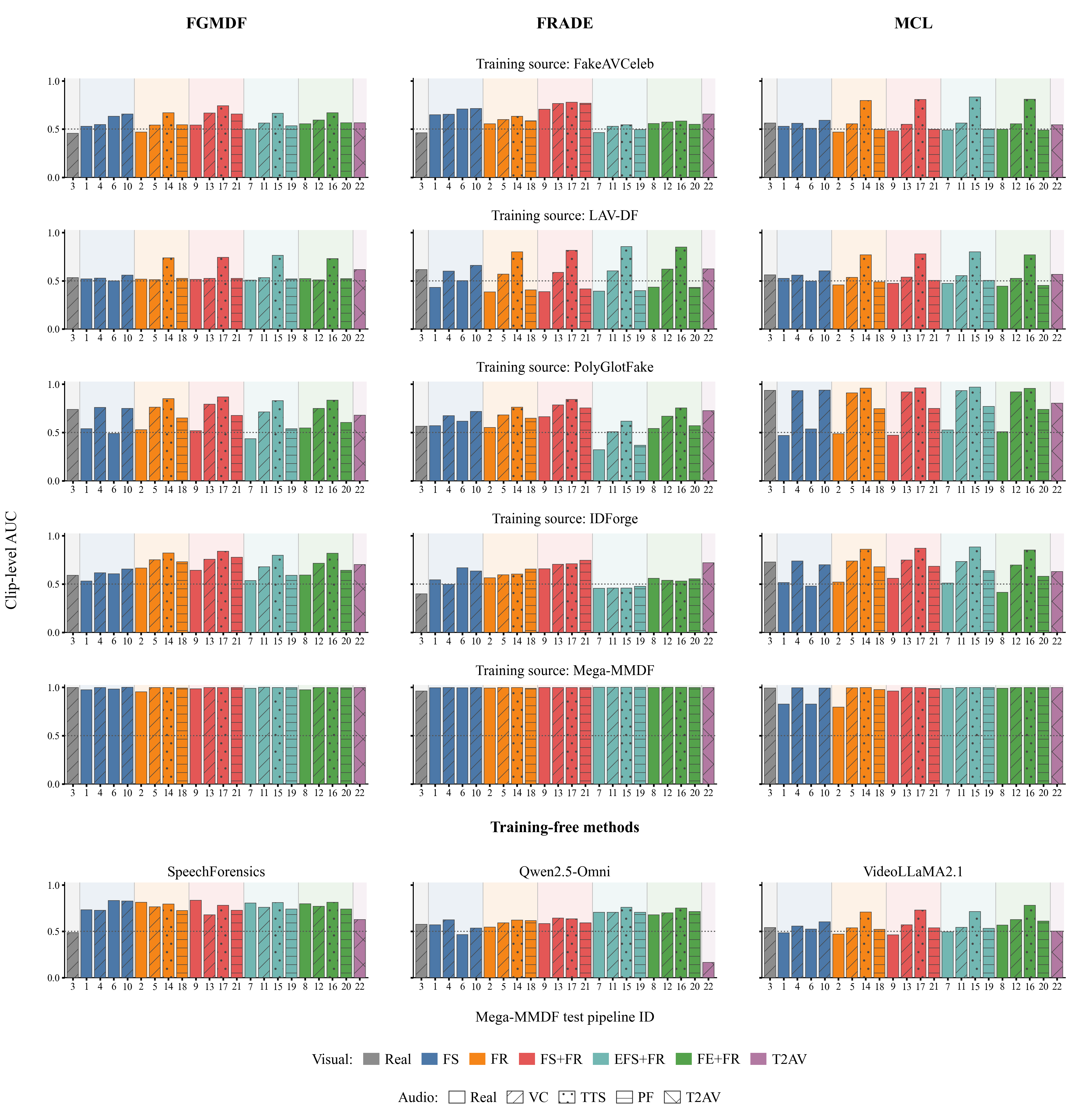}
    \caption{Pipeline-wise clip-level AUC for FGMDF, FRADE, MCL, and three training-free methods. The first five rows indicate training datasets for FGMDF, FRADE, and MCL; the final row presents the training-free methods. Colors encode visual forgery groups and hatches encode audio forgery groups. The FS color includes EFS+FS pipelines P6 and P10. Dotted lines indicate AUC = 0.5.}
    \label{fig:appendix-pipelines-02}
\end{figure*}

Figures~\ref{fig:appendix-pipelines-01} and~\ref{fig:appendix-pipelines-02} provide the complete pipeline-wise clip-level AUC results, extending the representative AVH results in the main paper. Across detectors, training on existing datasets produces uneven performance across pipelines, with the weakest pipelines depending on the detector and training source. Mega-MMDF training yields stronger and more consistent performance across these pipelines. The training-free methods also exhibit pipeline-dependent detection gaps.

\begin{figure*}[t]
    \centering
    \includegraphics[width=\textwidth]{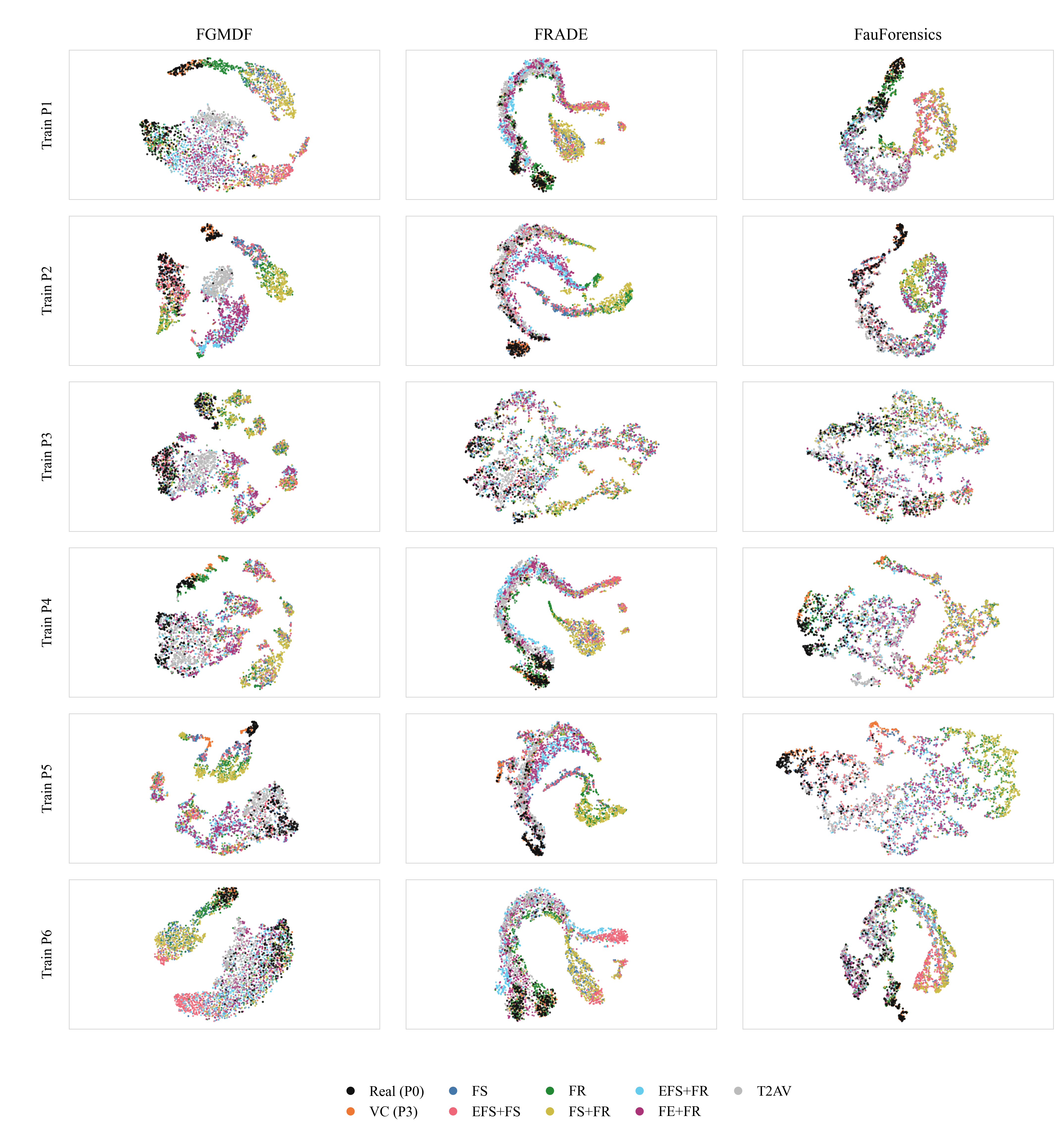}
    \caption{Instance-level fusion-feature t-SNE after training on P1--P6. Columns correspond to FGMDF, FRADE, and FauForensics; rows identify training pipelines. Colors distinguish real P0, audio-only VC P3, and the visual forgery groups. Each graph includes instances from all 22 fake test pipelines and P0.}
    \label{fig:appendix-tsne-01}
\end{figure*}

\begin{figure*}[t]
    \centering
    \includegraphics[width=\textwidth]{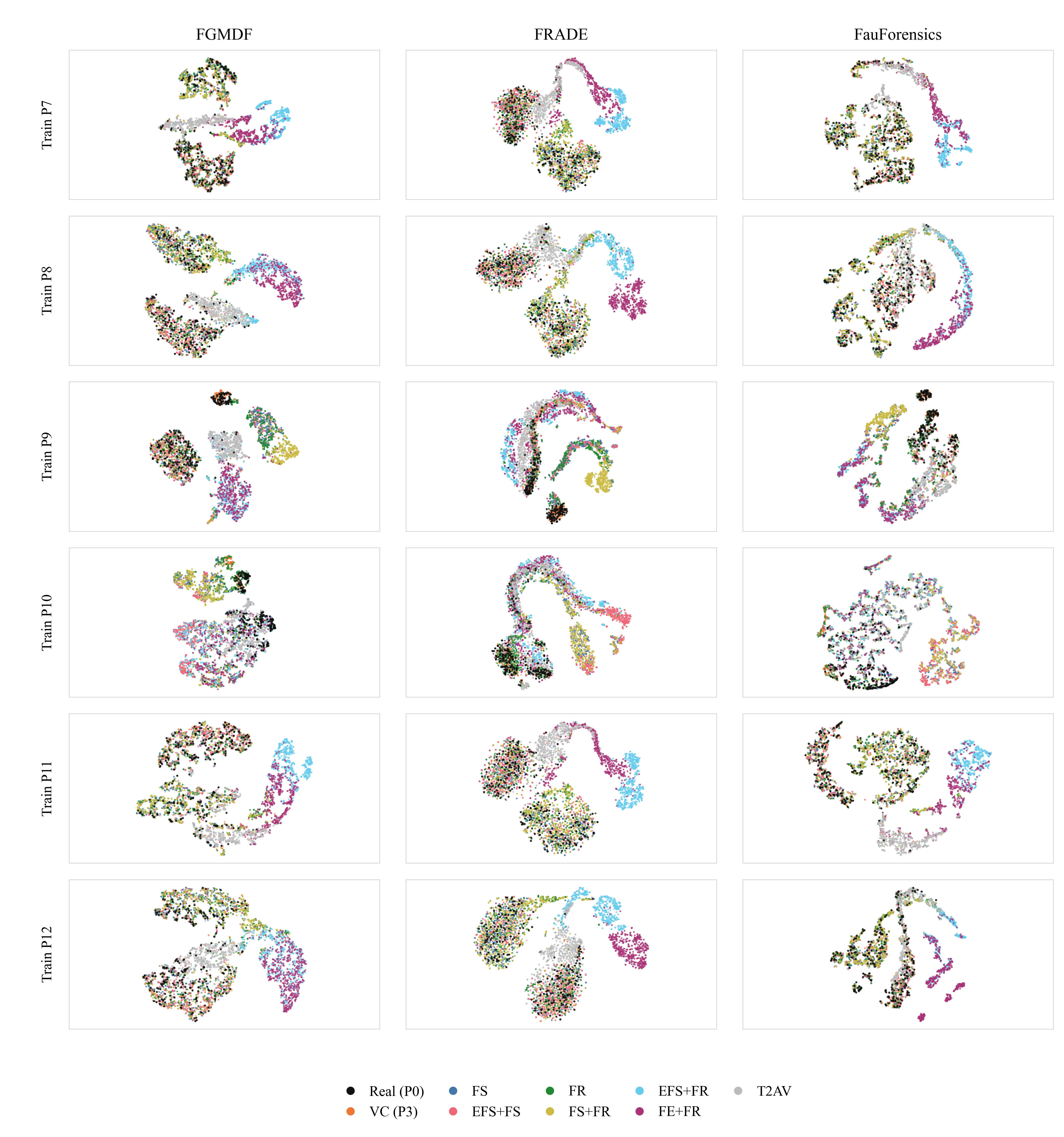}
    \caption{Instance-level fusion-feature t-SNE after training on P7--P12. Columns correspond to FGMDF, FRADE, and FauForensics; rows identify training pipelines. Colors distinguish real P0, audio-only VC P3, and the visual forgery groups. Each graph includes instances from all 22 fake test pipelines and P0.}
    \label{fig:appendix-tsne-02}
\end{figure*}

\begin{figure*}[t]
    \centering
    \includegraphics[width=\textwidth]{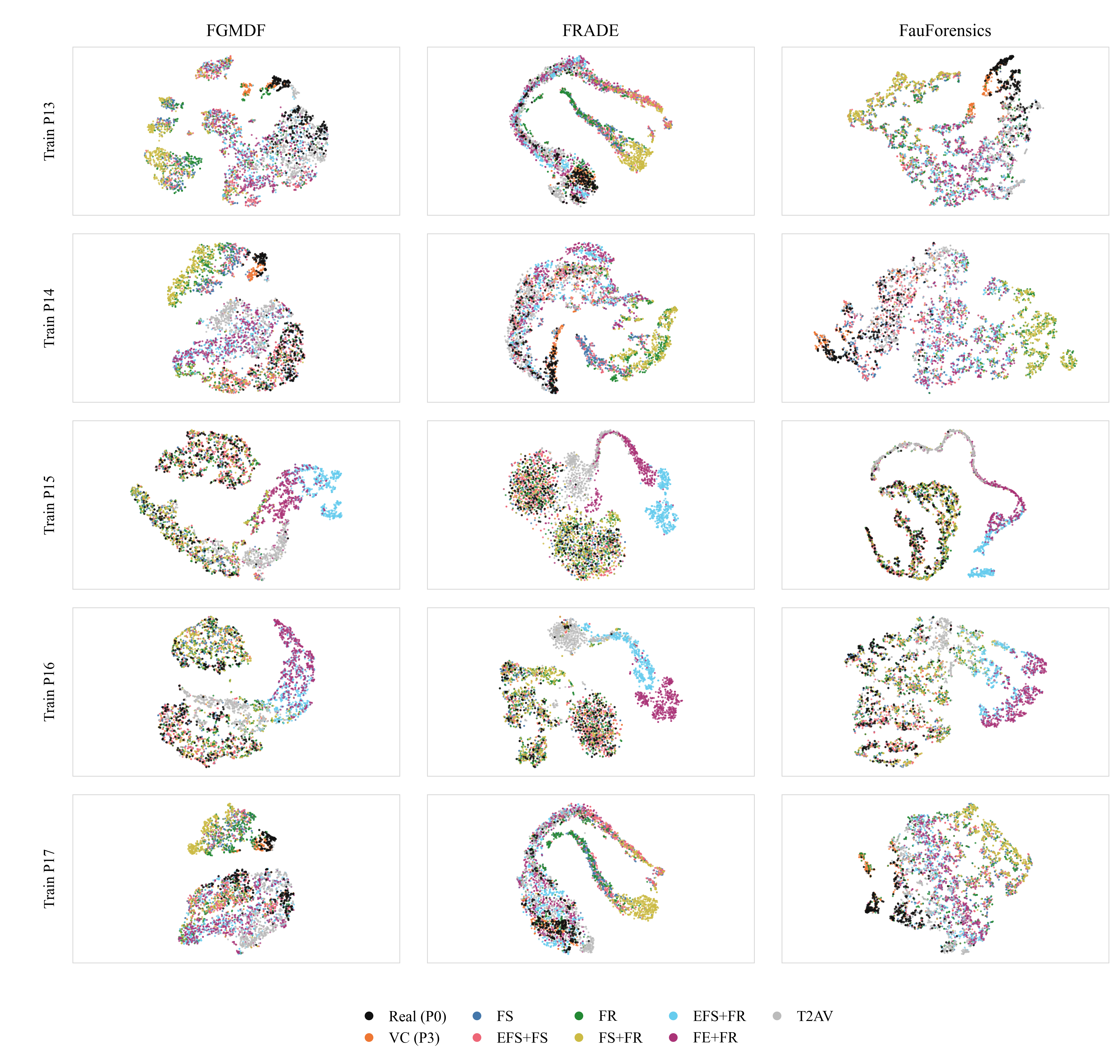}
    \caption{Instance-level fusion-feature t-SNE after training on P13--P17. Columns correspond to FGMDF, FRADE, and FauForensics; rows identify training pipelines. Colors distinguish real P0, audio-only VC P3, and the visual forgery groups. Each graph includes instances from all 22 fake test pipelines and P0.}
    \label{fig:appendix-tsne-03}
\end{figure*}

\begin{figure*}[t]
    \centering
    \includegraphics[width=\textwidth]{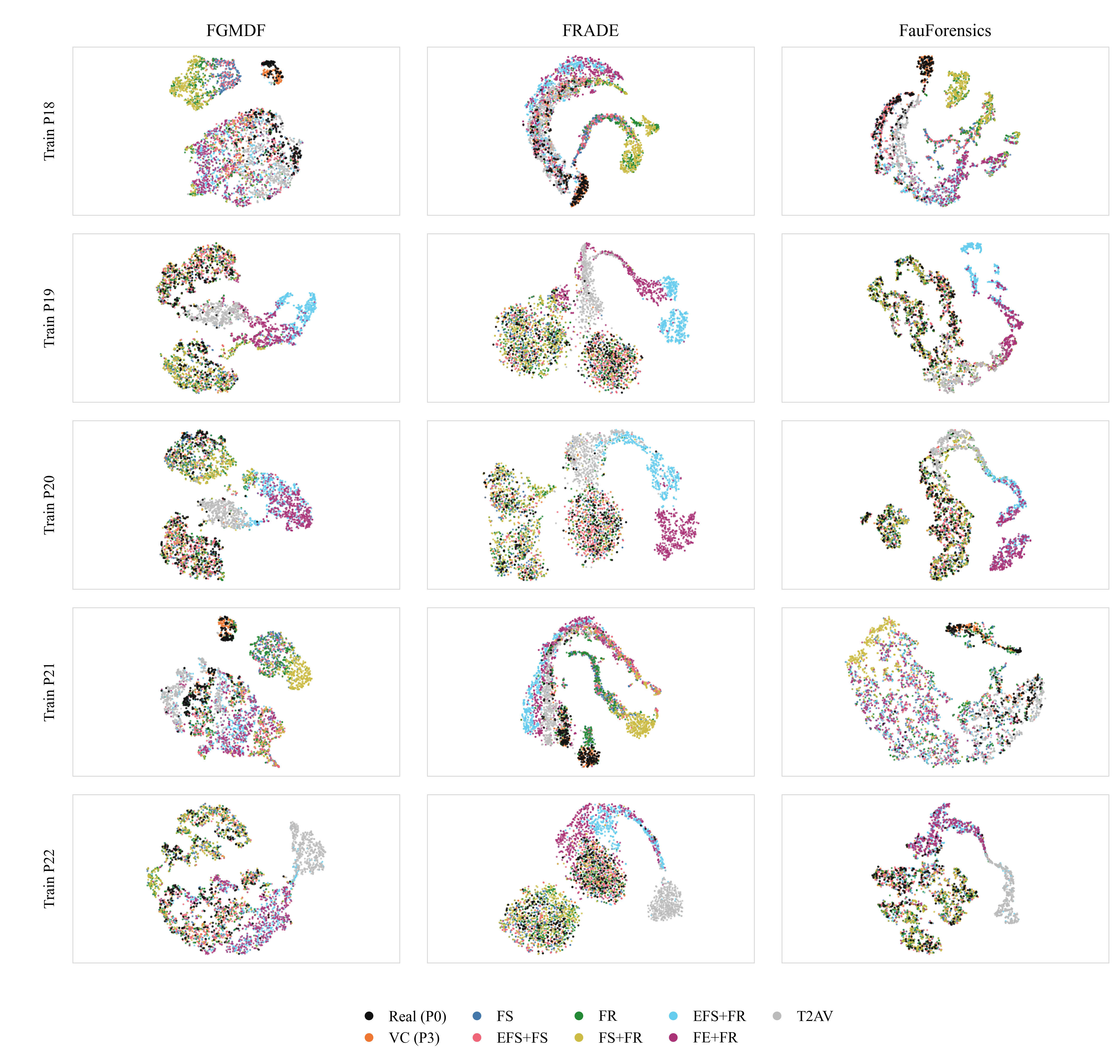}
    \caption{Instance-level fusion-feature t-SNE after training on P18--P22. Columns correspond to FGMDF, FRADE, and FauForensics; rows identify training pipelines. Colors distinguish real P0, audio-only VC P3, and the visual forgery groups. Each graph includes instances from all 22 fake test pipelines and P0.}
    \label{fig:appendix-tsne-04}
\end{figure*}

\paragraph{Additional Cross-Pipeline Feature Visualizations}
Figures~\ref{fig:appendix-tsne-01}--\ref{fig:appendix-tsne-04} provide fusion-feature t-SNE visualizations for FGMDF, FRADE, and FauForensics trained separately on P1--P22, covering all 66 checkpoints. The degree of separation between fake and real instances varies across forgery groups, training pipelines, and detectors. These feature distributions qualitatively complement the selective and asymmetric cross-pipeline generalization patterns discussed in the main paper.

\clearpage

\clearpage
% Full-width pipeline samples from pipeline_sample; no added row numbers.
% 10 panels/page at the current IEEEtran text width; 11 exceed the page.
% Preserve all source pixels and use the images' own white margins as separators.
\begingroup
\makeatletter
\@dblfptop=0pt
\@dblfpbot=0pt plus 1fil
\makeatother
\newcommand{\MegaSampleRow}[1]{%
    \hbox{\includegraphics[width=\textwidth]{images/pipeline_sample/#1_audio_video_presentation_rgb.jpg}}%
}
\setlength{\abovecaptionskip}{2pt}
\begin{figure*}[p]
    \centering
    \vbox{\offinterlineskip
        \MegaSampleRow{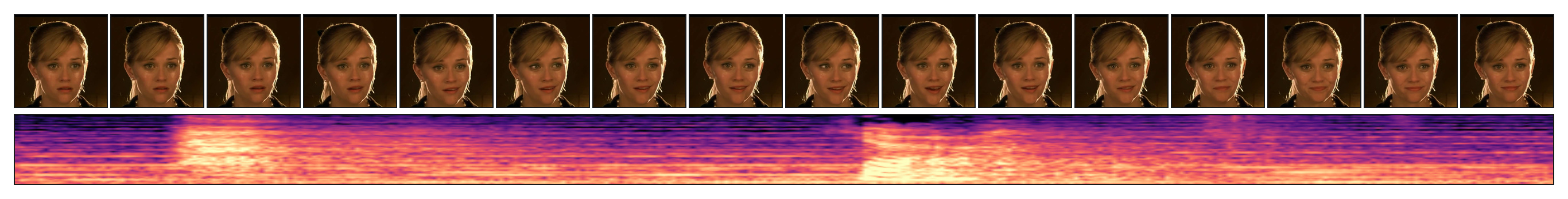}
        \MegaSampleRow{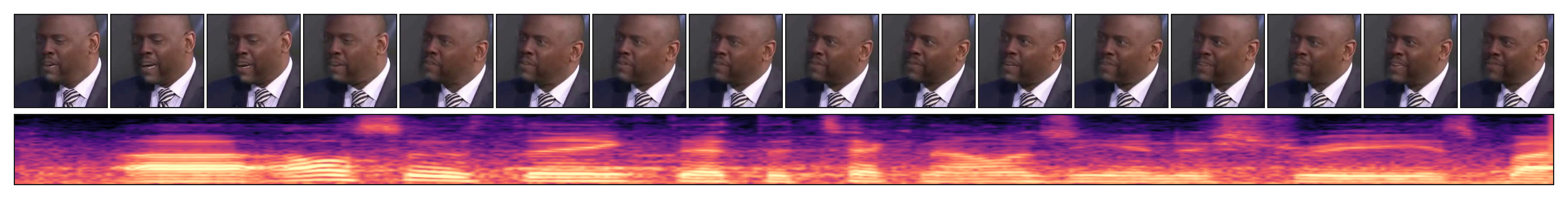}
        \MegaSampleRow{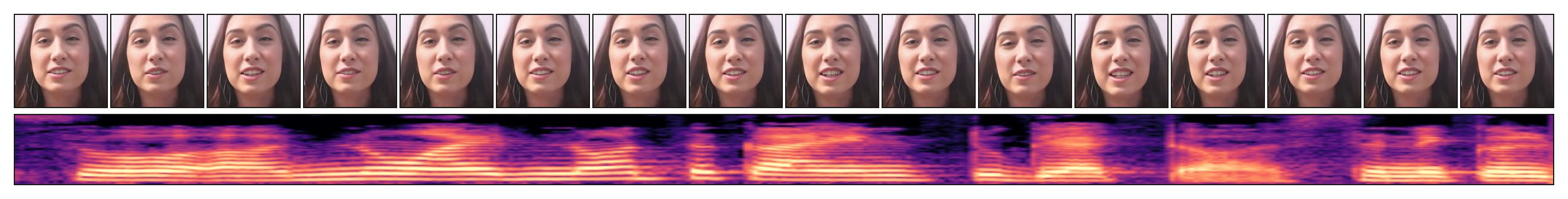}
        \MegaSampleRow{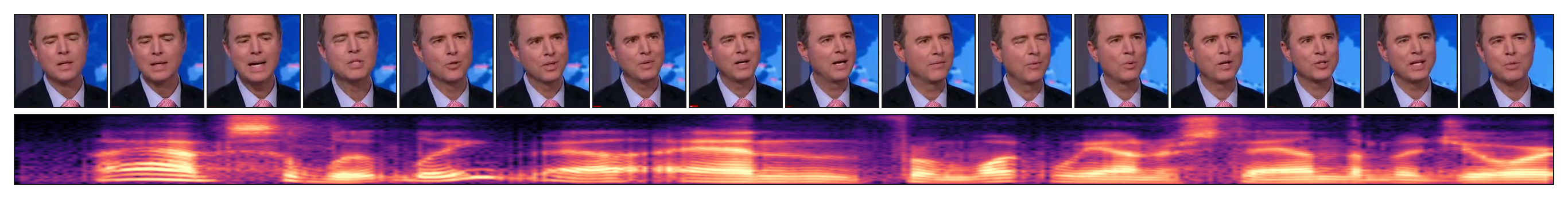}
        \MegaSampleRow{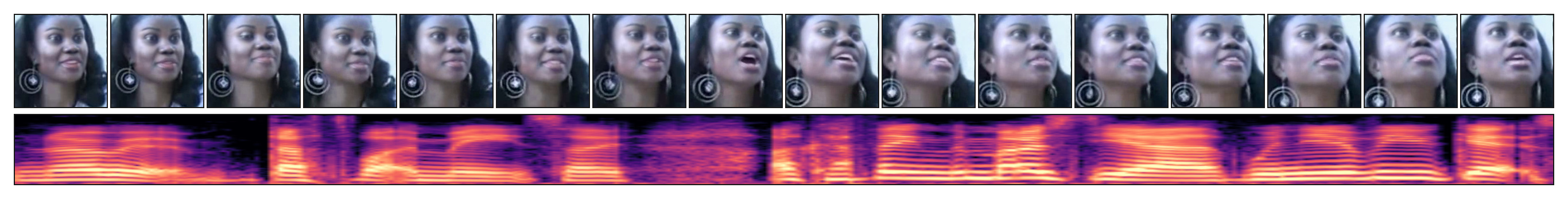}
        \MegaSampleRow{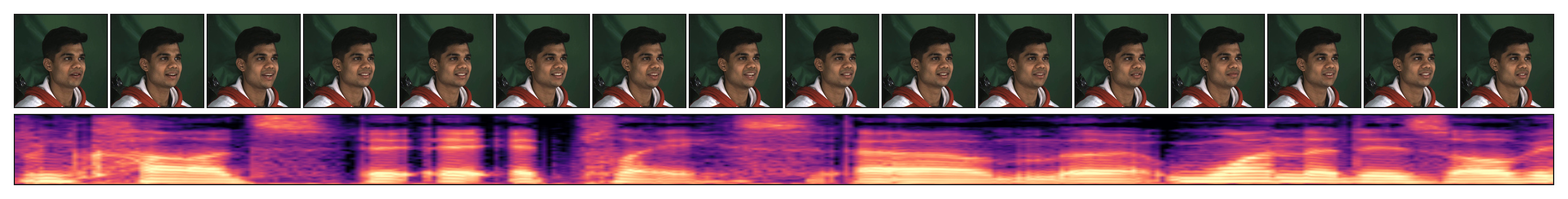}
        \MegaSampleRow{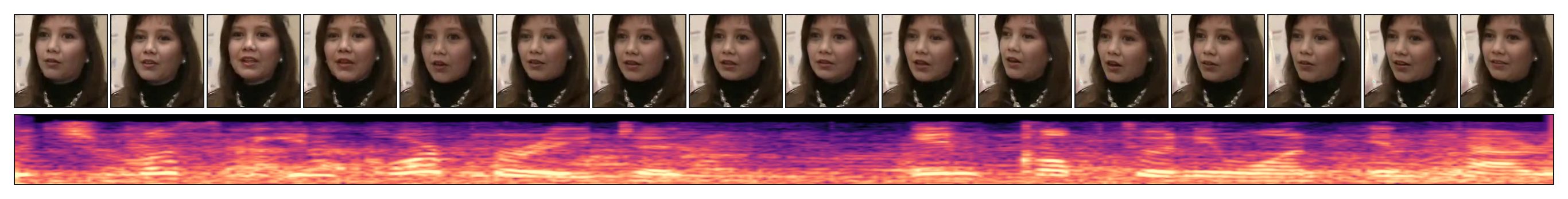}
        \MegaSampleRow{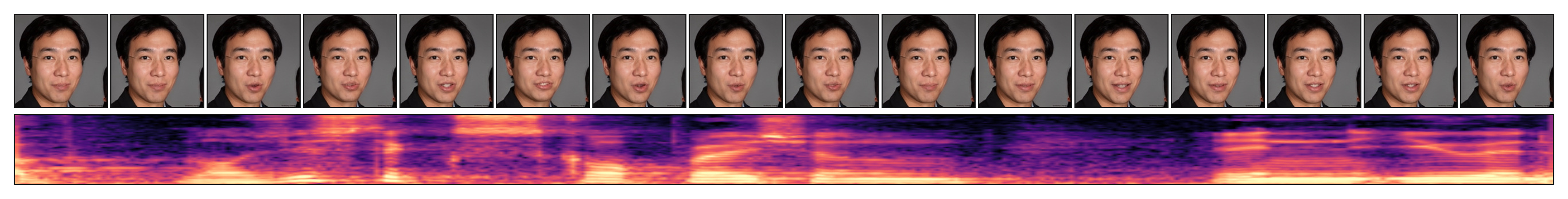}
        \MegaSampleRow{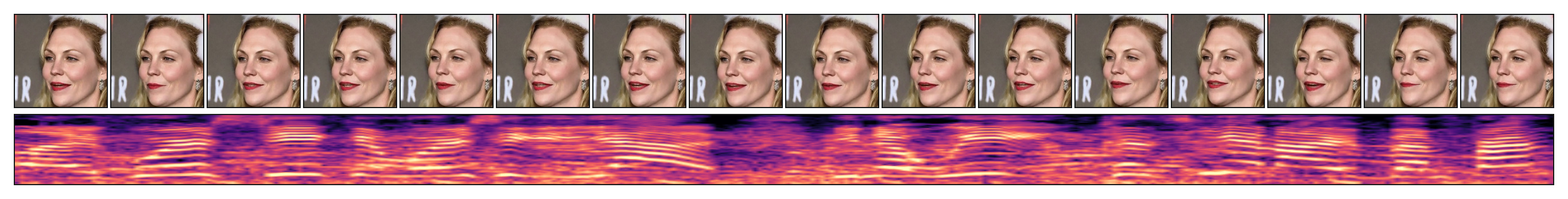}
        \MegaSampleRow{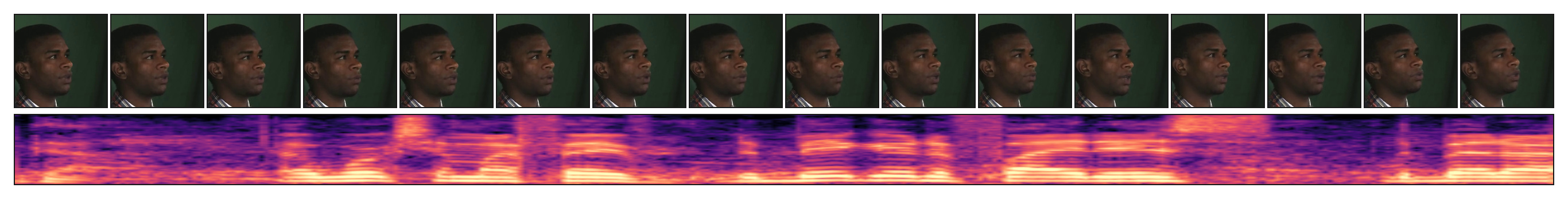}
    }
    \caption{Visualization of clips from our Mega-MMDF dataset (pipelines 0--9, top to bottom).}
    \label{fig:pipeline_samples_0_9}
\end{figure*}

\begin{figure*}[p]
    \centering
    \vbox{\offinterlineskip
        \MegaSampleRow{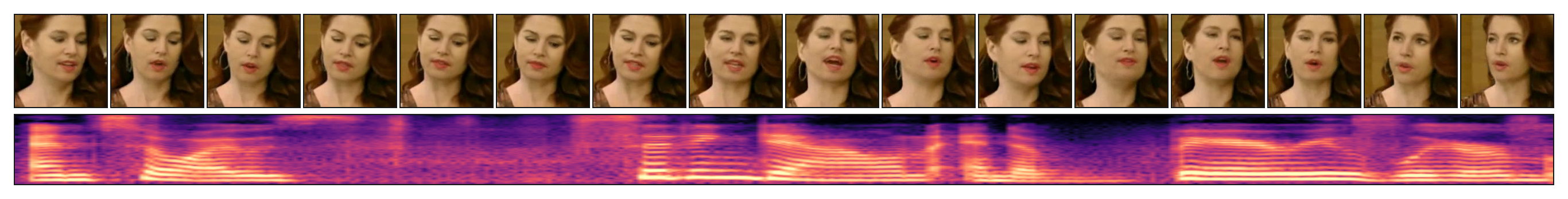}
        \MegaSampleRow{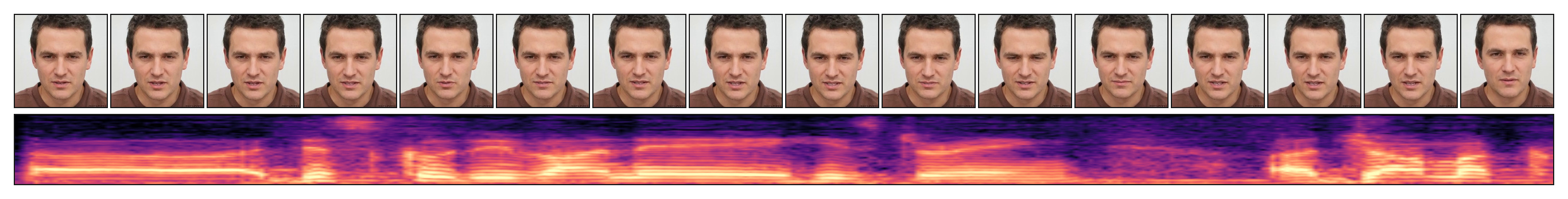}
        \MegaSampleRow{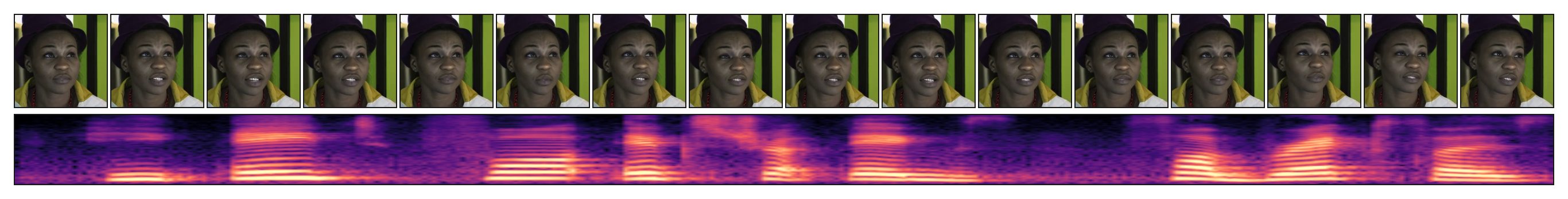}
        \MegaSampleRow{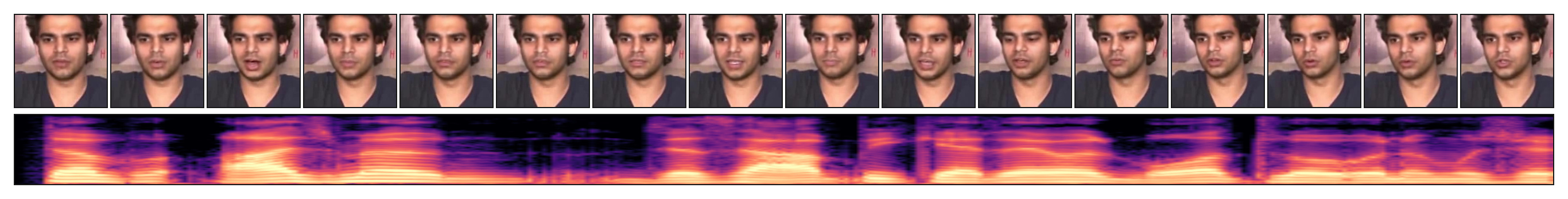}
        \MegaSampleRow{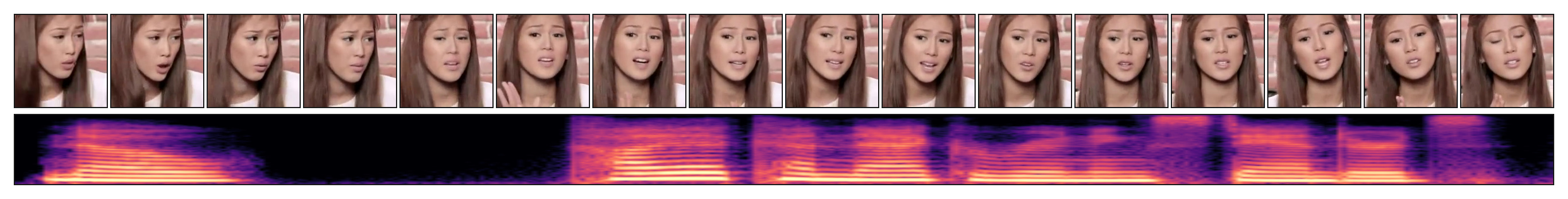}
        \MegaSampleRow{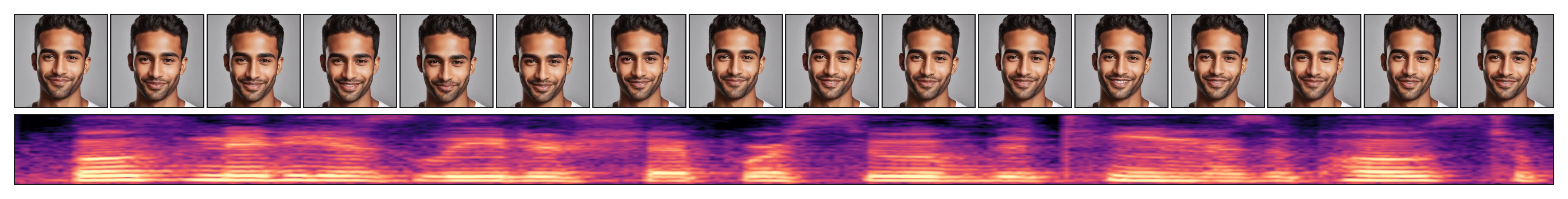}
        \MegaSampleRow{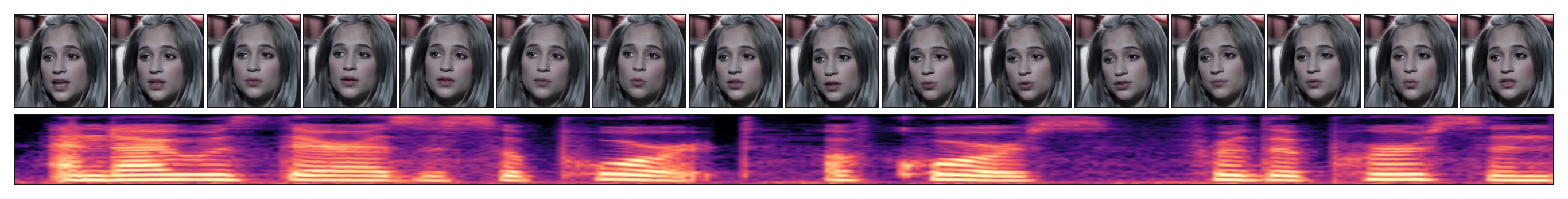}
        \MegaSampleRow{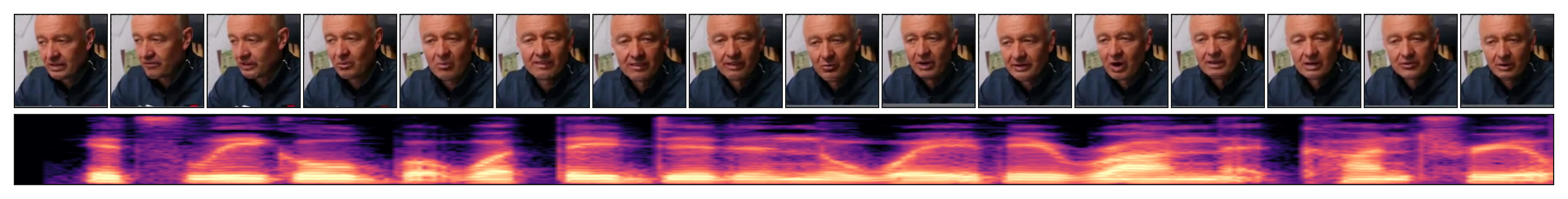}
        \MegaSampleRow{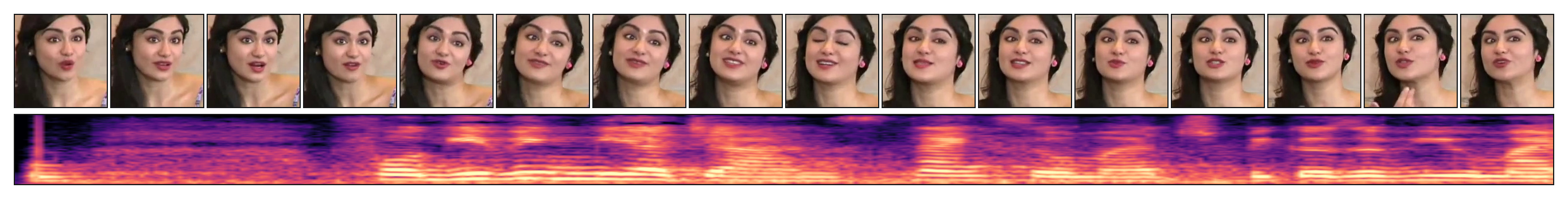}
        \MegaSampleRow{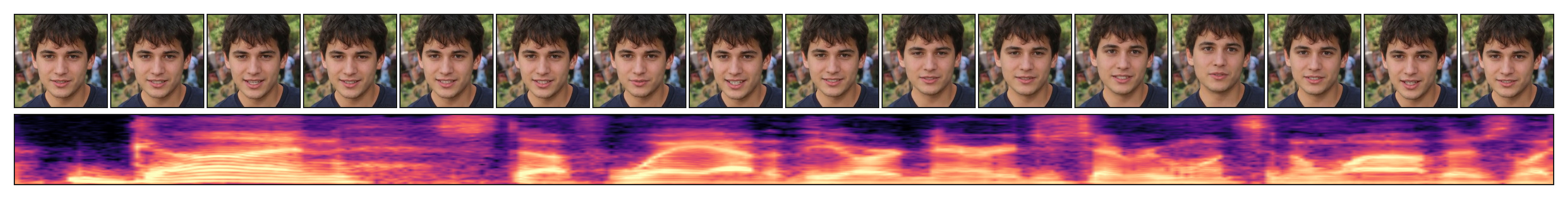}
    }
    \caption{Visualization of clips from our Mega-MMDF dataset (pipelines 10--19, top to bottom).}
    \label{fig:pipeline_samples_10_19}
\end{figure*}

\begin{figure*}[p]
    \centering
    \vbox{\offinterlineskip
        \MegaSampleRow{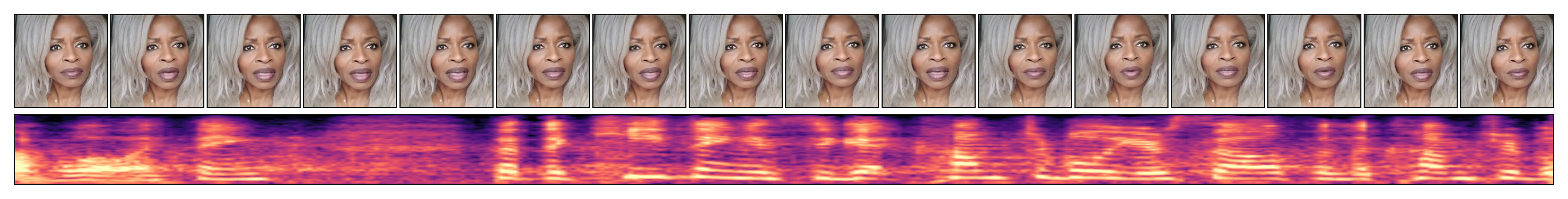}
        \MegaSampleRow{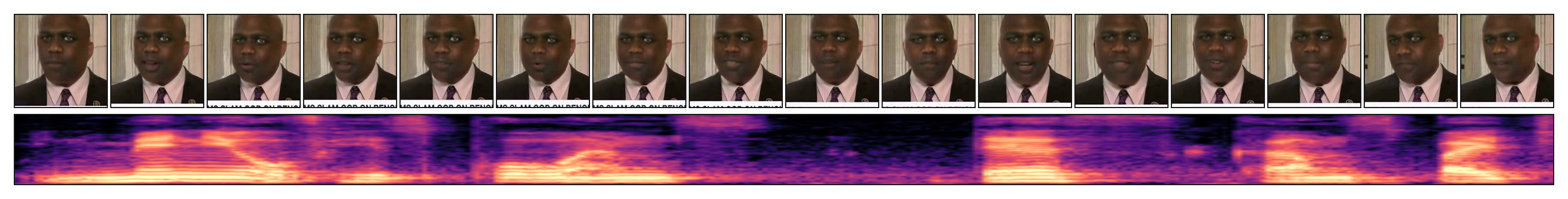}
        \MegaSampleRow{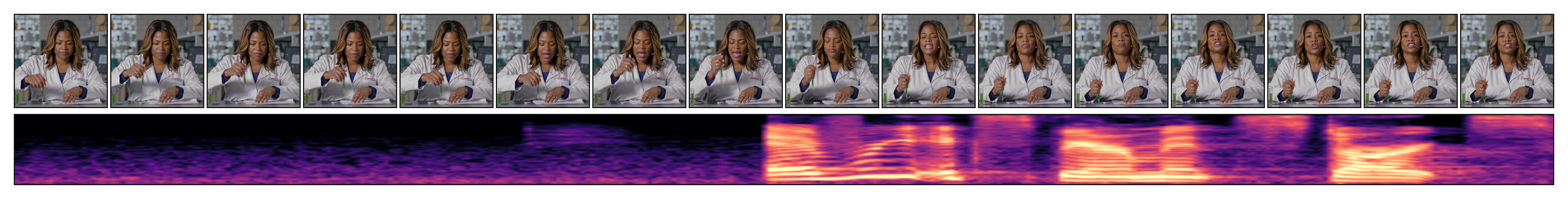}
    }
    \caption{Visualization of clips from our Mega-MMDF dataset (pipelines 20--22, top to bottom).}
    \label{fig:pipeline_samples_20_22}
\end{figure*}

\clearpage
\endgroup

\end{document}